\documentclass[a4paper,11pt]{article}
\usepackage[colorlinks=false]{hyperref}

\usepackage{amsmath, bm, amsthm, amssymb, mathrsfs}
\usepackage{amsfonts}
\usepackage{xcolor}
\usepackage{graphicx}
\usepackage{booktabs, multirow}
\usepackage{enumitem}
\usepackage{array} 
\usepackage{caption} 

\usepackage[subrefformat=parens,labelformat=parens]{subfig}
\usepackage{float}
\usepackage{bm}
\usepackage{algpseudocode,algorithm}
\usepackage{xfrac}

\graphicspath{{Graphics/}}

\title{Bayesian model inversion using stochastic spectral embedding}

\author{P.-R. Wagner, S. Marelli, B. Sudret}

\date{14.05.2020}

\usepackage{natbib}
\usepackage[left=25mm,right=25mm,top=1.5cm,bottom=1.5cm,includeheadfoot]{geometry}
\usepackage{bayesnotations}

\newcommand{\trans}{^{\intercal}}
\newcommand{\card}[1]{\mathrm{card}(#1)}
\newcommand{\expc}[1]{\mathbb{E}\left[#1\right]}

\DeclareMathOperator*{\argmax}{arg\,max}

\DeclareMathOperator*{\jsd}{JSD}
\newcommand{\cA}{\ensuremath{\mathcal{A}}}
\newcommand{\cD}{\ensuremath{\mathcal{D}}}
\newcommand{\cR}{\ensuremath{\mathcal{R}}}
\newcommand{\cK}{\ensuremath{\mathcal{K}}}
\newcommand{\cT}{\ensuremath{\mathcal{T}}}
\newcommand{\cV}{\ensuremath{\mathcal{V}}}
\newcommand{\cL}{\ensuremath{\mathcal{L}}}
\newcommand{\cE}{\ensuremath{\mathcal{E}}}

\newcommand{\cM}{\ensuremath{\mathcal{M}}}

\newcommand{\cX}{\ensuremath{\mathcal{X}}}
\newcommand{\cY}{\ensuremath{\mathcal{Y}}}

\newcommand{\vx}{\ve{x}}
\newcommand{\vX}{\ve{X}}

\newcommand{\indfun}[1]{\ve{1}_{#1}}
\newcommand{\Ntot}{\ensuremath{N_{\mathrm{ED}}}}

\newcommand{\NRefine}{\ensuremath{N_{\mathrm{ref}}}}

\newcommand{\wrt}{{\em w.r.t.~}}

\newcommand{\wmk}{\text{W/m/K}}
\newcommand{\wmq}{\text{W/m}^2}
\newcommand{\K}{\text{K}}

\begin{document}
	
\maketitle

\abstract{In this paper we propose a new \emph{sampling-free} approach to solve Bayesian model inversion problems that is an extension of the previously proposed \emph{spectral likelihood expansions} (SLE) method. Our approach, called \emph{stochastic spectral likelihood embedding} (SSLE), uses the recently presented \emph{stochastic spectral embedding} (SSE) method for \emph{local} spectral expansion refinement to approximate the likelihood function at the core of Bayesian inversion problems.
	
	We show that, similar to SLE, this approach results in analytical expressions for key statistics of the Bayesian posterior distribution, such as \emph{evidence, posterior moments} and \emph{posterior marginals}, by direct post-processing of the expansion coefficients. Because SSLE and SSE rely on the direct approximation of the likelihood function, they are in a way independent of the computational/mathematical complexity of the forward model. We further enhance the efficiency of SSLE by introducing a likelihood specific adaptive sample enrichment scheme.
	
	To showcase the performance of the proposed SSLE, we solve three problems that exhibit different kinds of complexity in the likelihood function: multimodality, high posterior concentration and high nominal dimensionality. We demonstrate how SSLE significantly improves on SLE, and present it as a promising alternative to existing inversion frameworks.\\

\textbf{Keywords}: Bayesian model inversion, inverse problems, polynomial chaos expansions, spectral likelihood expansions, stochastic spectral likelihood embedding, sampling-free inversion.
}

\section{Introduction}
Computational models are an invaluable tool for decision making, scientific advances and engineering breakthroughs. They establish a connection between a set of input parameters and output quantities with wide-ranging applications. \emph{Model inversion} uses available experimental observations of the output to determine the set of input parameters that maximize the predictive potential of a model. The importance of efficient and reliable model inversion 
frameworks can hardly be overstated, considering that they establish a direct connection between models and the real world. Without it, the most advanced model predictions might lack physical meaning and, consequently, be useless for their intended applications.

Bayesian model inversion is one way to formalize this problem 
\citep{Bayesian:Jaynes2003, Bayesian:Gelman2014:3rd}. It is based on Bayesian 
inference and poses the problem in a 
probabilistic setting by capitalizing on Bayes' theorem. 
In this setting a so-called \emph{prior} ({\em i.e.}, before 
observations) probability distribution about the model parameters is updated 
to a so-called \emph{posterior} ({\em i.e.}, after observations) distribution. 
The posterior distribution is the 
probability distribution of the input parameters conditioned on the available
observations, and the main outcome of the Bayesian inversion process. 

In Bayesian model inversion, the connection between the model output and the 
observations is established 
through a probabilistic \emph{discrepancy model}. This model, which is a function 
of the input parameters, leads to the so-called \emph{likelihood} function. 
The specific form of the likelihood function depends on the problem at hand, 
but typically it has a global maximum for the input parameters with the model 
output that is closest 
to the available observations (\wrt some metric), and rapidly goes to zero with 
increasing distance to those parameters. 

Analytical expressions for the posterior distribution can only be found in few academic examples ({\em e.g.}, conjugate priors with a linear forward model, \citet{ML:Bishop2006, Bayesian:Gelman2014:3rd}). 
In general model inversion problems, such analytical 
solutions are not available though. Instead, it is common practice to resort 
to sampling methods to generate a sample distributed according to the 
posterior distribution. 
The family of \emph{Markov chain  Monte Carlo} (MCMC) algorithms are particularly suitable for generating such a \emph{posterior sample} \citep{MCMC:Beck2002, Robert2004}.

While MCMC and its extensions are extensively used in model inversion, and new 
algorithms are continuously being developed \citep{MCMC:Haario2001, MCMC:Ching2007, MCMC:Goodman2010, MCMC:Neal2011}, it has a few notable shortcomings that hinder its application in many practical cases. 
It is well known that there are no robust convergence criteria for MCMC algorithms, and that their performance is particularly sensitive to their tuning parameters. 
Additionally, samples generated by MCMC algorithms are often highly correlated, 
thus requiring extensive heuristic post-processing and 
empirical rules \citep{MCMC:Gelman1992, MCMC:Brooks1998:Gelman}. 
MCMC algorithms are also in general not well suited for sampling multimodal posterior distributions.

When considering complex engineering scenarios, the models subject to inversion are often computationally expensive. 
Because MCMC algorithms usually require a significant number of forward model 
evaluations, it has been proposed to accelerate the procedure by using \emph{surrogate models} in lieu of the original models. 
These surrogate models are either constructed non-adaptively before sampling from the 
posterior distribution \citep{Marzouk2007, Marzouk2009} or adaptively during 
the sampling procedure \citep{Li2014, 	Birolleau2014, Cui2014, Conrad2016, Conrad2018, Yan2019}. 
Adaptive techniques can be of great benefit with posterior distributions that 
are concentrated in a small subspace of the prior domain, as the surrogate only 
needs to be accurate near high density areas of the posterior distribution.

\emph{Polynomial chaos expansions} (PCE) are a widely used 
surrogate modelling technique based on expanding the forward model onto a 
suitable polynomial basis \citep{Ghanembook1991, Xiu2002}. In 
other words, it provides a spectral representation of the computational forward model. 
Thanks to the introduction of sparse regression (see, e.g. \cite{BlatmanJCP2011}), its computation has become feasible even in the presence of complex and computationally expensive engineering models.
This technique has been successfully used in conjunction with MCMC to 
reduce the total computational costs associated with sampling from the posterior 
distribution \citep{Marzouk2007, Calibration:Wagner2020}.

Alternative approaches to compute the posterior distribution or its statistics 
include the \emph{Laplace approximation} at a posterior mode 
\citep{Bayesian:Tierney1986,Bayesian:Tierney1989:a,Bayesian:Tierney1989:b}, 
approximate Bayesian 
computations (ABC) \citep{Bayesian:Marin2012, Bayesian:Sisson2018}, optimal transport 
approaches \citep{Mapping:ElMoselhy2012, Mapping:Parno2015:PhD, 
	Mapping:Marzouk2016} and embarrassingly parallel quasi-Monte Carlo sampling 
\citep{QMC:Dick2017, QMC:Gantner2016}.

\emph{Stochastic spectral embedding} (SSE) is a metamodelling technique suitable for approximating functions with complex localized features recently developed in \citet{Marelli2020}. 
In this paper we propose to extend this technique with an ad-hoc adaptive sample enrichment strategy that makes it suitable to efficiently approximate likelihood functions in Bayesian model inversion problems. This method can be seen as a generalization of the previously proposed \emph{spectral likelihood expansions}
(SLE) approach \citep{NagelJCP2016}.

Due to its deep connection to SLE, we call the application of SSE to likelihood functions \emph{stochastic spectral likelihood embedding} (SSLE). 
We show that, due to its local spectral characteristics, this approach allows us to analytically derive expressions for the posterior marginals and general posterior moments by post-processing its expansion coefficients.

The paper is organized as follows: In Section~\ref{sec:modelCal} we establish the basics of Bayesian inference and particularly Bayesian model inversion. We then give an introduction into spectral function decomposition with a focus on polynomial chaos expansions and their application to likelihood functions (SLE) in Section~\ref{sec:specRepres}. In Section~\ref{sec:sse} we present the main contribution of the paper, namely the derivation of Bayesian posterior quantities of interest through SSLE and the extension of the SSE algorithm with an adaptive sampling strategy. Finally, in Section~\ref{sec:applications} we showcase the performance of our approach on three case studies of varying complexity.  
\section{Model inversion}
\label{sec:modelCal}
The problem of \emph{model inversion} occurs whenever the predictions of a model 
are to be brought into agreement with available observations or \emph{data}. 
This is achieved by properly  adjusting a set of \emph{input parameters} of the model. 
The goal of inversion can be twofold: on the one hand the inferred input parameters
might be used 
to predict new realizations of the model output. 
On the other hand, the inferred input parameters might be the  
main interest.
Model inversion is a common problem in many engineering disciplines, that in some 
cases is still routinely solved 
manually, {\em i.e.} by simply changing the input parameters until some, often qualitative, 
\emph{goodness-of-fit} criterion is met. 
More quantitative inversion approaches aim at automatizing this process, 
by establishing a metric ({\em e.g.}, $L^2$-distance) between the data and 
the model response, which is then minimized through suitable 
optimization 
algorithms.

While such approaches can often be used in practical applications, they tend not to provide measures of the 
\emph{uncertainties} associated with the inferred model input or predictions.
These uncertainties are useful in judging the accuracy of the inversion, as 
well as indicating non-informative measurements. In fact, the lack of 
uncertainty quantification in the context of model inversion can lead to 
erroneous results that have far-reaching consequences in subsequent 
applications. One approach to consider uncertainties in inverse problems is 
the \emph{Bayesian framework} for model inversion that will be presented hereinafter. 

\subsection{Bayesian inference}
Consider some \emph{non-observable} parameters $\BParams\in\cd_{\BParams}$ and 
the \emph{observables} $\BData\in\cd_{\BData}$. Furthermore, 
let ${\Bdata=\{\ve{y}^{(1)},\dots,\ve{y}^{(N)}\}}$ be a set 
of $N$ measurements, {\em i.e.}, noisy observations of a set of realizations of 
$\BData$. 
Statistical inference consists in drawing conclusions about $\BParams$ using the 
information from $\Bdata$ \citep{Bayesian:Gelman2014:3rd}. These measurements 
can be direct observations of the parameters ($\BData = \BParams$) or some 
quantities indirectly 
related to $\BParams$ through a function or \emph{model}
$\cm:\cd_{\BParams}\to\cd_{\BData}$. 
One way to conduct this inference is through Bayes' theorem of conditional probabilities, a process known as  \emph{Bayesian inference}.

Denoting by $\pi(\cdot)$ a probability density function (PDF) and by 
$\pi(\cdot\vert\Bparams)$ a PDF conditioned on $\Bparams$, Bayes' theorem can be written as
\begin{equation}
	\label{eq:modelCal:BayesTheorem}
	\pi(\Bparams\vert\Bdata) = 
	\frac{\pi(\Bdata\vert\Bparams)\Bprior}{\pi(\Bdata)},
\end{equation}
where $\Bprior$ is known as the \emph{prior} distribution of the parameters, 
{\em i.e.}, the distribution of $\BParams$ before observing the data $\Bdata$. 
%
The conditional distribution $\pi(\Bdata\vert\Bparams)$, known as \textit{likelihood}, establishes a connection 
between the observations $\Bdata$ and a realization of the parameters $\BParams = \Bparams$. 
For a given realization $\Bparams$, it returns the probability 
density of observing the data $\Bdata$. 
Under the common assumption of \emph{independence} between individual observations, 
$\{\ve{y}^{(i)}\}_{i=1}^N$, the likelihood function takes the form:
\begin{equation}
	\label{eq:modelCal:LikelihoodDef}
	\cl:\Bparams\mapsto\Blikelifun\eqdef 
	\pi(\Bdata\vert\Bparams) = \prod_{i=1}^N\pi(\ve{y}^{(i)}\vert\Bparams).
\end{equation}
The likelihood function is a map $\cd_{\BParams} \rightarrow \mathbb{R}_+$, and it
attains its maximum for the parameter set with the highest probability of 
yielding $\Bdata$. With this,  Bayes' theorem from 
Eq.~\eqref{eq:modelCal:BayesTheorem} can be rewritten as:
\begin{equation}
	\label{eq:modelCal:PosteriorDist}
	\Bpost = \frac{\Blikelifun\Bprior}{\Bevi}, \quad \text{with} \quad \Bevi = 
	\int_{\cd_{\BParams}}\Blikelifun\Bprior\,\di{\Bparams},
\end{equation}
where $Z$ is a  normalizing constant often called \emph{evidence} or 
\emph{marginal likelihood}. On the left-hand side, $\Bpost$ is the 
\emph{posterior} PDF, {\em 
	i.e.}, the distribution of $\BParams$ after observing data $\Bdata$. In this 
sense, Bayes' theorem establishes a general expression for updating the prior 
distribution using a likelihood function to incorporate information from the 
data.

\subsection{Bayesian model inversion}
\emph{Bayesian model inversion} describes the 
application of the Bayesian inference framework to the problem of model 
inversion \citep{Bayesian:Beck1998, Bayesian:Kennedy2001, 
	Bayesian:Jaynes2003, Inversion:Tarantola2005}. 
The two main ingredients needed to infer model parameters within the 
Bayesian framework are a prior distribution $\Bprior$ of the model 
parameters and a likelihood function $\cl$. 
%
In practical applications, \emph{prior} information about the model parameters 
is often readily available. 
Typical sources of such information are physical parameter 
constraints or expert knowledge. Additionally, prior inversion attempts can 
serve as guidelines to assign \emph{informative} prior distributions. In cases 
where no prior information about the parameters is available, so-called 
\emph{non-informative} or \emph{invariant} prior distributions 
\citep{Bayesian:Jeffreys1946, Bayesian:Harney2016} can also be assigned. 
%
The likelihood function serves instead as the link 
between model parameters $\BParams$ and observations of the model 
output $\Bdata$. To connect these two quantities, it is necessary to choose a 
so-called \emph{discrepancy model} that gives the relative probability that the 
model response to a realization of $\BParams = \Bparams$ describes the observations. 
One common assumption for this probabilistic model is that the measurements are perturbed by a Gaussian additive discrepancy term 
$\ve{E}\sim\cn(\ve{\varepsilon}\vert\ve{0},\mat{\Sigma})$, with covariance 
matrix $\mat{\Sigma}$. For a single 
measurement $\ve{y}^{(i)}$ it reads:
\begin{equation}
	\ve{y}^{(i)} = \cm(\Bparams) + \ve{\varepsilon}.
\end{equation}
This discrepancy between the model output 
$\cm(\BParams)$ and the observables $\BData$ can result from 
\emph{measurement error} or \emph{model inadequacies}. 
By using this additive discrepancy model, the distribution of the observables  
conditioned on the parameters $\ve{Y}\vert\Bparams$ is written as:
\begin{equation}
	\pi(\ve{y}^{(i)}\vert\Bparams) = 
	\cn(\ve{y}^{(i)}\vert\cm(\Bparams),\mat{\Sigma}),
\end{equation} 
where $\cn(\cdot\vert\ve{\mu},\mat{\Sigma})$ denotes the multivariate Gaussian PDF with mean value $\ve{\mu}$ and covariance matrix $\mat{\Sigma}$. The likelihood function $\cl$ is then constructed using this probabilistic 
model $\pi(\ve{y}^{(i)}\vert\Bparams)$ and 
Eq.~\eqref{eq:modelCal:LikelihoodDef}. For a given set of measurements $\Bdata$ 
it thus reads:
\begin{equation}
	\Blikelifun \eqdef 
	\prod_{i=1}^N\cn(\ve{y}^{(i)}\vert\cm(\Bparams),\ve{\Sigma}).
\end{equation} 
%
With the fully specified Bayesian model inversion problem, 
Eq.~\eqref{eq:modelCal:PosteriorDist} directly gives the posterior distribution 
of the model parameters $\Bpost$. In the setting 
of model inversion, the posterior distribution represents therefore the state of belief
about the true \emph{data-generating} model parameters, considering all available information:
computational forward model, discrepancy model and measurement data 
\citep{Bayesian:Beck1998, Bayesian:Jaynes2003}. 

Often, the ultimate goal of model inversion is to provide a set of inferred parameters, 
with associate confidence measures/intervals. This is often achieved by computing posterior statistics ({\em e.g.}, moments, mode, etc.). Propagating the posterior through secondary models is also of interest. 
So-called \emph{quantites of interest} (QoI) can be expressed by calculating the 
posterior expectation of suitable functions of the parameters $h(\Bparams):\mathbb{R}^M\to\mathbb{R}$, with 
$\BParams\vert\Bdata\sim\Bpost$, as in:
\begin{equation}
	\label{eq:modelCal:QuantitiesOfInterest}
	\Esp{h(\BParams)\vert\Bdata} = 
	\int_{\cd_{\BParams\vert\cy}}h(\Bparams)\Bpost\,\di{\Bparams}.
\end{equation}
Depending on $h$, this formulation encompasses \emph{posterior moments} 
($h(\Bparams)=x_i$ or $h(\Bparams)=(x_i-\Esp{X_i})^2$ for the first and second moments, respectively), \emph{posterior covariance} ($h(\Bparams)=x_ix_j-\Esp{X_i}\Esp{X_j}$) or 
expectations of secondary models ($h(\Bparams)=\cm^{\star}(\Bparams)$).
\section{Spectral function decomposition}
\label{sec:specRepres}
To pose a Bayesian inversion problem, the specification of a prior distribution and a likelihood function described in the previous section is sufficient. Its solution, however, is not available in closed form in the  general case. 

\emph{Spectral likelihood expansion} (SLE) is a recently proposed method that aims at solving the Bayesian inversion problem by finding a \emph{polynomial chaos expansion} (PCE) of the likelihood function in a basis orthogonal {\em w.r.t.}\ the prior distribution \citep{NagelJCP2016}. 
This representation allows one to derive analytical expressions for the evidence $Z$, the posterior distribution, the posterior marginals, and many types of QoIs, including the posterior moments. 

We offer here a brief introduction to \emph{regression-based, sparse PCE} before introducing SLE, but refer the 
interested reader to more exhaustive resources on PCE \citep{Ghanembook1991, 
	Xiu2002} and sparse PCE \citep{XiuBook2010, BlatmanPEM2010, BlatmanJCP2011}.

Let us consider a random variable $\BParams$ with \emph{independent} components 
$\{X_i,i=1,\dots,M\}$ and associated probability density functions 
$\pi_i(x_i)$ so that $\pi(\Bparams) = \prod_{i=1}^M\pi_i(x_i).$ Assume further 
that $\cM:\cd_{\BParams} = \prod_{i=1}^M\cd_{X_i}
\subseteq\mathbb{R}^M\to\mathbb{R}$ is a scalar function of 
$\BParams$ which fulfills the \emph{finite variance} 
condition ($\Esp{\cM(\BParams)^2}<+\infty$).
Then it is possible to find a so-called \emph{truncated polynomial chaos} 
approximation 
of $\cM$ 
such that
\begin{equation}
	\label{eq:solution:practical:expansion}
	\cM(\BParams) \approx \cM_{\mathrm{PCE}}(\BParams) \eqdef 
	\sum_{\ve{\alpha}\in\ca} 
	a_{\ve{\alpha}}\Psi_{\ve{\alpha}}(\BParams)
\end{equation}
where $\ve{\alpha}$ is an $M$-tuple $(\alpha_1,\dots,\alpha_M)\in\mathbb{N}^M$ 
and $\ca\subset\mathbb{N}^M$. For most parametric distributions, well-known classical orthonormal polynomials $\{\Psi_{\ve{\alpha}}\}_{\ve{\alpha}\in\mathbb{N}^M}$ satisfy the necessary orthonormality condition {\em w.r.t.}\ $\pi(\Bparams)$ \citep{Xiu2002}. For more general distributions, arbitrary orthonormal polynomials can be constructed numerically through the Stieltjes procedure \citep{Math:Gautschi2004}. If additionally, $\ca$ is a sparse subset of $\mathbb{N}^M$, the truncated expansion in Eq.~\eqref{eq:solution:practical:expansion} is called a \emph{sparse PCE}.

In this contribution, we restrict the discussion to independent inputs, because it is computationally challenging, albeit possible, to construct an orthogonal basis for dependent inputs (e.g. through Gram-Schmidt decomposition of ). Furthermore, \citet{Torre2019} demonstrated that for purely predictive purposes, ignoring input dependence can significantly improve the predictive performance of PCE, at the cost of losing basis orthonormality. 
The latter, however, is required to derive analytical post-processing quantities such as the moments of the posterior distribution (see Sections~\ref{sec:sol:SLE} and \ref{sec:sse:inversion}).

There exist different algorithms to produce a sparse PCE in practice, {\em i.e.} select a sparse basis $\ca$ and compute the corresponding coefficients. A powerful class of methods are \emph{regression-based} approaches that rely on an initial input sample $\cX$, called experimental design, and corresponding model evaluations $\cM(\cX)$ (See, \emph{e.g.} \citet{PCE:Luethen} for a recent survey). Additionally, it is possible to design adaptive algorithms that choose the truncated basis size \citep{BlatmanJCP2011, Jakeman2015}.

To assess the accuracy of PCE, the 
so-called \emph{generalization error} 
$\expc{(\cM(\BParams)-\cM_{\mathrm{PCE}}(\BParams))^2}$ shall be evaluated. A 
robust generalization error estimator is given by the 
\emph{leave-one-out} (LOO) cross validation technique. This estimator is 
obtained by
\begin{equation}
	\label{eq:LOO}
	\varepsilon_{\mathrm{LOO}} = 
	\frac{1}{K}\sum_{i=1}^{K}\left( 
	\cM(\Bparams^{(i)}) - 
	\cM_{\mathrm{PCE}}^{\sim i}(\Bparams^{(i)}) \right)^{2},
\end{equation}
where $\cM_{\mathrm{PCE}}^{\sim i}$ is constructed by leaving out the $i$-th 
point from 
the experimental design $\cx$. For methods based on linear regression, it can be 
shown \citep{Chapelle2002, BlatmanPEM2010} that the LOO error is available analytically by post-processing the regressor matrix.
\subsection{Spectral likelihood expansions}
\label{sec:sol:SLE}
The idea of SLE is to use sparse PCE to find a spectral representation 
of the likelihood function $\cl$ occurring in Bayesian model inversion problems (see 
Eq.~\eqref{eq:modelCal:LikelihoodDef}). We present here a brief 
introduction to the method and the main results of \citet{NagelJCP2016}.

Likelihood functions can be seen as scalar functions of the input random vector 
$\BParams \sim \Bprior$. In this work we assume priors of the type $\Bprior = \prod_{i=1}^M\pi_i(x_i)$, \emph{i.e.} with independent marginals, their spectral expansion then reads:
\begin{equation}
	\label{eq:solution:SLE:expansion}
	\cl(\BParams) \approx \cl_{\mathrm{SLE}}(\BParams) \eqdef  
	\sum_{\ve{\alpha}\in\ca}a_{\ve{\alpha}}\Psi_{\ve{\alpha}}(\BParams),
\end{equation}
where the explicit dependence on $\cy$ was dropped for notational simplicity. 

Upon computing the basis and coefficients in 
Eq.~\eqref{eq:solution:practical:expansion}, the 
solution to the inverse problem is converted to merely post-processing the 
coefficients $a_{\ve{\alpha}}$. The following expressions can be 
derived for the individual quantities:
\begin{description}
	\item[Evidence]
	The evidence emerges as the coefficient of the constant polynomial $a_{
		\ve{0}}$
	\begin{equation}
		\Bevi = \int_{\cd_{\BParams}} \cl(\Bparams)\Bprior\,\di{\Bparams} \approx 
		\left<\cl_{\mathrm{SLE}},1\right>_{\pi} = a_{\ve{0}}.
	\end{equation}
	\item[Posterior]
	Upon computing the evidence $Z$, the posterior can be evaluated directly 
	through 
	\begin{equation}
		\Bpost \approx \frac{\cl_{\mathrm{SLE}}(\Bparams)\Bprior}{Z} = 
		\frac{\Bprior}{a_{\ve{0}}}\sum_{\ve{\alpha}\in\ca}a_{\ve{\alpha}} 
		\Psi_{\ve{\alpha}}(\Bparams).
	\end{equation}
	\item[Posterior marginals] 
	We split the random vector $\BParams$ into two vectors 
	$\BParams_{\iu}$ with components $\{X_i\}_{i\in\iu}\in\cD_{\vX_{\iu}}$ 
	and $\BParams_{\iv}$ with components $\{X_i\}_{i\in\iv}\in\cD_{\vX_{\iv}}$, where $\iu$ and $\iv$ are two non-empty disjoint index sets such 
	that $\iu\cup\iv = \{1,\dots,M\}$. Denote further by $\pi_{\iu}(\vx_{\iu})\eqdef\prod_{i\in\iu}\pi_i(x_i)$ and $\pi_{\iv}(\vx_{\iv})\eqdef\prod_{i\in\iv}\pi_i(x_i)$ the prior marginal density functions of $\vX_{\iu}$ and $\vX_{\iv}$ respectively. The posterior marginals then read:
	\begin{equation}
		\label{eq:solution:SLE:postMarginal}
		\pi_{\iu}(\Bparams_{\iu}\vert\cy) = 
		\int_{\cd_{\BParams_{\iv}}}\Bpost\,\di{\Bparams_{\iv}} 
		\approx 
		\frac{\pi_{\iu}(\Bparams_{\iu})}{a_{\ve{0}}} 
		\sum_{\ve{\alpha}\in\ca_{\iv=0}} 
		a_{\ve{\alpha}}\Psi_{\ve{\alpha}}(\Bparams_{\iu}),
	\end{equation}
	where $\ca_{\iv=0}=\{\ve{\alpha}\in\cA:\alpha_i= 0 \Leftrightarrow 
	i\in\iv\}$. 
	The series in the above equation
	constitutes a 
	subexpansion that contains non-constant polynomials only in the 
	directions $i\in\iu$.
	\item[Quantities of interest] Finally, it is also possible to analytically 
	compute posterior expectations of functions that admit a polynomial chaos expansion in the same basis of the form $	h(\BParams) \approx \sum_{\ve{\alpha}\in\ca}b_{\ve{\alpha}}\Psi_{\ve{\alpha}}(\BParams)$.
	Eq.~\eqref{eq:modelCal:QuantitiesOfInterest} then reduces to the spectral product:
	\begin{equation}
		\Esp{h(\BParams)\vert\cy} = 
		\frac{1}{a_{\ve{0}}}\sum_{\ve{\alpha}\in\ca}a_{\ve{\alpha}}b_{\ve{\alpha}}.
	\end{equation}
\end{description}

The quality of these results depends only on the approximation error 
introduced in Eq.~\eqref{eq:solution:SLE:expansion}. The latter, in turn, depends 
mainly on the chosen PCE truncation strategy 
\citep{BlatmanJCP2011,NagelJCP2016} and the number of points 
used to compute the coefficients ({\em i.e.}, the experimental design). It is 
known that likelihood functions 
typically have quasi-compact supports ({\em i.e.}, $\cl(\BParams)\approx0$ on a 
majority of $\cd_{\BParams}$). Such functions require a very high polynomial 
degree to be approximated accurately, which in turn can lead to the need for 
prohibitively large experimental designs. 
\section{Stochastic spectral embedding}
\label{sec:sse}
Stochastic spectral embedding (SSE) is  a multi-level approach to surrogate modeling originally proposed in \citet{Marelli2020}. It attempts to approximate a given square-integrable function $\cM$ with independent inputs $\vX$ through
\begin{equation}
	\label{eq:SSE}
	\cM \approx \cM_\text{SSE}(\vX) = \sum\limits_{k\in\cK}  \indfun{\cD_{\vX}^{k}}(\vX)\, \widehat{\cR}_S^{k}(\vX),
\end{equation}
where $\cK\subseteq\mathbb{N}^2$ is a set of multi-indices with elements $k=(\ell,p)$ for which $\ell = 0,\dots, L$ and $p = 1,\dots,P_\ell$ where $L$ is the number of levels and $P_\ell$ is the number of subdomains at a specific level $\ell$. We call $\widehat{\cR}_S^{k}(\vX)$ a \emph{residual expansion} given by
\begin{equation}
	\label{eq:SSE residual expansion}
	\widehat{\cR}^{k}_S(\vX) = \sum\limits_{\ve{\alpha} \in \cA^{k}} a_{\ve{\alpha}}^{k} \Psi_{\ve{\alpha}}^{k}(\vX).
\end{equation}

In the present paper the term $\sum_{j \in \cA^{k}} a_j^{k} \Psi_j^{k}(\vX)$ denotes a polynomial chaos expansion (see Eq.~\eqref{eq:solution:practical:expansion}) constructed in the subdomain $\cd_{\vX}^{k}$, but in principle it can refer to any spectral expansion ({\em e.g.}, Fourier series). A schematic representation of the summation in Eq.~\eqref{eq:SSE} is given in Figure~\ref{fig:SSE partitioning}. The detailed notation and the algorithm to sequentially construct an SSE are given in the sequel.

\subsection{Stochastic spectral likelihood embedding}
\label{sec:sse:inversion}

Viewing the likelihood as a function of a random variable $\vX$ with independent marginals, 
we can directly use Eq.~\eqref{eq:SSE} to write down its SSLE 
representation
\begin{equation}
	\cl(\vX) \approx \cl_{\mathrm{SSLE}}(\vX) \eqdef  
	\sum\limits_{k\in\cK} \indfun{\cD_{\vX}^{k}}(\vX)\, \widehat{\cR}_S^{k}(\vX),
\end{equation}
where the variable $\vX$ is distributed according to the 
prior distribution $\Bprior$ and, consequently, the local basis used to 
compute $\widehat{\cR}_S^{k}(\vX)$ is orthonormal {\em w.r.t.}\ that distribution.

Due to the local spectral properties of the residual expansions, the SSLE representation of the likelihood function retains all of the post-processing properties of SLE (Section~\ref{sec:sol:SLE}):

\begin{description}
	\item[Evidence] The normalization constant $Z$ emerges as the sum of the 
	constant polynomial coefficients weighted by the prior mass:
	\begin{equation}
		\Bevi = \sum_{k\in\ck}
		\sum_{\ve{\alpha}\in\ca^k}a_{\ve{\alpha}}^k\int_{\cd_{\vX}^k}
		\Psi_{\ve{\alpha}}^k(\vx)\Bprior\,\di{\vx}
		= \sum_{k\in\ck} \cV^k a_{\ve{0}}^k, \quad \text{where} \quad 
		\cV^k = \int_{\cd_{\vX}^{k}}\Bprior\,\di{\vx}.
	\end{equation}
	\item[Posterior] This allows us to write the posterior as
	\begin{equation}
		\Bpost \approx \frac{\cl_{\mathrm{SSLE}}(\vx)\Bprior}{Z} = 
		\frac{\Bprior}{\sum_{k\in\ck} \cV^k 
			a_{\ve{0}}^k}\sum_{k\in\ck}\indfun{\cD_{\vX}^{k}}(\vx)
		\widehat{\cR}^{k}_S(\vx).
	\end{equation}
	\item[Posterior marginal]
	Utilizing again the disjoint sets $\iu$ and $\iv$ from Eq.~\eqref{eq:solution:SLE:postMarginal} it is also possible to analytically derive posterior marginal PDFs as
	\begin{equation}
		\label{eq:SSE:algorithm:postMarginal}
		\pi_{\iu}(\vx_{\iu}\vert\cy) = 
		\int_{\cd_{\vX_{\iv}}}\Bpost\,\di{\vx_{\iv}}
		\approx 
		\frac{\pi_{\iu}(\vx_{\iu})}{\sum_{k\in\ck} \cV^k 
			a_{\ve{0}}^k}\sum_{k\in\ck}\ve{1}_{\cd_{\vX_{\iu}}^k}(\vx_{\iu})
		\widehat{\cR}^{k}_{S,\iu}(\vx_{\iu})\cV^{k}_{\iv}
	\end{equation}
	where 
	\begin{equation}
		\widehat{\cR}^{k}_{S,\iu}(\vx_{\iu}) = 
		\sum_{\ve{\alpha}\in\ca_{\iv=0}^k}a_{\ve{\alpha}}^k\Psi_{\ve{\alpha}}^k
		(\vx_{\iu})
		\quad \text{and} \quad \cV^{k}_{\iv} = 
		\int_{\cd_{\vX_{\iv}}^{k}}\pi_{\iv}(\vx_{\iv})
		\,\di{\vx_{\iv}}.
	\end{equation}
	$\widehat{\cR}^{k}_{S,\iu}(\vx_{\iu})$ is a subexpansion of $\widehat{\cR}^{k}_{S}(\vx)$ that contains only non-constant polynomials 
	in the directions $i\in\iu$. Note that, as we assumed that the prior distribution has independent components, the constants $\cV^{k}$ and $\cV^{k}_{\iv}$ are obtained as products of univariate integrals which are available analytically from the prior marginal cumulative distribution functions (CDFs). 
	\item[Quantities of interest]
	Expected values of function 
	$h(\vx)=\sum_{\ve{\alpha}\in\ca^{k}}b_{\ve{\alpha}}^{k}
	\Psi_{\ve{\alpha}}^{k}(\vx)$ for 
	$k\in\ck$ under the posterior can 
	be approximated by:
	\begin{align}
		\begin{split}
			\label{eq:SSE:SSEforBI:QoI}
			\mathbb{E} \left[h(\vX)\vert\Bdata\right] &= 
			\int_{\cd_{\vX}}h(\vx)\Bpost\,\di{\vx}\\ 
			&= 
			\frac{1}{\Bevi}\sum_{k\in\ck} 
			\sum_{\ve{\alpha}\in\ca^k}a_{\ve{\alpha}}^k\int_{\cd_{\vX}^{k}}
			h(\vx)\Psi_{\ve{\alpha}}^k(\vx)\Bprior\,\di{\vx}\\
			&=
			\frac{1}{\Bevi}\sum_{k\in\ck} 
			\sum_{\ve{\alpha}\in\ca^{k}}a_{\ve{\alpha}}^{k}
			b_{\ve{\alpha}}^{k},
		\end{split}
	\end{align}
	where $b_{\ve{\alpha}}^k$ are the coefficients of the PCE 
	of $h$ in the $\card{\ck}$ bases 
	$\{\Psi_{\ve{\alpha}}^k\}_{\ve{\alpha}\in\ca^k}$.
	This can also be used for computing posterior moments like mean, variance or covariance.
\end{description}

These expressions can be seen as a generalization of the ones for SLE detailed 
in Section~\ref{sec:sol:SLE}. For a single-level global expansion ({\em 
	i.e.}, $\card{\ck}=1$) and consequently $\cV^{(0,1)} = 1$, they are identical.

\subsection{Modifications to the original algorithm}
\label{sec:SSE:modifications}
The original algorithm for computing an SSE was presented in \citet{Marelli2020}. It recursively partitions the input domain $\cD_{\vX}$ and constructs truncated expansions of the residual. We reproduce it below for reference but replace the model $\cM$ with the likelihood function $\cL$. We further simplify the algorithm by choosing a partitioning strategy with $N_S=2$.

\begin{enumerate}\itemsep0pt
	\item \textbf{Initialization:}
	\begin{enumerate}\itemsep0pt\small
		\item $\ell = 0$, $p = 1$
		\item $\cD_{\vX}^{\ell,p} = \cD_{\vX}$
		\item ${\cR}^{\ell}(\vX) = \cL(\vX)$
	\end{enumerate}
	\item \textbf{For each subdomain $\cD_{\vX}^{\ell,p}$, $p = 1,\cdots,P_\ell$:}
	\begin{enumerate}\itemsep0pt\small
		\item \label{algo:2a} Calculate the truncated expansion $\widehat{\cR}_S^{\ell,p}(\vX^{\ell,p})$ of the residual ${\cR}^{\ell}(\vX^{\ell,p})$ in the current subdomain
		\item \label{algo:2b} Update the residual in the current subdomain ${\cR}^{\ell+1}(\vX^{\ell,p}) = {\cR}^{\ell}(\vX^{\ell,p}) - \widehat{\cR}_S^{\ell,p}(\vX^{\ell,p})$
		\item \label{algo:2c} Split the current subdomain $\cD_{\vX}^{\ell,p}$ in $2$ subdomains $\cD_{\vX}^{\ell+1,\{s_1,s_{2}\}}$ based on a partitioning strategy
		\item If $\ell < L$, $\ell \leftarrow \ell + 1$, go back to  \ref{algo:2a}, otherwise terminate the algorithm
	\end{enumerate}
	\item \textbf{Termination}
	\begin{enumerate}\itemsep0pt\small
		\item Return the full sequence of $\cD_{\vX}^{\ell,p}$ and  $\widehat{\cR}_S^{\ell,p}(\vX^{\ell,p})$ needed to compute Eq.~\eqref{eq:SSE}.
	\end{enumerate}
\end{enumerate}

In practice, the residual expansions $\widehat{\cR}_S^{\ell,p}(\vX^{\ell,p})$ are computed using a \emph{fixed experimental design} $\cX$ and corresponding model evaluations $\cL(\cX)$. The algorithm then only requires the specification of a partitioning strategy and a termination criterion, as detailed in \citet{Marelli2020}.

Likelihood functions are typically characterized by a localized behaviour: Close to the data-generating parameters they peak while quickly decaying to $0$ in the remainder of the prior domain. This means that in a majority of the domain the likelihood evaluation is non-informative. Directly applying the original algorithm is then expected to \emph{waste} many likelihood evaluations.

We therefore modify the original algorithm by adding an 
adaptive sampling scheme (Section~\ref{sec:sse:mod:adaptiveSampling}) that includes the termination criterion and introducing an improved partitioning strategy (Section~\ref{sec:sse:mod:splitDomain}) that is especially suitable for finding compact support features. The rationale for these modifications is presented next.

\subsubsection{Adaptive sampling scheme}
\label{sec:sse:mod:adaptiveSampling}

The proposed algorithm has two parameters: the experimental design size for the residual expansions $\NRefine$ and the final experimental design size corresponding to the available computational budget $\Ntot$. At the initialization of the algorithm $\NRefine$ points are sampled as a first experimental design. At every further iteration, additional points are then sampled from the prior distribution. 
These samples are generated in a space-filling way (\emph{e.g.} through latin hypercube sampling) in the newly created subdomains $\cD_{\vX}^{\ell+1,s}$ to have always \emph{exactly} $\NRefine$ points available for constructing the residual expansions. The algorithm is terminated, once the computational budget $\Ntot$ has been exhausted.

Adding new samples points requires the evaluation of the likelihood function. Because multiple points are added at every iteration of the algorithm, this step can be easily performed simultaneously on multiple computational nodes.

At every step, the proposed algorithm chooses a single \emph{refinement domain} from the set of unsplit, \emph{i.e.} \emph{terminal domains}, creates two new subdomains by splitting the refinement domain and constructs residual expansions after enriching the experimental design. The selection of this refinement domain is based on the  error estimator $\cE^{k}$ that is defined by 
\begin{equation}
	\label{eq:algo:error}
	\cE^{\ell+1,s} = 
	\begin{cases}
		E^{\ell+1,s}_{\mathrm{LOO}}\cV^{\ell+1,s}, \quad &\text{if} \quad \exists~ \widehat{\cR}_S^{\ell+1,s},\\
		E^{\ell,s}_{\mathrm{LOO}}\cV^{\ell+1,s}, \quad &\text{otherwise}.
	\end{cases}
\end{equation} 

This estimator incorporates the subdomain size through the prior mass $\cV^{\ell+1,s}$, and the approximation accuracy, through the leave-one-out estimator.
The distinction is necessary to assign an error estimator also to domains that have too few points to construct a residual expansion, in which case the error estimator of the previous level $E^{\ell,s}_{\mathrm{LOO}}$ is reused. 

The algorithm sequentially splits and refines subdomains with large approximation errors. Because likelihood functions typically have the highest complexity close to their peak, these regions tend to have larger approximation errors and are therefore predominantly picked for refinement. The proposed way of adaptive sampling then ends up placing more points near the likelihood peak, thereby reducing the number of non-informative likelihood evaluations. 

The choice of a constant $\NRefine$ is simple and could in principle be replaced by a more elaborate strategy (\emph{e.g.}, based on the approximation error of the current subdomain relative to the total approximation error). A benefit of this enrichment criterion is that all residual expansions are computed with experimental designs of the same size. Upon choosing the domain with the maximum approximation error among the terminal domains, the error estimators then have a more comparable estimation accuracy. 


\subsubsection{Partitioning strategy}
\label{sec:sse:mod:splitDomain}
The partitioning strategy determines how a selected refinement domain is split. As described in \citet{Marelli2020}, it is easy to define the splitting in the uniformly distributed quantile space $\ve{U}$ and map the resulting split domains $\cD_{\ve{U}}^{\ell,p}$ to the (possibly unbounded) real space $\vX$ through an appropriate \emph{isoprobabilistic transform} ({\em e.g.}, the Rosenblatt transform \citep{Rosenblatt1952}). 

Similar to the original SSE algorithm presented in \citet{Marelli2020}, we split the refinement domain in half \emph{w.r.t.} its prior mass. The original algorithm chooses the splitting direction based on the partial variance in the refinement domain. This approach is well suited for generic function approximation problems. For the approximation of likelihood functions, however, we propose a partitioning strategy that is more apt for dealing with their \emph{compact support}.

\begin{figure}
	\centering
	\subfloat[Refinement domain]{
		\includegraphics[width=0.23\linewidth]{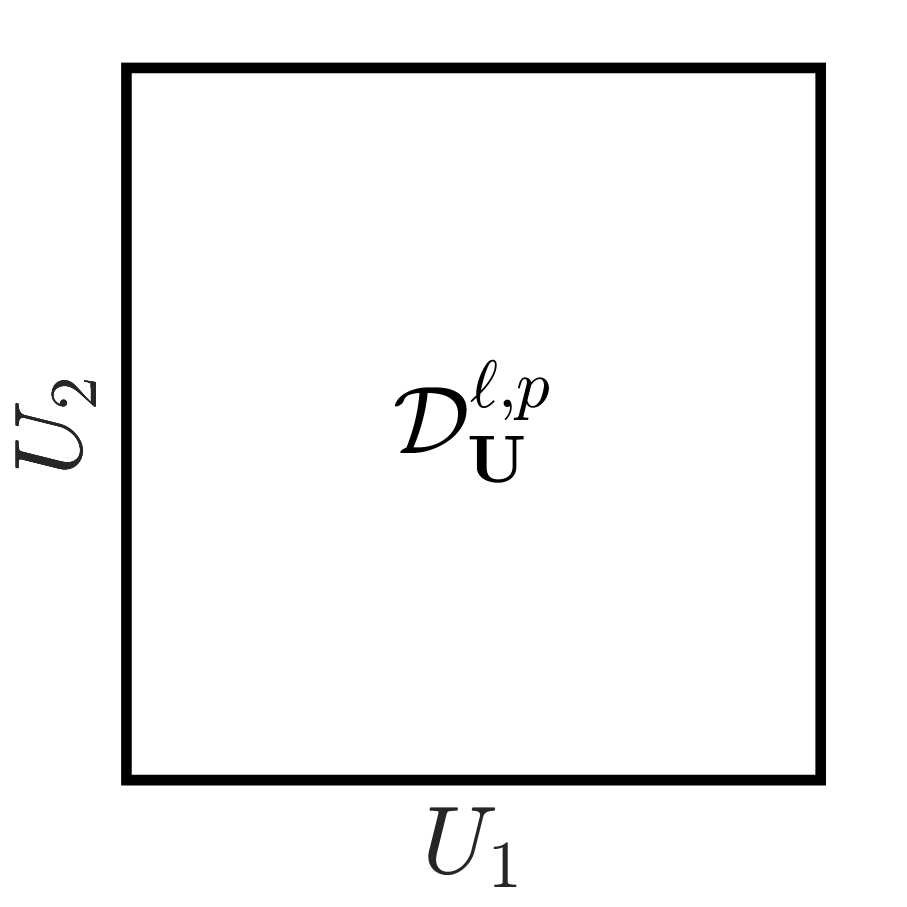}
		\label{fig:SSE:algo:splitting:1}
	}%
	\subfloat[Split along $d = 1$]{
		\includegraphics[width=0.23\linewidth]{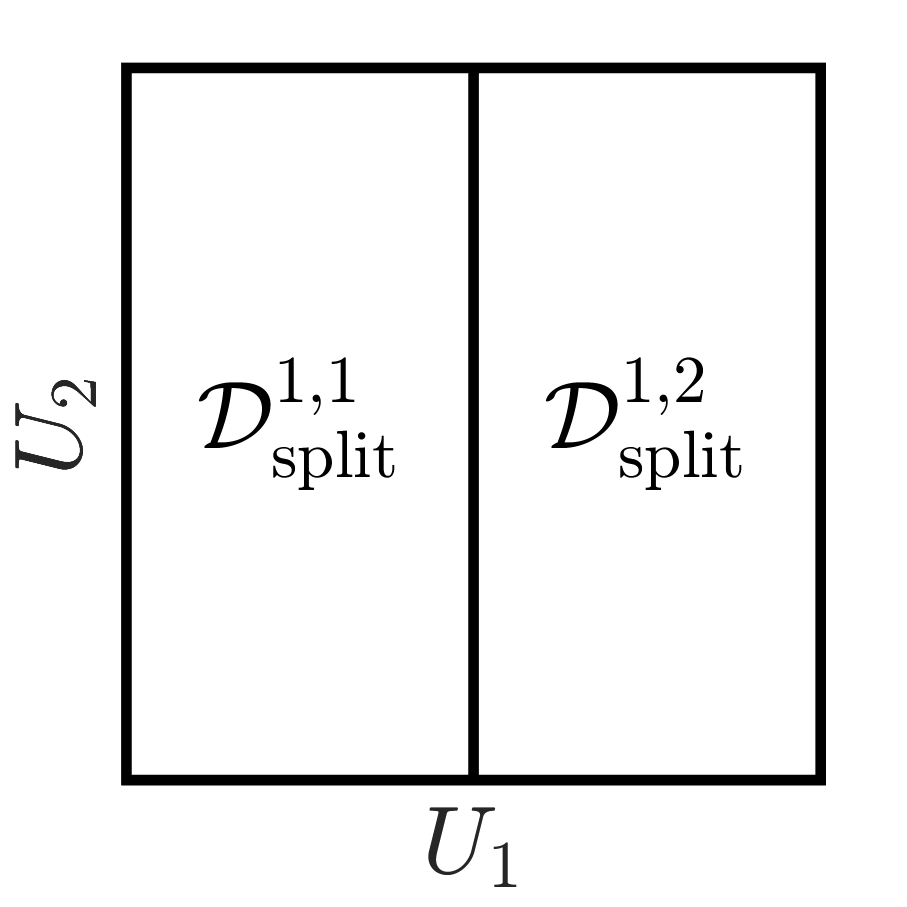}
		\label{fig:SSE:algo:splitting:2}
	}%
	\subfloat[Split along $d = 2$]{
		\includegraphics[width=0.23\linewidth]{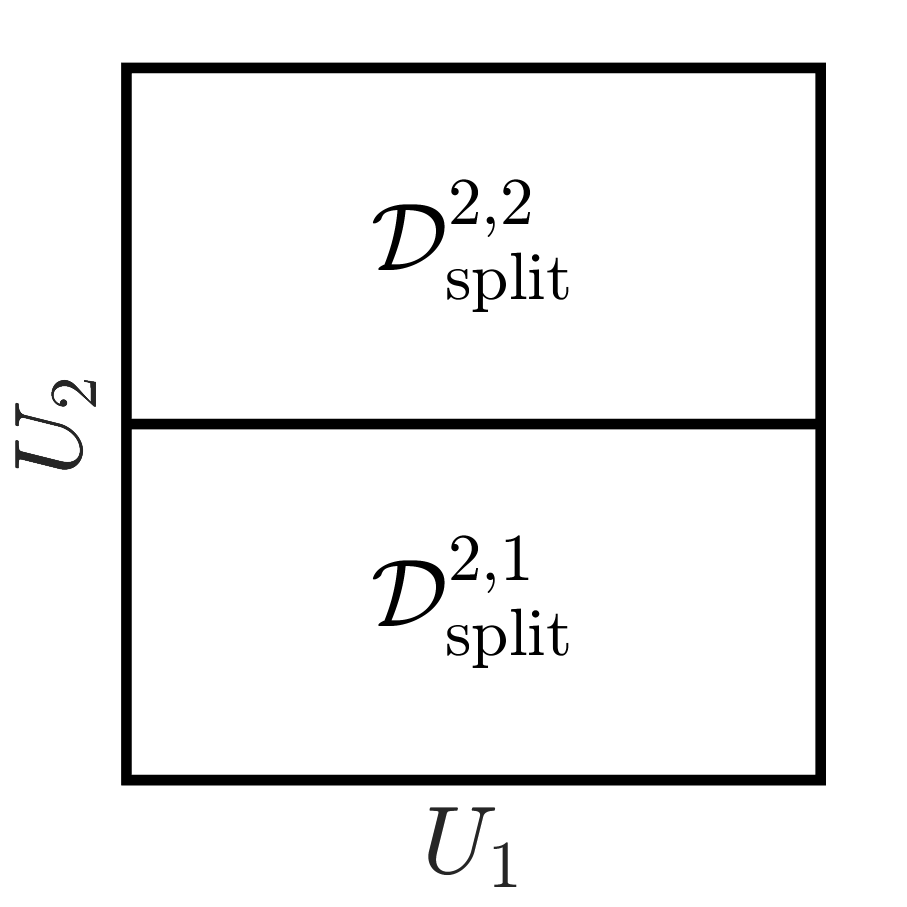}
		\label{fig:SSE:algo:splitting:3}
	}%
	\subfloat[Selected pair]{
		\includegraphics[width=0.23\linewidth]{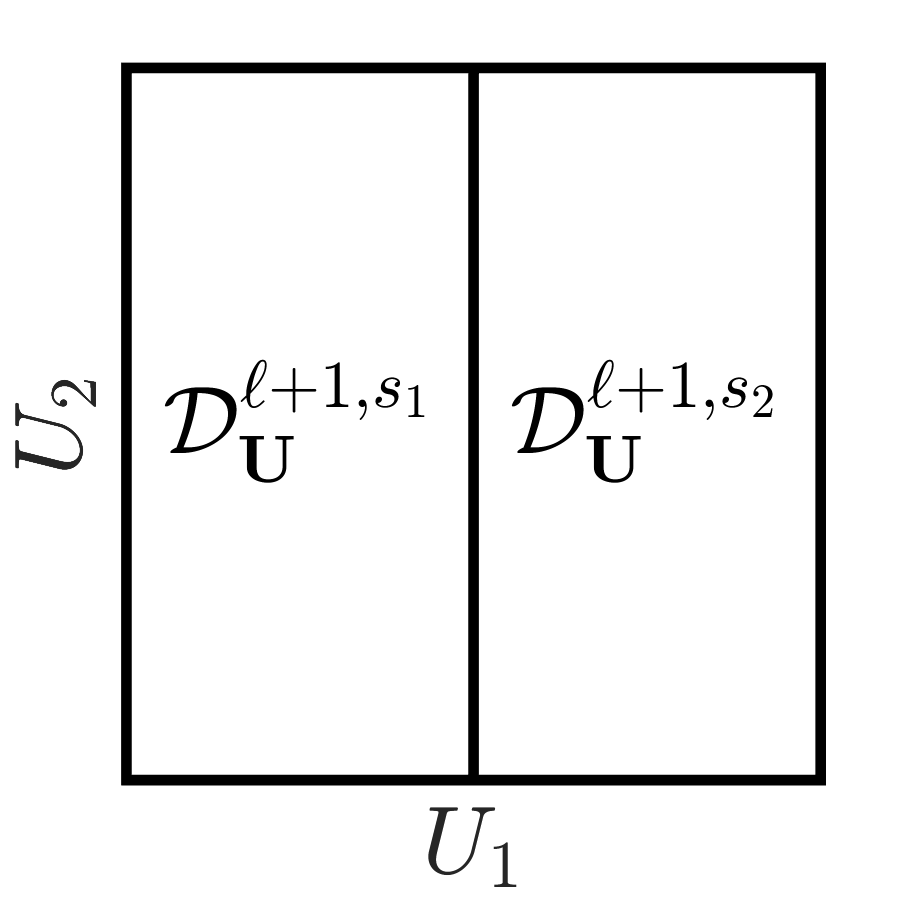}
		\label{fig:SSE:algo:splitting:4}
	}%
	\label{fig:SSE:algo:splitting}
	\caption{Partitioning strategy for a 2D example visualized in the quantile space $\ve{U}$. The refinement domain	
		$\cd_{\ve{U}}^{\ell,p}$ is split into two subdomains $\cd_{\ve{U}}^{\ell+1,s_1}$ 
		and $\cd_{\ve{U}}^{\ell+1,s_2}$.}
\end{figure}

We propose to pick the split direction along which a split yields a maximum difference in the residual empirical variance between the two candidate subdomains created by the split. This can easily be visualized with an example 
given by the $M=2$ dimensional domain 
$\cd_{\vX}^{\ell,p}$ in Figure~\subref*{fig:SSE:algo:splitting:1}. Assume this 
subdomain was selected as the refinement domain. To decide along which dimension 
to 
split, we construct the $M$ candidate subdomain pairs 
$\{\cd_{\mathrm{split}}^{i,1},\cd_{\mathrm{split}}^{i,2}\}_{i=1,\dots,M}$ and 
estimate the corresponding
$\{E_{\mathrm{split}}^{i}\}_{i=1,\dots,M}$
in those subdomains defined by
\begin{equation}
	\label{eq:SSE:algorithmMod:splitError}
	E_{\mathrm{split}}^{i} \eqdef 
	\left\vert\Var{\cR^{\ell+1}(\cX^{i,1}_{\mathrm{split}})}
	- 
	\Var{\cR^{\ell+1}(\cX^{i,2}_{\mathrm{split}})}\right\vert.
\end{equation}

In this expression, $\cX^{i,1}_{\mathrm{split}}$ and $\cX^{i,2}_{\mathrm{split}}$ denote subsets of the experimental design $\cX$ inside the subdomains $\cd_{\mathrm{split}}^{i,1}$ and $\cd_{\mathrm{split}}^{i,2}$ respectively. The occurring variances can be easily estimated with the 
empirical variance of the residuals in the respective candidate subdomains.

After computing the residual variance differences, the split is carried out 
along the 
dimension
\begin{equation}
	\label{eq:SSE:algorithmMod:splitDirection}
	d = 
	\arg\max_{i\in\{1,\dots,M\}} E_{\mathrm{split}}^{i},
\end{equation}
{\em i.e.}, to keep the subdomains $\cd_{\mathrm{split}}^{d,1}$ and 
$\cd_{\mathrm{split}}^{d,2}$ that introduce the largest difference in variance. 
For $d=1$, the resulting split can be seen in 
Figure~\subref*{fig:SSE:algo:splitting:4}.

The choice of this partitioning strategy can be justified
heuristically with the goal of approximating compact support functions. Assume 
that the likelihood function has compact support, this criterion will 
avoid \emph{cutting through} its support and instead identify a split 
direction that results in one subdomain with large variance (expected to contain the likelihood support) and a subdomain with small variance. In 
subsequent steps, the algorithm will 
proceed by \emph{cutting away} low variance subdomains, until the likelihood support is isolated.

\subsubsection{The adaptive SSLE algorithm}
\label{sec:sseAdaptive:algorithm}

\begin{figure}
	\centering
	\subfloat[Initialization]{
		\begin{minipage}{0.33\linewidth}
			\includegraphics[width=\linewidth,clip=true,trim=0 0 0 0]{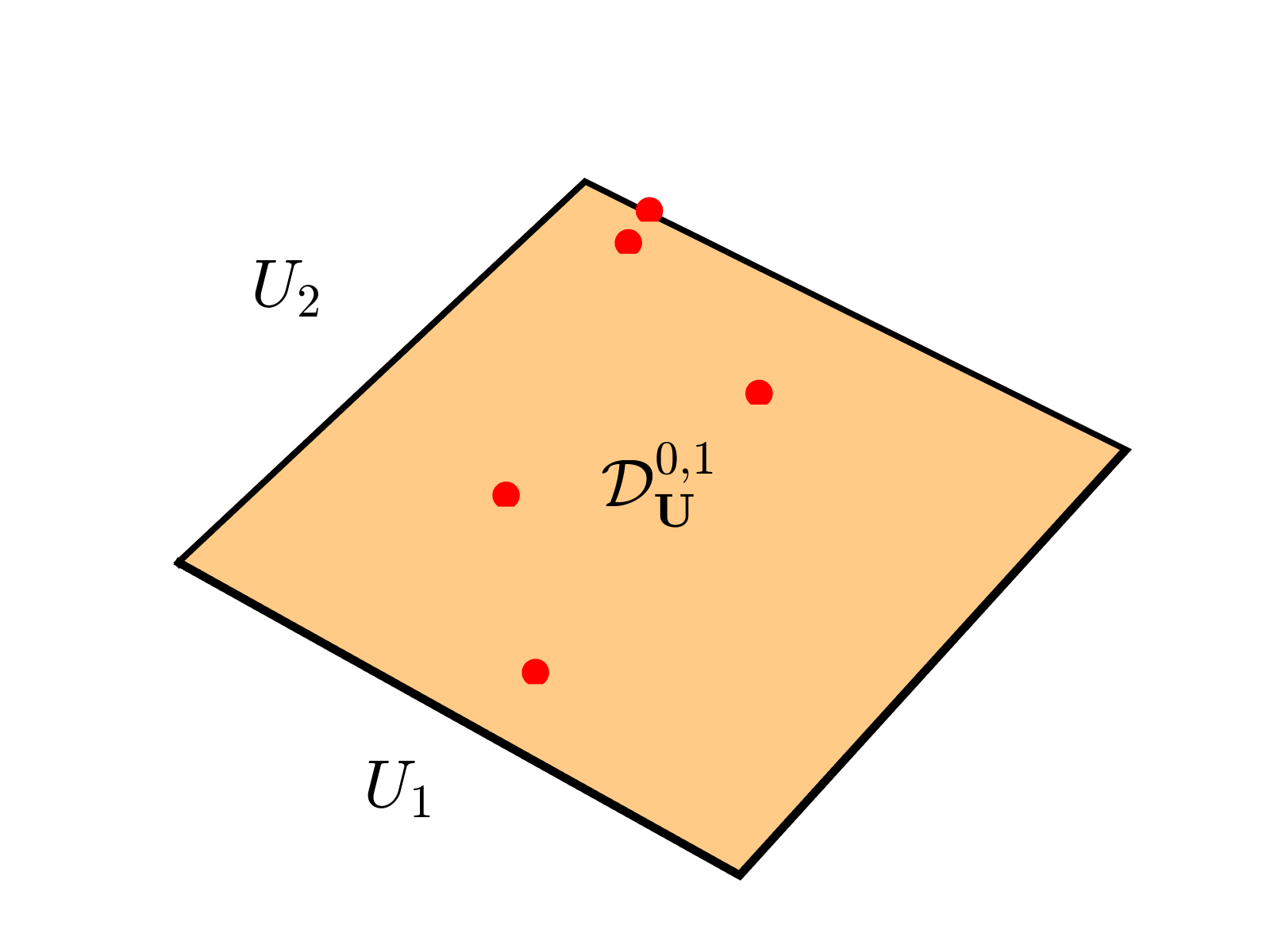}
			
			\includegraphics[width=\linewidth,clip=true,trim=0 0 0 0]{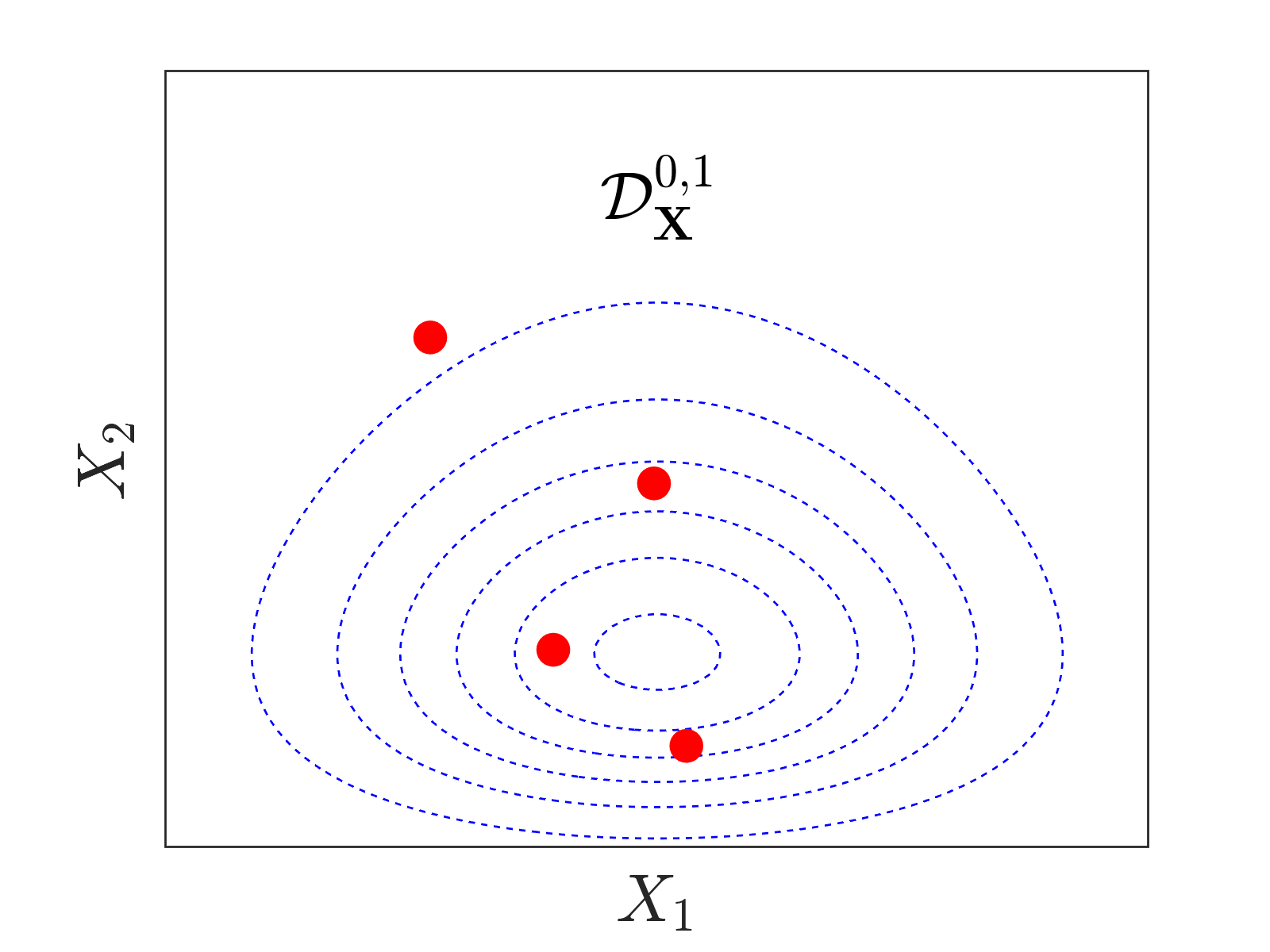}
		\end{minipage}
	}%
	\subfloat[First iteration]{
		\begin{minipage}{0.33\linewidth}
			\includegraphics[width=\linewidth,clip=true,trim=0 0 0 0]{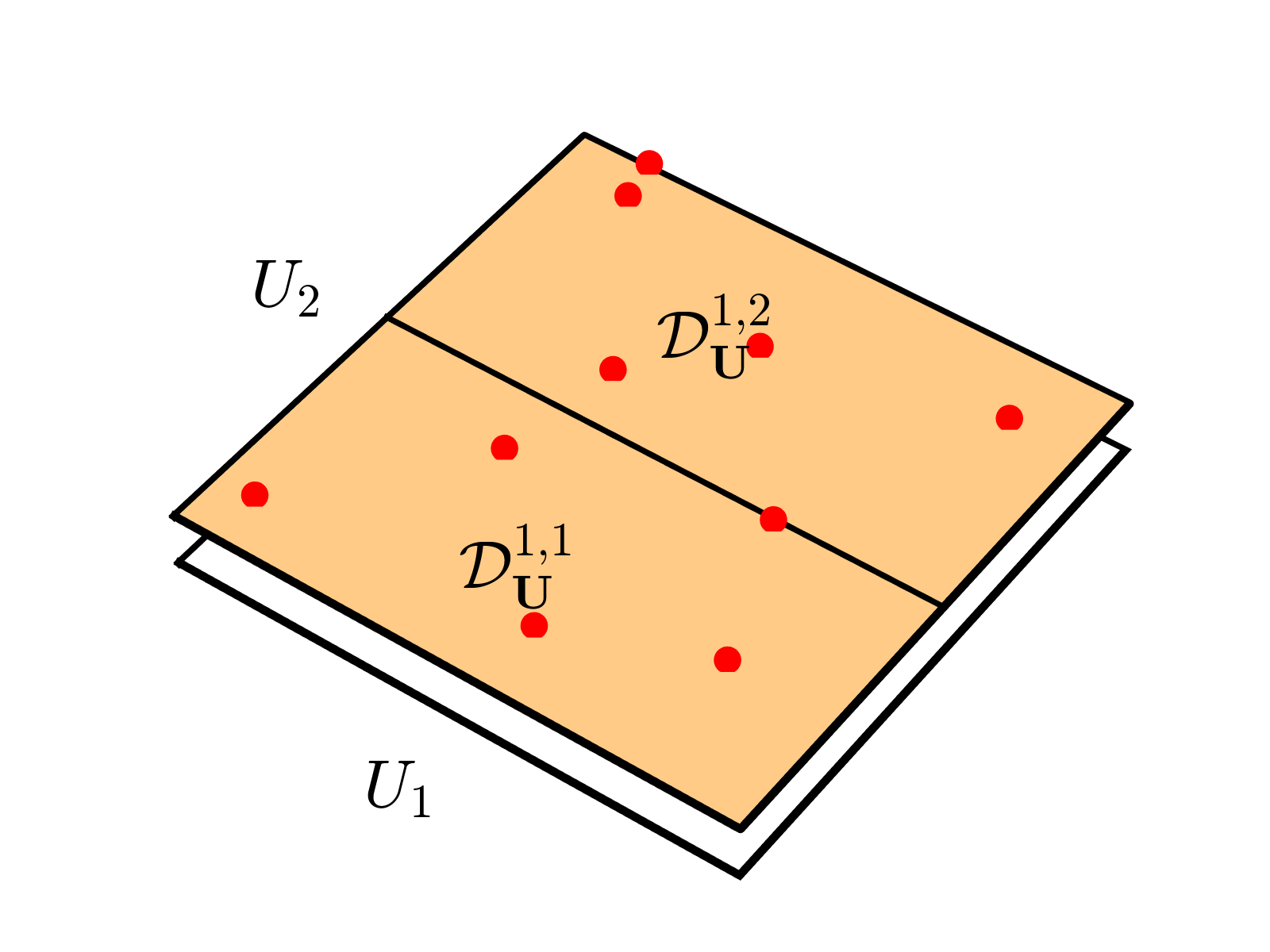}
			
			\includegraphics[width=\linewidth,clip=true,trim=0 0 0 0]{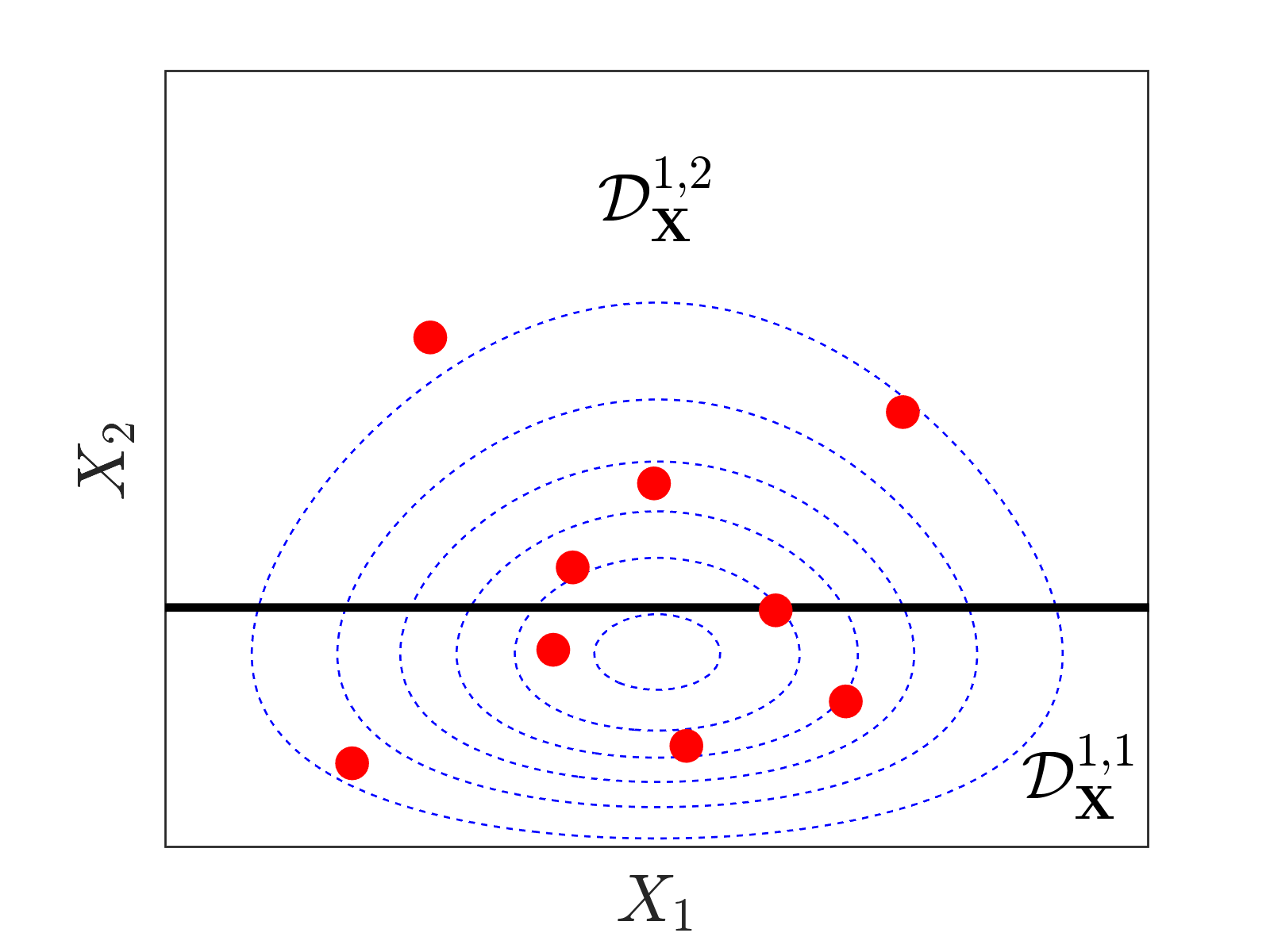}
		\end{minipage}
		\label{fig:SSE partitioning:b}
	}%
	%
	\subfloat[$3$rd iteration]{
		\begin{minipage}{0.33\linewidth}
			\includegraphics[width=\linewidth,clip=true,trim=0 0 0 0]{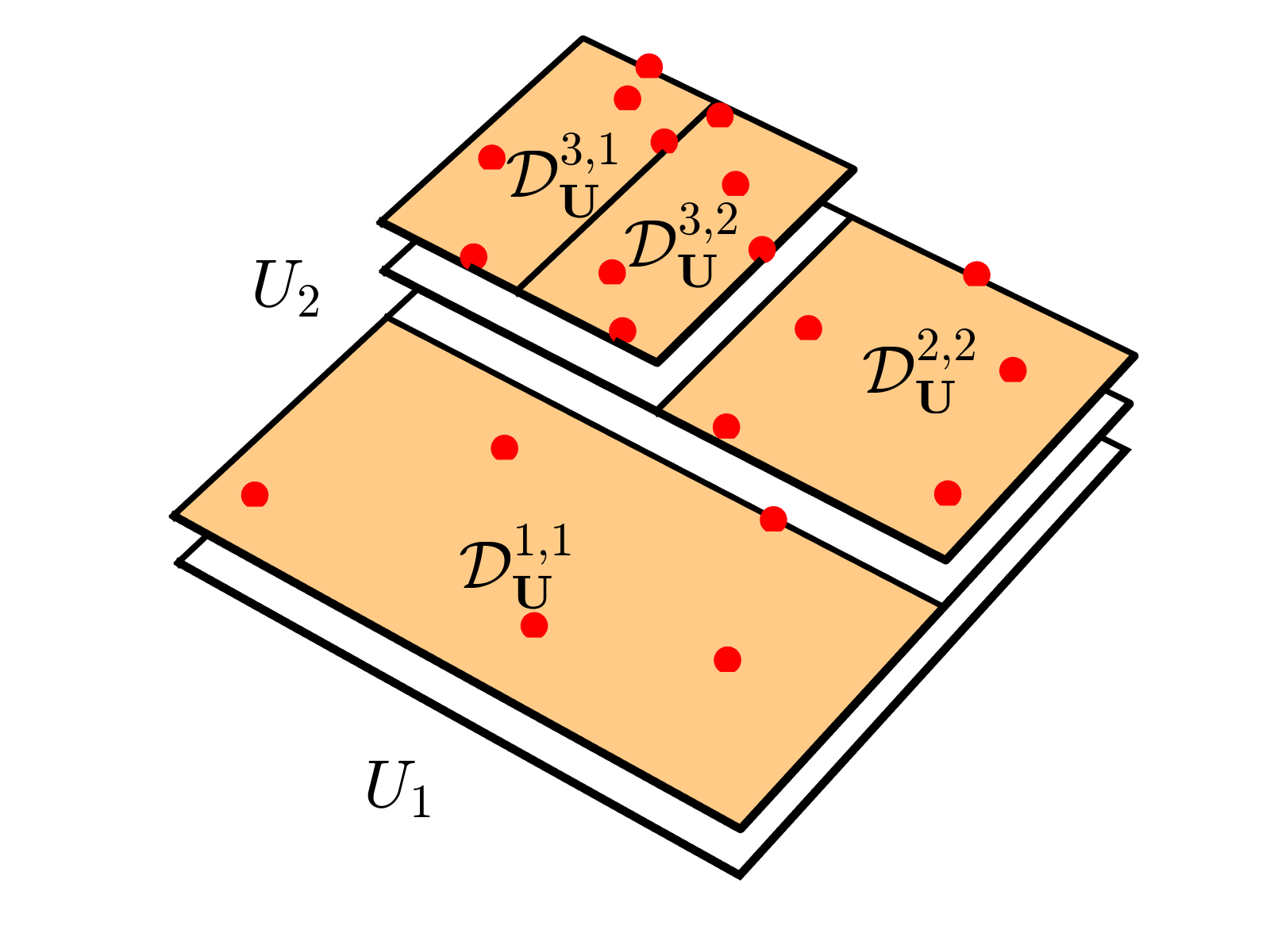}
			
			\includegraphics[width=\linewidth,clip=true,trim=0 0 0 0]{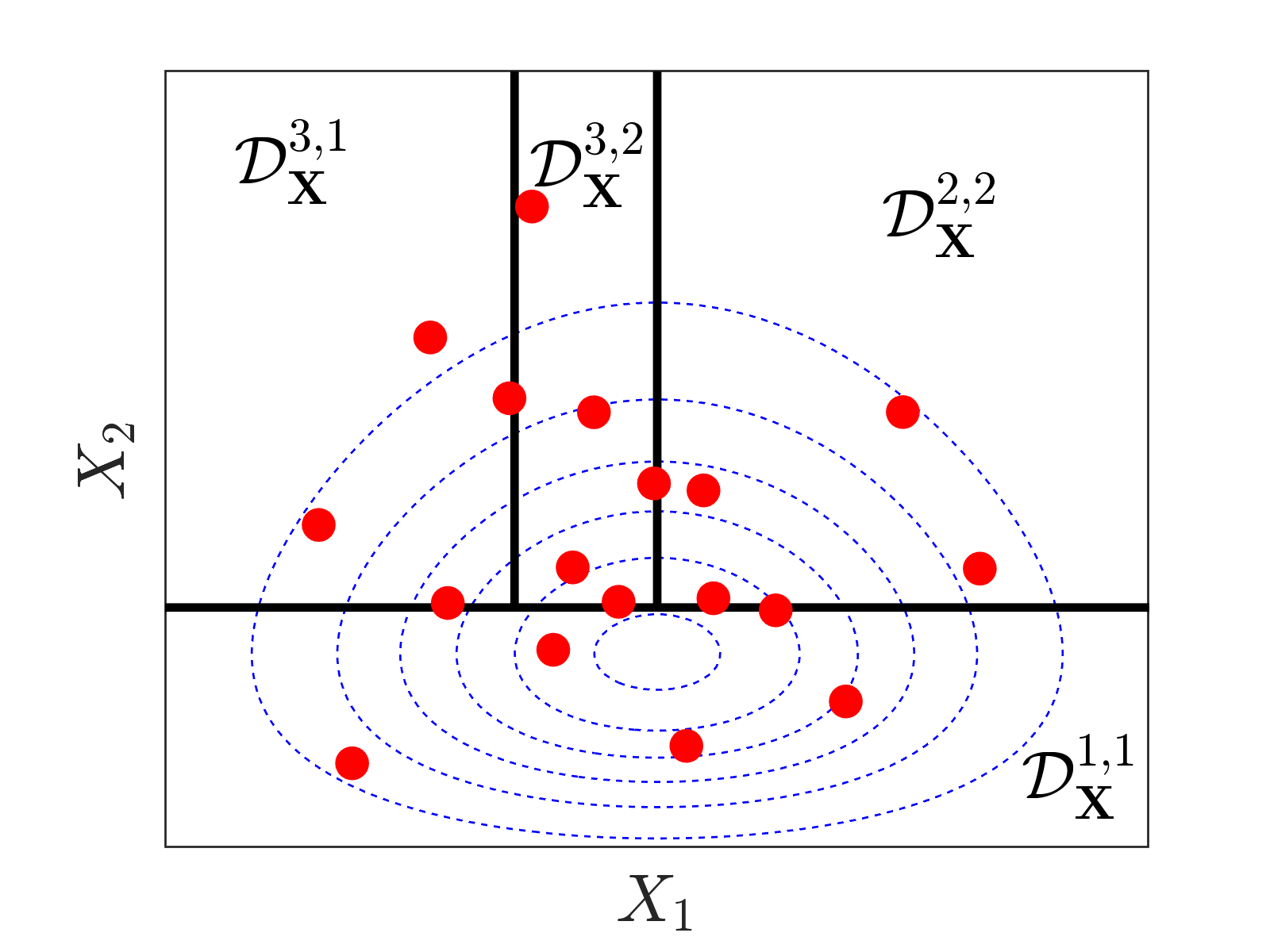}
		\end{minipage}
	}%
	\caption{Graphical representation of the first steps of the adaptive SSLE algorithm described in Section~\ref{sec:sseAdaptive:algorithm} for a two-dimensional problem with independent prior distribution. Upper row: partitioning in the quantile space; Lower row: partitioning in the unbounded real space with $\Bprior$ contour lines in dashed blue. Red dots show the adaptive experimental design that has a constant size of $\NRefine=5$ in each created subdomain. The terminal domains $\cT$ are highlighted in orange. The splitting direction in each subdomain is determined randomly in this example.}
	\label{fig:SSE partitioning}
\end{figure}

The algorithm is presented here with its two parameters $\NRefine$, the minimum experimental design size needed to expand a residual, and $\Ntot$, the final experimental design size. The sample $\cx^{\ell,p}$ refers to $\cx \cap \cd_{X}^{\ell,p}$, \emph{i.e.} the subset of $\cX$ inside $\cd_{X}^{\ell,p}$. Further, the multi-index set $\cT\in\mathbb{N}^2$ at each step of the algorithm gathers all indices $(\ell,p)$ of unsplit subdomains. It thus denotes the terminal domains: $\cD_{\vX}^k, k\in\cT$. 
For visualization purposes we show the first iterations of the algorithm for a two-dimensional example in Figure~\ref{fig:SSE partitioning}.

\begin{enumerate}\itemsep0pt
	\item \textbf{Initialization:}
	\begin{enumerate}\itemsep0pt\small
		\item $\cD_{\vX}^{0,1} = \cD_{\vX}$
		\item Sample from prior distribution $\cX = \{\vx^{(1)},\cdots,\vx^{(\NRefine)}\}$
		\item Calculate the truncated expansion $\widehat{\cR}_S^{0,1}(\vX)$ of $\cL(\vX)$ in the full domain $\cX^{0,1}$, retrieve its approximation error $\cE^{0,1}$ and initialize $\cT = \{(0,1)\}$
		\item ${\cR}^{1}(\vX) = \cL(\vX) - \widehat{\cR}_S^{0,1}(\vX)$
	\end{enumerate}
	\item \label{algoAdapt:2} \textbf{For $(\ell, p)=\argmax_{k\in\cT}\cE^{k}$:}
	\begin{enumerate}\itemsep0pt\small
		\item \label{algoAdapt:updateT} Split the current subdomain $\cD_{\vX}^{\ell,p}$ in $2$ sub-parts  $\cD_{\vX}^{\ell+1,\{s_1,s_2\}}$ and update $\cT$
		\item \textbf{For each split $s = \{s_1, s_2\}$}
		\begin{enumerate}
			\item \textbf{If} $\vert\cX^{\ell+1,s}\vert < \NRefine$ and $\NRefine - \vert\cX^{\ell+1,s}\vert < \Ntot - \vert\cX\vert$
			\begin{enumerate}
				\item \label{algoAdapt:enrichED} Enrich sample $\cX$ with $\NRefine - \vert\cX^{\ell+1,s}\vert$ new points inside $\cD_{\vX}^{\ell+1,s}$
			\end{enumerate}
			\item \textbf{If} $\vert\cX^{\ell+1,s}\vert = \NRefine$
			\begin{enumerate}
				\item \label{algoAdapt:createResExp} Create the truncated expansion $\widehat{\cR}_S^{\ell+1,s}(\vX^{\ell+1,s})$ of the residual ${\cR}^{\ell+1}(\vX^{\ell+1,s})$ in the current subdomain using $\cX^{\ell+1,s}$
				\item Update the residual in the current subdomain ${\cR}^{\ell+2}(\vX^{\ell+1,s}) = {\cR}^{\ell+1}(\vX^{\ell+1,s}) - \widehat{\cR}_S^{\ell+1,s}(\vX^{\ell+1,s})$
			\end{enumerate}
			\item \label{algoAdapt:Error} Retrieve the approximation error $\cE^{\ell+1,s}$ from Eq.~\eqref{eq:algo:error}
		\end{enumerate}
		\item \textbf{If} no new expansions were created, terminate the algorithm, otherwise go back to \ref{algoAdapt:2}
	\end{enumerate}
	\item \textbf{Termination}
	\begin{enumerate}\itemsep0pt\small
		\item Return the full sequence of $\cD_{\vX}^{\ell,p}$ and  $\widehat{\cR}_S^{\ell,p}(\vX^{\ell,p})$ needed to compute Eq.~\eqref{eq:SSE}
	\end{enumerate}
\end{enumerate}

The updating of the multi-index set in Step~\ref{algoAdapt:updateT} refers to removing the current index $(\ell,p)$ from the set and adding to it the newly created indices $(\ell+1,s_1)$ and $(\ell+1,s_2)$.

\subsubsection{Convergence of the adaptive SSLE algorithm}
Convergence of the original SSE algorithm is guaranteed in the mean-square sense by the spectral convergence in each subdomain \citet{Marelli2020}. This convergence property directly applies to the SSLE of any likelihood function as, per existence of a finite maximum likelihood value, they fulfil the finite variance condition \citet{NagelJCP2016}.

This result cannot be directly extended to the present adaptive (greedy) setting, because a combination of parameters might in general lead to a whole subdomain not being explored further. 
This can only happen, however, if the error estimator $\cE^{k}$ tremendously underestimates the actual error in a terminal domain. The choice of the leave-one-out cross-validation error estimator (see Eq.~\eqref{eq:algo:error}), with its tendency to avoid overfitting, reduces significantly the probability of this scenario. Further investigations in this direction are ongoing.
\section{Case studies}
\label{sec:applications}

To showcase the effectiveness of the proposed SSLE approach, we present three case studies with different types of likelihood complexity: (i) a one-dimensional vibration problem with a bimodal posterior, (ii) a six-dimensional heat transfer problem that exhibits high posterior concentrations (i.e. highly informative likelihood) and (iii) a $62$-dimensional diffusion problem with low active dimensionality that models concentration-driven diffusion in a one-dimensional domain.

For all case studies, we adopt the adaptive sparse-PCE based on LARS approach developed in \cite{BlatmanJCP2011} through its numerical implementation in UQLab \citep{MarelliUQLab2014,UQdoc_13_104}.
Each $\widehat{\cR}_S^{k}$ is therefore a degree- and $q$-norm-adaptive polynomial chaos expansion. 
We further introduce a rank truncation of $r=2$, \textit{i.e.} we limit the maximum number of input interactions \citep{UQdoc_13_104} to two variables at a time. The truncation set for each spectral expansion (Eq.~\eqref{eq:solution:practical:expansion}) thus reads:
\begin{equation}
	\cA^{M,p,q,r} = \{\ve{\alpha} \in \mathbb{N}^{M} : \vert\vert\ve{\alpha}\vert\vert_q \le p, \vert\vert\ve{\alpha}\vert\vert_0 \le r\}.
\end{equation}
where
\begin{equation}
	\vert\vert\ve{\alpha}\vert\vert_q = \left(\sum_{i=1}^M \alpha_i^q\right)^{\frac{1}{q}}, q\in (0,1]; \quad \vert\vert\ve{\alpha}\vert\vert_0 = \sum_{i=1}^M 1_{\{\alpha_i>0\}}.
\end{equation}

The $q$-norm is adaptively increased between $q = \{0.5,\cdots,0.8\}$ while the maximum polynomial degree is adaptively increased in the interval $p = \{0,1,\cdots,p\}$, where the maximum degree $p=20$ for case study (i) and (ii) and $p=3$ for case study (iii) due to its high dimensionality. 

In case study (ii) and (iii), the performance of SLE \citep{NagelJCP2016}, the original non-adaptive SSLE \citep{Marelli2020} and the proposed adaptive SSLE approach presented in Section~\ref{sec:SSE:modifications} is compared. The comparison was omitted for case study (i), because only adaptive SSLE succeeded in solving the problem. For clarity, we henceforth abbreviate the adaptive SSLE algorithm to adSSLE.

To simplify the comparison, the same partitioning strategy employed for adSSLE (Section~\ref{sec:sse:mod:splitDomain}) was employed for the non-adaptive SSLE approach. Also, the same experimental designs were used for the non-adaptive SSLE and the SLE approaches. Finally, the same parameter $\NRefine$ was used to define the enrichment samples in adSSLE and the termination criterion in non-adaptive SSLE.

Because the considered case studies do not admit a closed form solution, we validate the algorithm instead through MCMC reference solutions with long chain lengths to produce samples from the posterior distributions. In this respect, identifiability of the ``true'' underlying data-generating model, while clearly important in inverse problems in general, is not central in the present discussion. As a proxy for identifiability, however, we consider several different case studies with more or less informative (peaked) likelihood and posterior distributions.

To assess the performance of the three algorithms considered, we define an error 
measure that allows to quantitatively compare the similarity of the SSLE, adSSLE and SLE solution 
with the reference MCMC solution. This comparison is inherently difficult, as a 
sampling-based approach (MCMC) needs to be compared to a functional 
approximation (SSLE, adSSLE, SLE). We proceed to compare 
the \emph{univariate posterior marginals}, available analytically in 
SSLE, adSSLE and SLE (See Eq.~\eqref{eq:solution:SLE:postMarginal} and 
Eq.~\eqref{eq:SSE:algorithm:postMarginal}), to the reference posterior 
marginals estimated with kernel density estimation (KDE, \citet{Wand1995}) from 
the MCMC sample. 
Denoting by $\hat{\pi}_{i}(\kappa_i\vert\Bdata)$ the SSLE, adSSLE or SLE approximations 
and by ${\pi}_{i}(\kappa_i\vert\Bdata)$ the reference solution, we define the 
following error measure
\begin{equation}
	\label{eq:eq3:errorMeasure}
	\eta \eqdef \frac{1}{M}\sum_{i=1}^M 
	\jsd\left(\hat{\pi}_{i}(\kappa_i\vert\Bdata) \vert\vert 
	\pi_{i}(\kappa_i\vert\Bdata)\right)
\end{equation}
where $M$ is the dimensionality of the problem and $\jsd$ is the 
\emph{Jensen-Shannon divergence} \citep{Li1991}, a 
symmetric and bounded ($[0,\log(2)]$) distance measure for probability distributions based on the 
Kullback-Leibler divergence. 

The purpose of the error measure $\eta$ is to allow for a fair comparison between the different methods investigated. It is not a practical measure for engineering applications because it relies on the availability of a reference solution, and it its magnitude does not have a clear quantitative interpretation. 
However, it is considerably more comprehensive than a pure moment-based error measure. Because it is averaged over all $M$ marginals, it encapsulates the approximation accuracy of all univariate posterior marginals in a scalar value. 

As all algorithms (SSLE, adSSLE, SLE) depend on a randomly chosen experimental designs, we produce $20$ replications for case study (ii) and $5$ replications for case study (iii) by running them multiple times. The computational cost of all examples is given in terms of number of likelihood evaluations $\Ntot$, as they dominate the total computational cost in any engineering setting. All computations shown were carried out on a standard laptop, with the longest single runs taking $\approx 1~h$, which includes both forward model evaluations and the overall algorithmic overhead.

\subsection{1-dimensional vibration problem}

In this first case study, the goal is the inference of a single unknown parameter with a multimodal posterior distribution. This problem is difficult to solve with standard MCMC methods due to the presence of a \textit{probability valley}, \emph{i.e.} low probability region, between essentially disjoint posterior peaks. It also serves as an illustrative example of how the proposed adaptive algorithm constructs an adSSLE. The problem is fabricated and uses synthetic data, but is presented in the context of a relevant engineering problem. 

\begin{figure}
	\centering
	\includegraphics[width=5cm]{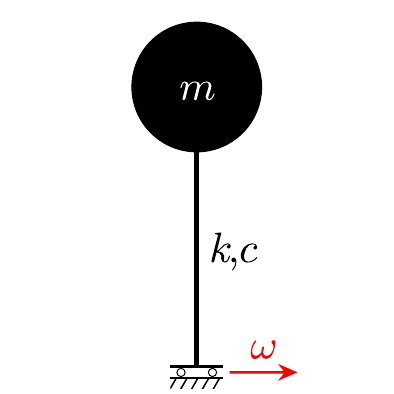}
	\caption{\emph{1-dimensional vibration problem}: Sketch of the linear oscillator}%
	\label{fig:cs1:sketch}%
\end{figure}

Consider the oscillator displayed in Figure~\ref{fig:cs1:sketch} subject to 
harmonic ({\em i.e.}, sinusoidal) excitation. Assume the prior information about its stiffness $X\eqdef k$ is that it follows a lognormal distribution with $\mu = 0.8~N/m$ and $\sigma = 0.1~N/m$. Its true value shall be determined using measurements of the oscillation amplitude at the location of the mass $m$. The known properties of the oscillator system are the oscillator mass $m=1~kg$, the excitation frequency $\omega = 1~rad/s$ and the viscous damping coefficient $c = 0.1~Ns/m$. The oscillation amplitude is measured in five 
independent oscillation events and normalized by the forcing 
amplitude yielding the measured amplitude ratios $\Bdata=\{9.01, 8.67, 8.84, 
9.22, 8.54\}$.

This problem is well known in mechanics and in the linear case ({\em i.e.}, 
assuming small deformations and linear material behavior) can be solved 
analytically with the \emph{amplitude of the frequency response function}. This 
function returns the ratio between the steady state amplitude of a linear 
oscillator and the amplitude of its excitation. It is given by
\begin{equation}
	\cm(X) = \frac{m\omega^2}{\sqrt{(X-m\omega^2)^2+(c\omega)^2}}.
\end{equation}

We assume a discrepancy model with known discrepancy standard deviation 
$\sigma$. 
In conjunction with the available measurements $\Bdata$, this leads to 
the following likelihood function:
\begin{equation}
	\Blikelifun = \prod_{i=1}^5 \cn(y^{(i)}\vert \cm(x),\sigma^2).
\end{equation}

\begin{figure}
	\centering
	\subfloat[$1$st iteration, $\Ntot=10$]{
		\label{fig:cs1:SSEBehaviour:1}
		\begin{minipage}{0.5\linewidth}
			\includegraphics[width=0.9\linewidth, trim = 0 10 0 12 clip]{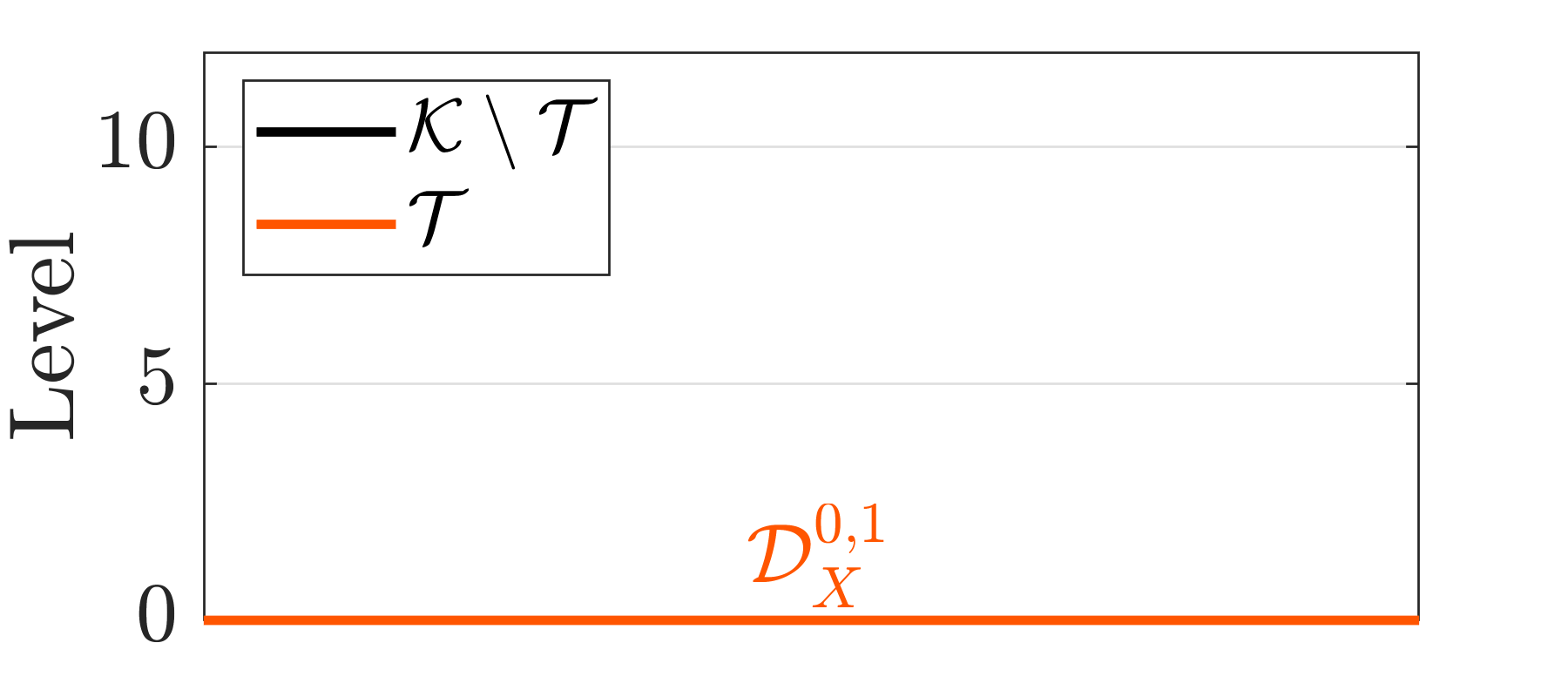}
			\includegraphics[width=0.9\linewidth, trim = 0 10 0 10, clip]{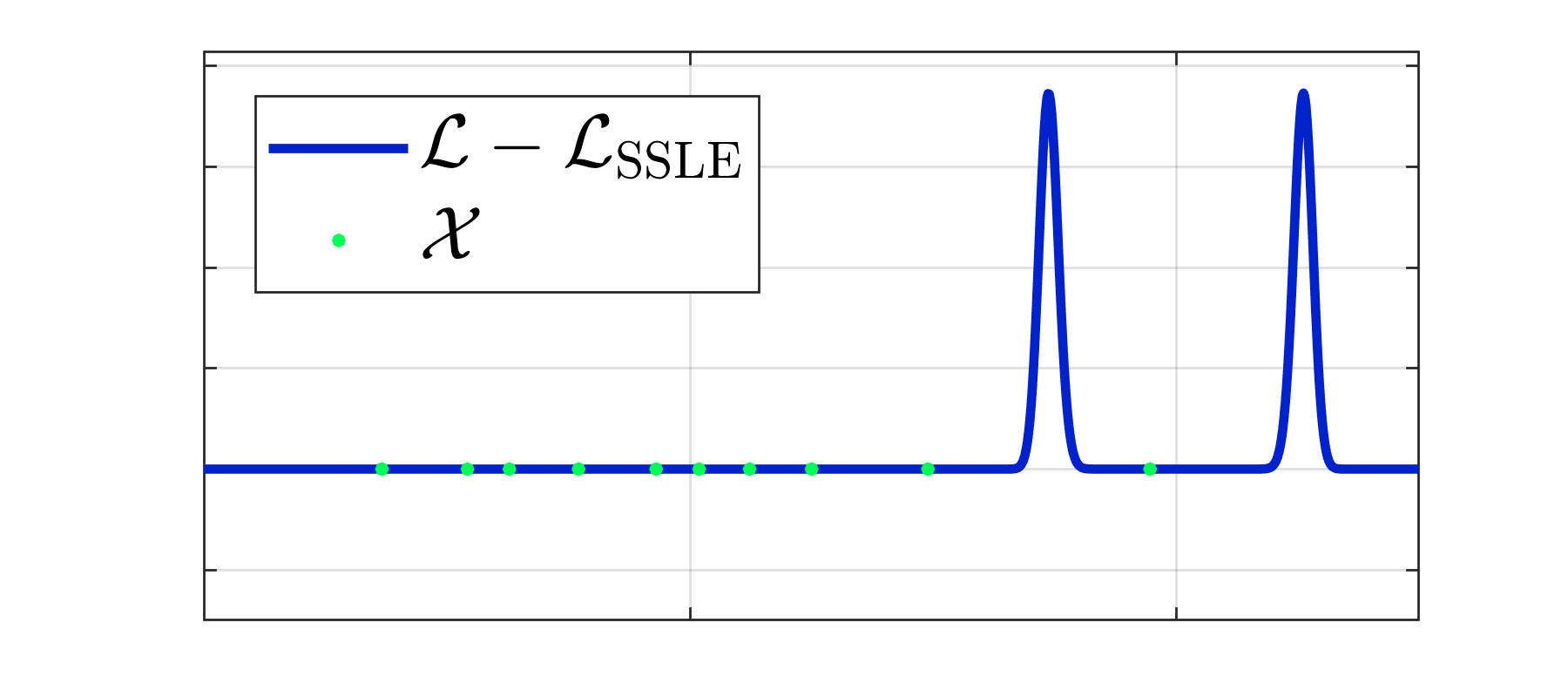}
			\includegraphics[width=0.9\linewidth, trim = 0 5 0 10, clip]{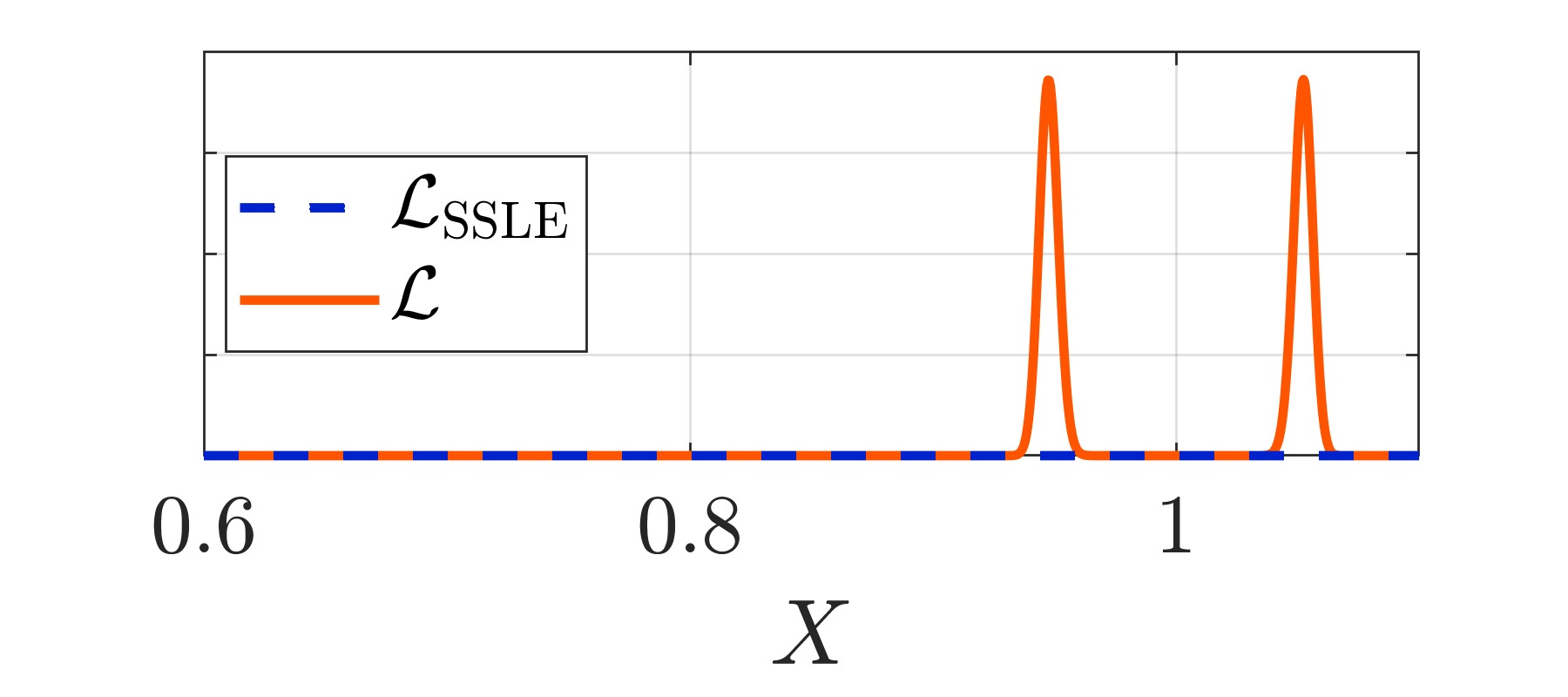}
		\end{minipage}
	}%
	\subfloat[$3$rd iteration, $\Ntot=30$]{
		\label{fig:cs1:SSEBehaviour:2}
		\begin{minipage}{0.5\linewidth}
			\includegraphics[width=0.9\linewidth, trim = 0 10 0 12 clip]{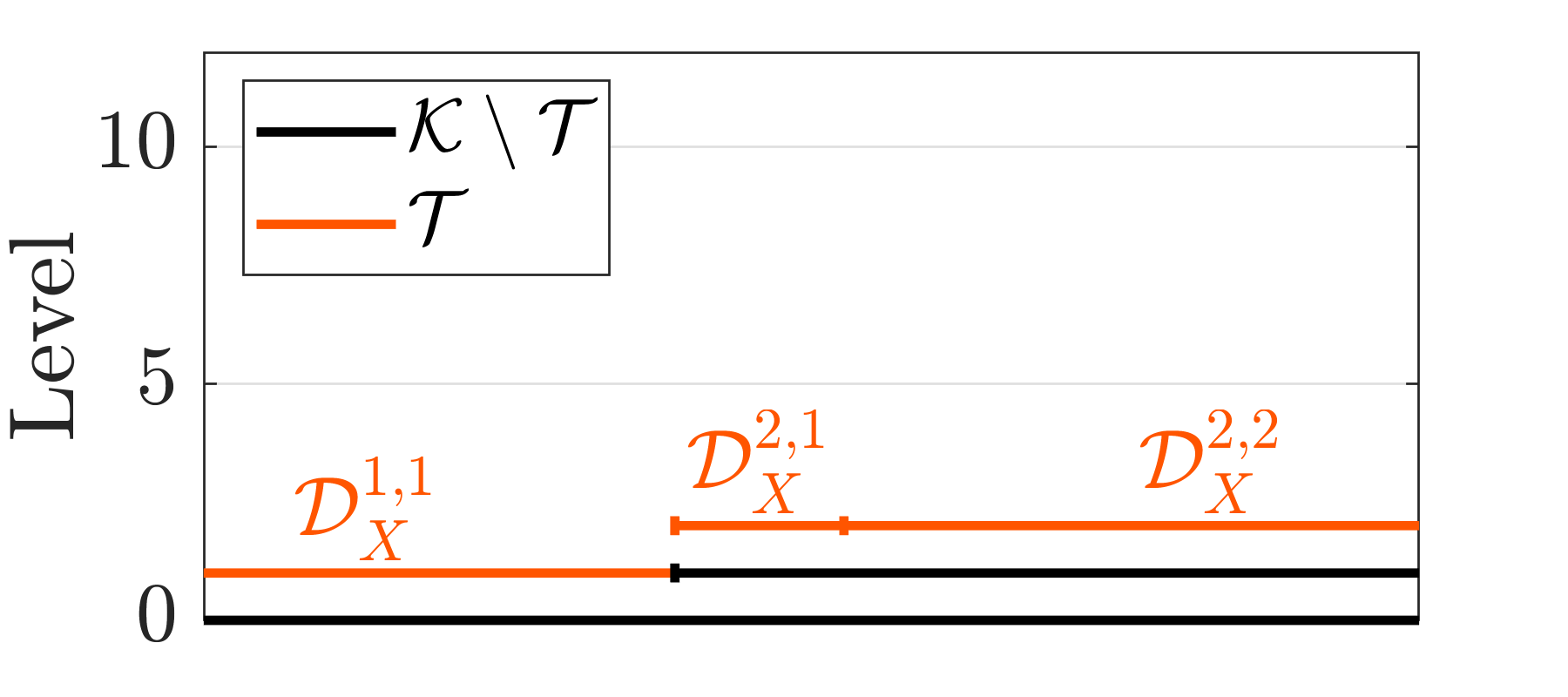}
			\includegraphics[width=0.9\linewidth, trim = 0 10 0 10, clip]{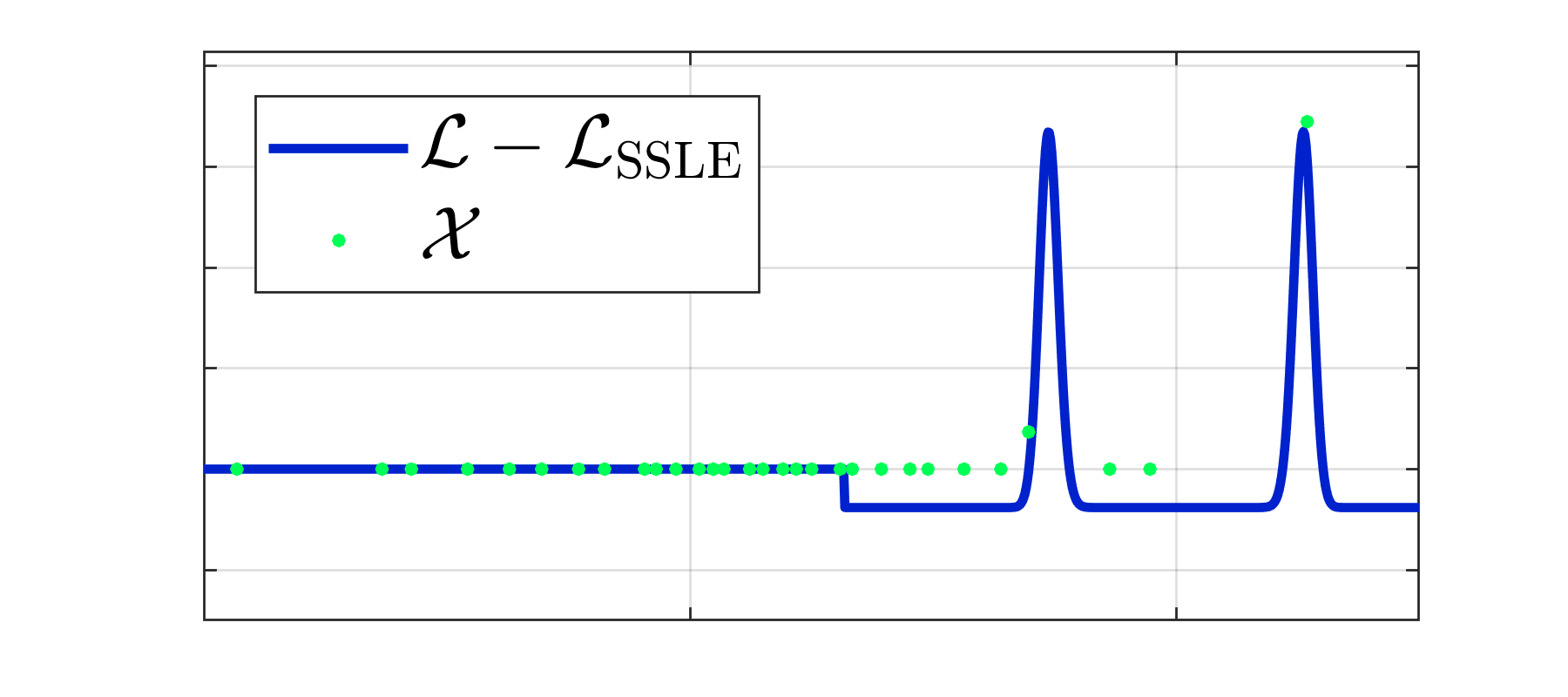}
			\includegraphics[width=0.9\linewidth, trim = 0 5 0 10, clip]{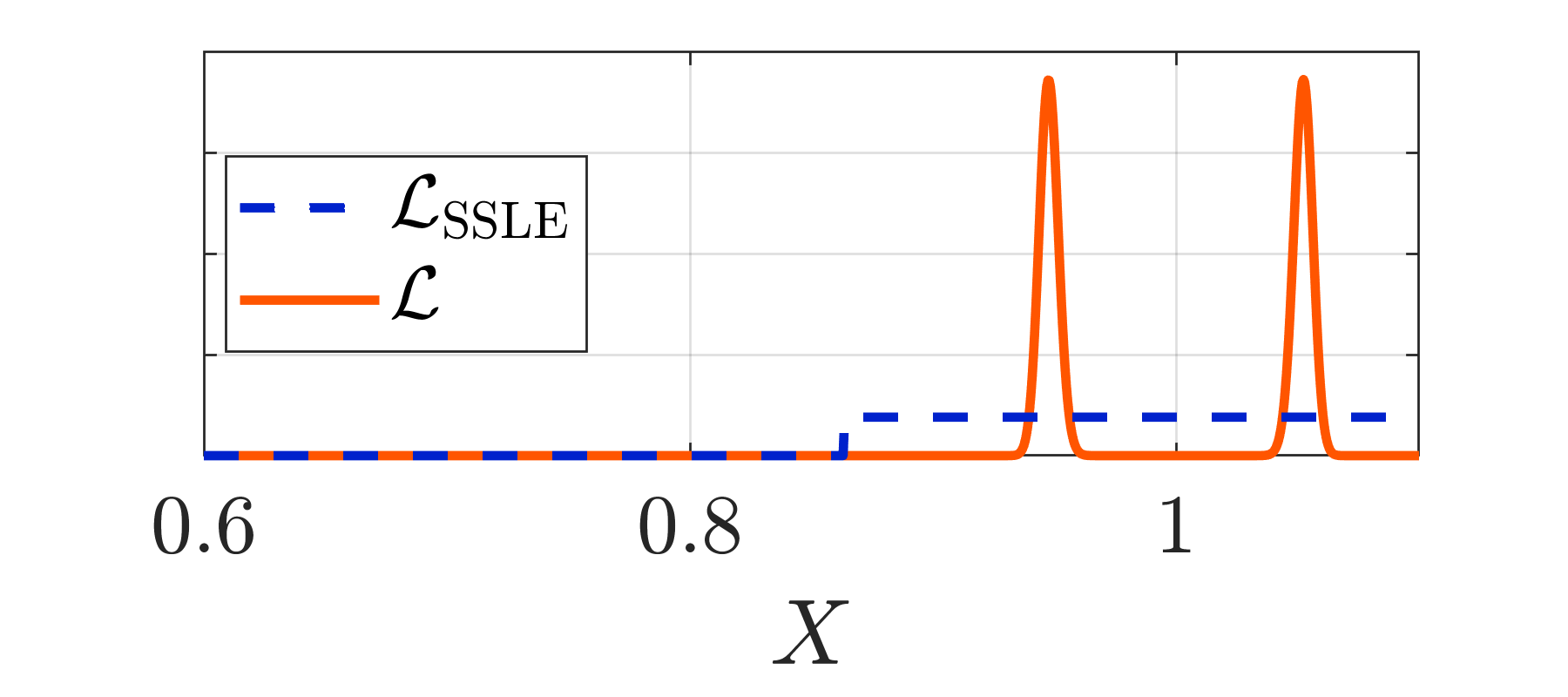}
		\end{minipage}
	}%
	\\
	\subfloat[$8$th iteration, $\Ntot=80$]{
		\label{fig:cs1:SSEBehaviour:3}
		\begin{minipage}{0.5\linewidth}
			\includegraphics[width=0.9\linewidth, trim = 0 10 0 12 clip]{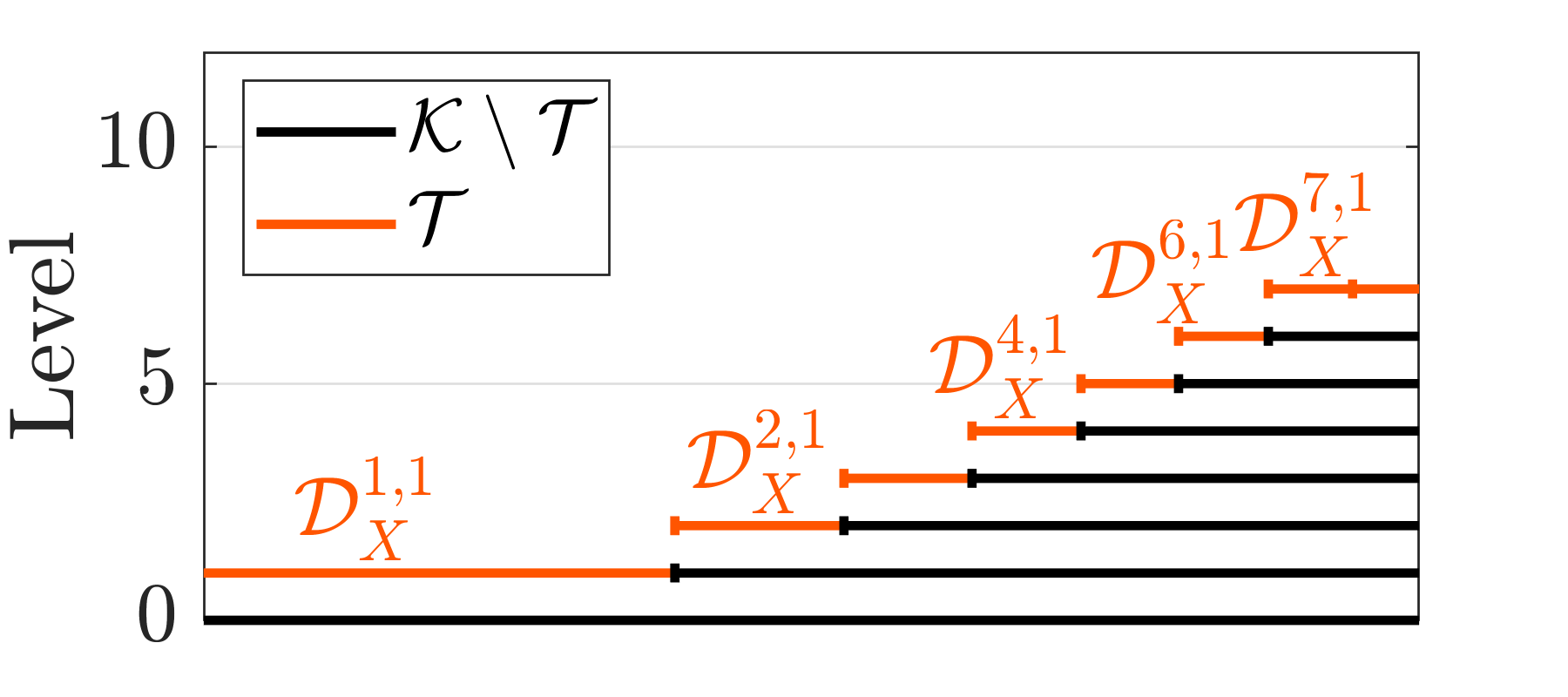}
			\includegraphics[width=0.9\linewidth, trim = 0 10 0 10, clip]{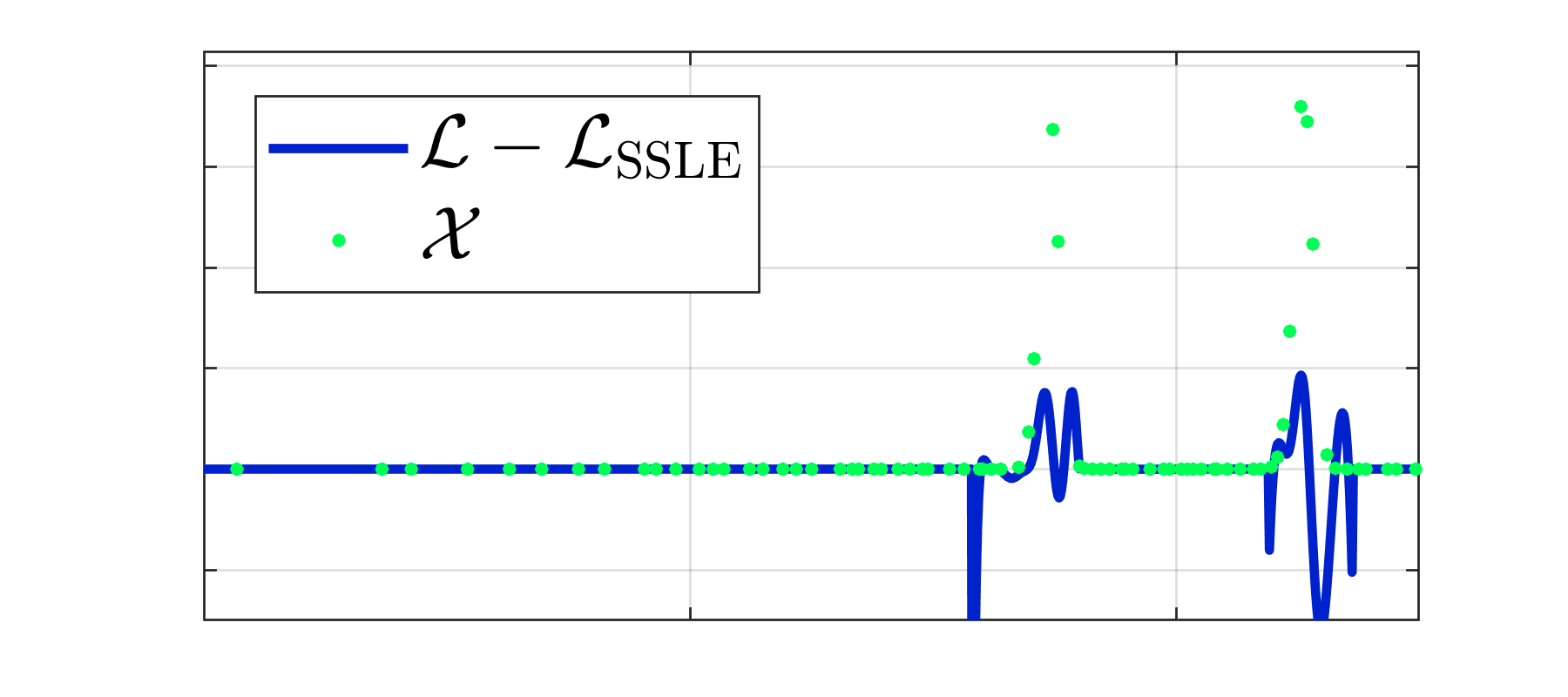}
			\includegraphics[width=0.9\linewidth, trim = 0 5 0 10, clip]{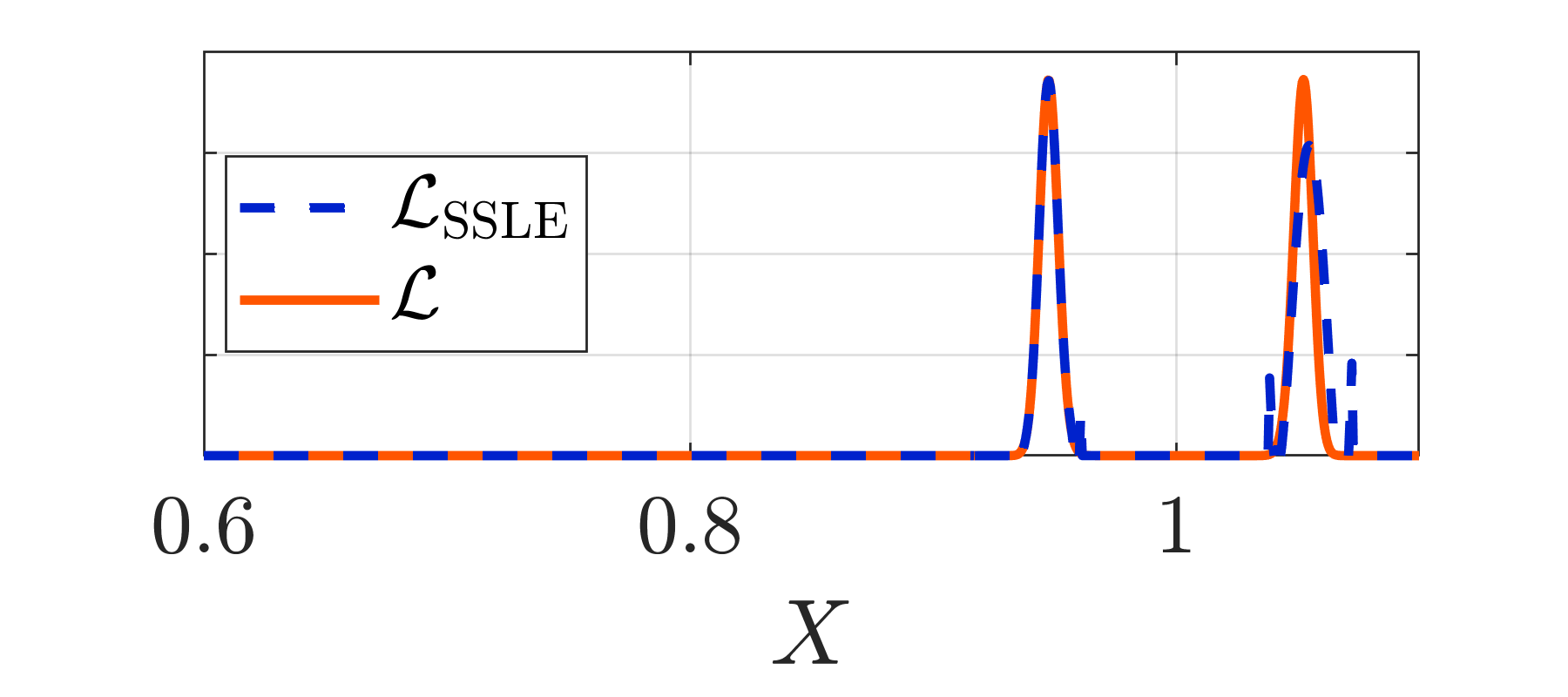}
		\end{minipage}
	}%
	\subfloat[$10$th iteration, $\Ntot=100$]{
		\label{fig:cs1:SSEBehaviour:4}
		\begin{minipage}{0.5\linewidth}
			\includegraphics[width=0.9\linewidth, trim = 0 10 0 12 clip]{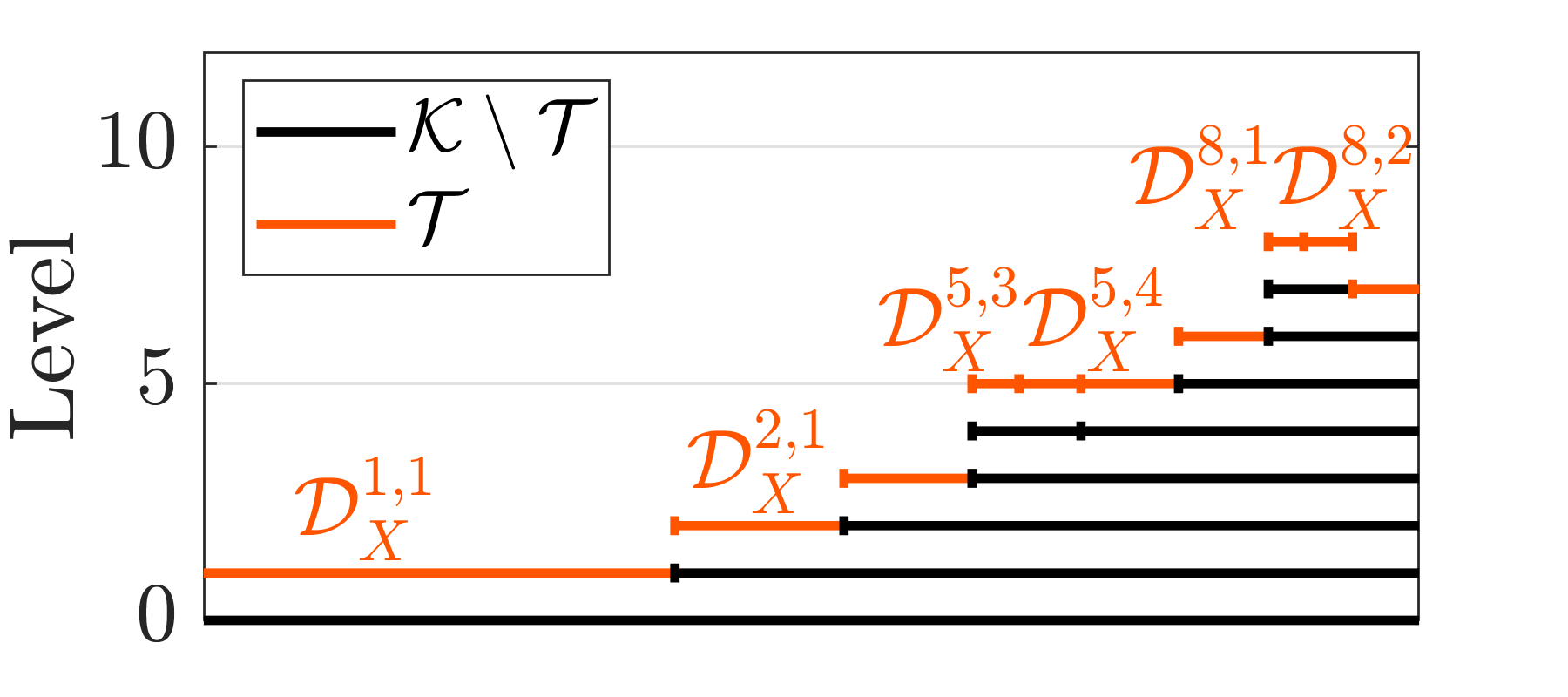}
			\includegraphics[width=0.9\linewidth, trim = 0 10 0 10, clip]{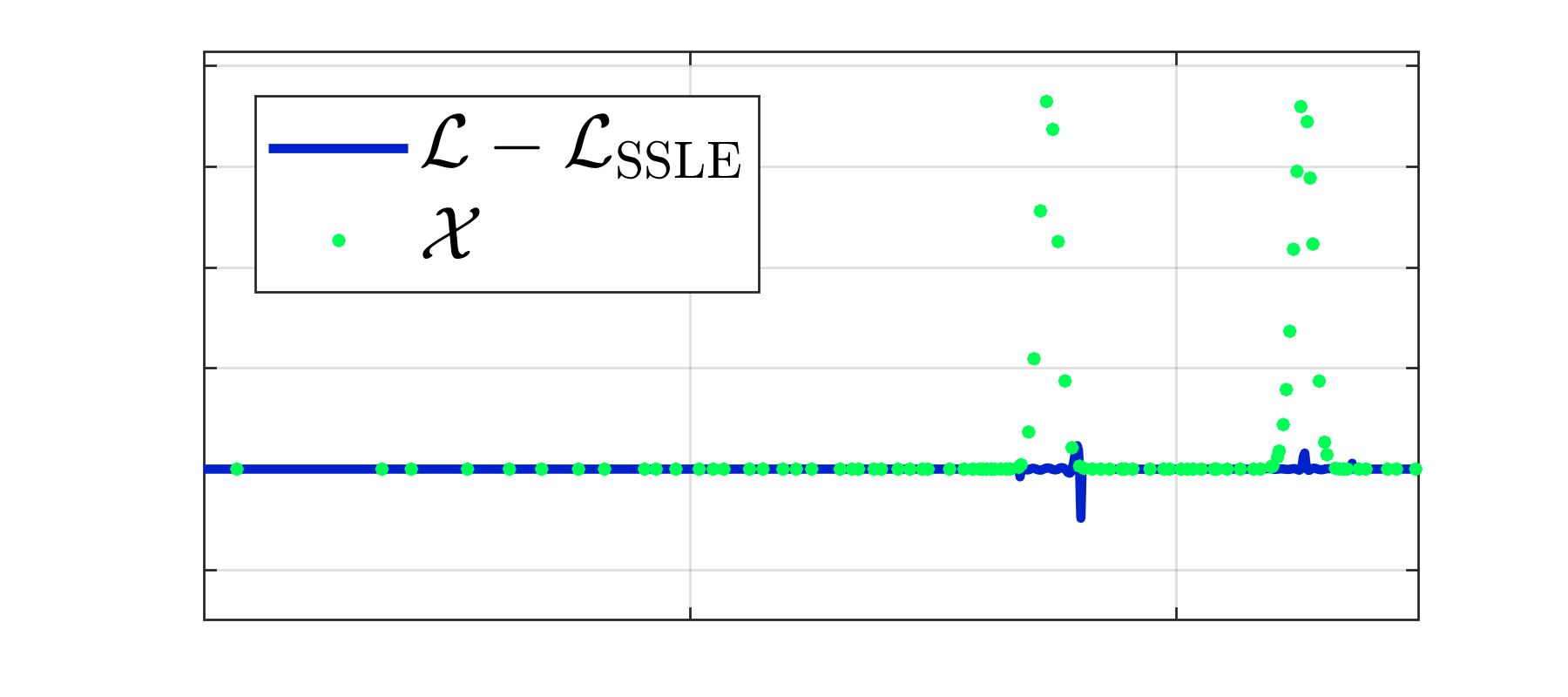}
			\includegraphics[width=0.9\linewidth, trim = 0 5 0 10, clip]{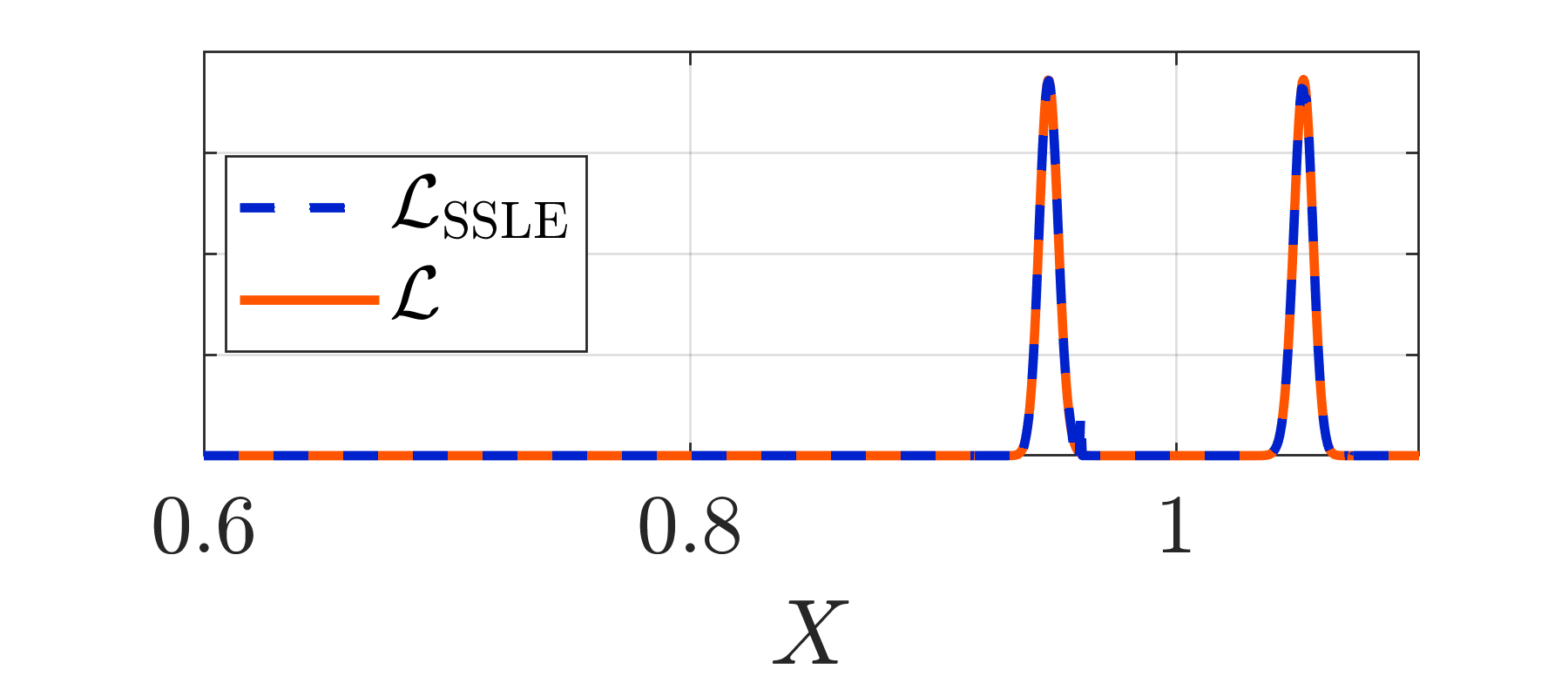}
		\end{minipage}
	}
	\caption{\emph{One-dimensional vibration problem}: Illustration of the adSSLE algorithm approximating the likelihood function $\cl$.}%
	\label{fig:cs1:SSEBehaviour}%
\end{figure}

We employ the adSSLE algorithm to approximate this likelihood function with 
$\NRefine = 10$. A few 
iterations from the solution process are shown in 
Figure~\ref{fig:cs1:SSEBehaviour}. 
The top plots show the subdomains $\cd_{X}^{\ell,p}$ constructed at each refinement step, highlighting the terminal domains $\ct$. 
The middle plots display the residual between the true likelihood and the approximation at the current iteration, as well as the adaptively chosen experimental design $\cX$. 
The bottom plots display the target likelihood function and its current approximation.

The initial global approximation of the 
first iteration in Figure~\subref*{fig:cs1:SSEBehaviour:1} is a constant 
polynomial based on the initial experimental design. By the third iteration, 
the algorithm has identified the subdomain $\cd_{X}^{2,2}$ as the one of 
interest and proceeds to refine it in subsequent steps. By the $8$th iteration 
both likelihood peaks have been identified. Finally, by the $10$th iteration 
in Figure~\subref*{fig:cs1:SSEBehaviour:4}, 
both likelihood peaks are approximated well by the adSSLE approach.

The last iteration shows how the algorithm splits domains and adds new sample points. There is a clear clustering of subdomains and sample points near the likelihood peaks at $X=0.95$ and $X=1.05$.

\begin{figure}
	\centering
	\includegraphics[width=0.8\linewidth]{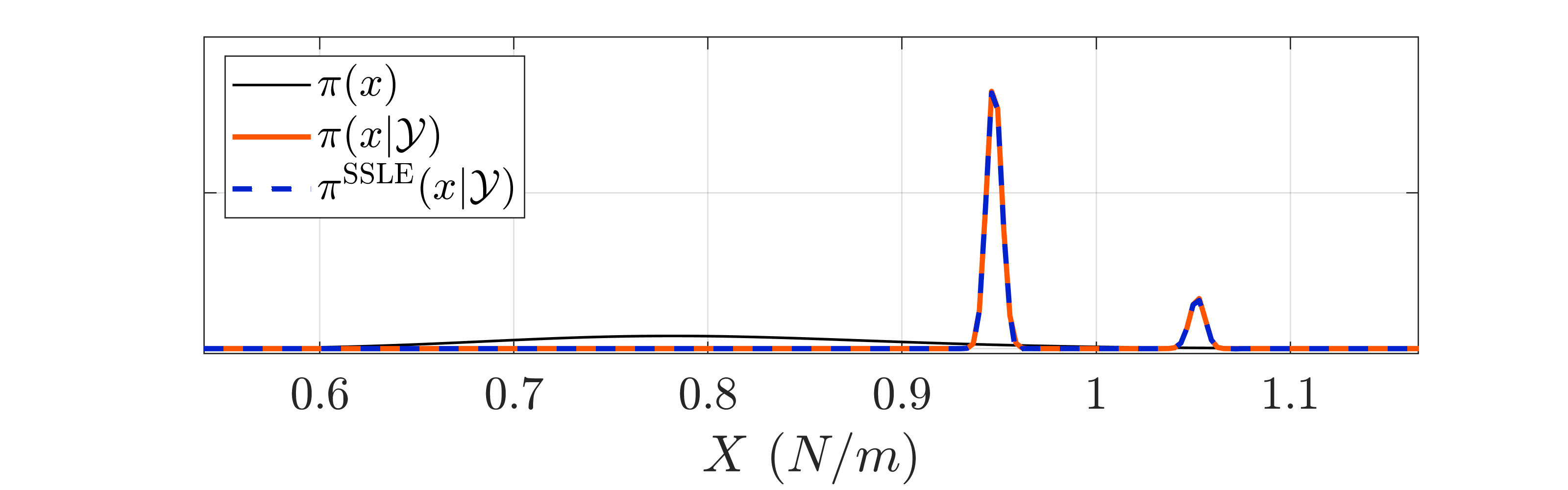}
	\caption{\emph{One-dimensional vibration problem}: Comparison of the true multimodal posterior and its adSSLE based 
		approximation.}%
	\label{fig:cs1:SSEResults}%
\end{figure}

The results from Eq.~\eqref{eq:SSE:SSEforBI:QoI} show that without further computations it would be possible to directly extract the posterior moments by post-processing the SSLE coefficients. In the present bimodal case, however, the posterior moments are not very meaningful. Instead, the available posterior approximation gives a full picture of the inferred parameter $X\vert\cy$. It is shown together with the true posterior and the original prior distribution in Figure~\ref{fig:cs1:SSEResults}. 

For this case study, non-adaptive experimental design approaches like the standard SSLE \citep{Marelli2020} and the original SLE algorithm \citep{NagelJCP2016} will almost surely fail for the considered experimental design of $\Ntot=100$. In numerous trial runs these approaches did not manage to accurately reconstruct the likelihood function due to a lack of \emph{informative samples} near the likelihood peaks.

\subsection{Moderate-dimensional heat transfer problem}
This case study is a complex engineering problem that can be modified to exhibit high posterior concentration in the prior domain. It was originally presented in \citet{NagelJCP2016} and solved there using SLE. We again solve the same problem with SSLE and compare the performance of SLE \citep{NagelJCP2016}, the original non-adaptive SSLE \citep{Marelli2020} and the proposed adSSLE approach presented in Section~\ref{sec:SSE:modifications}. To investigate the  performance of the algorithm in the case of high posterior concentrations (e.g. due to a large data-set), two instances of the problem with different discrepancy parameters are investigated.

\begin{figure}
	\centering
	\subfloat[Problem sketch]{
		\begin{minipage}{8cm}
			\label{fig:cs3:setup}
			\includegraphics[width=0.823\linewidth]{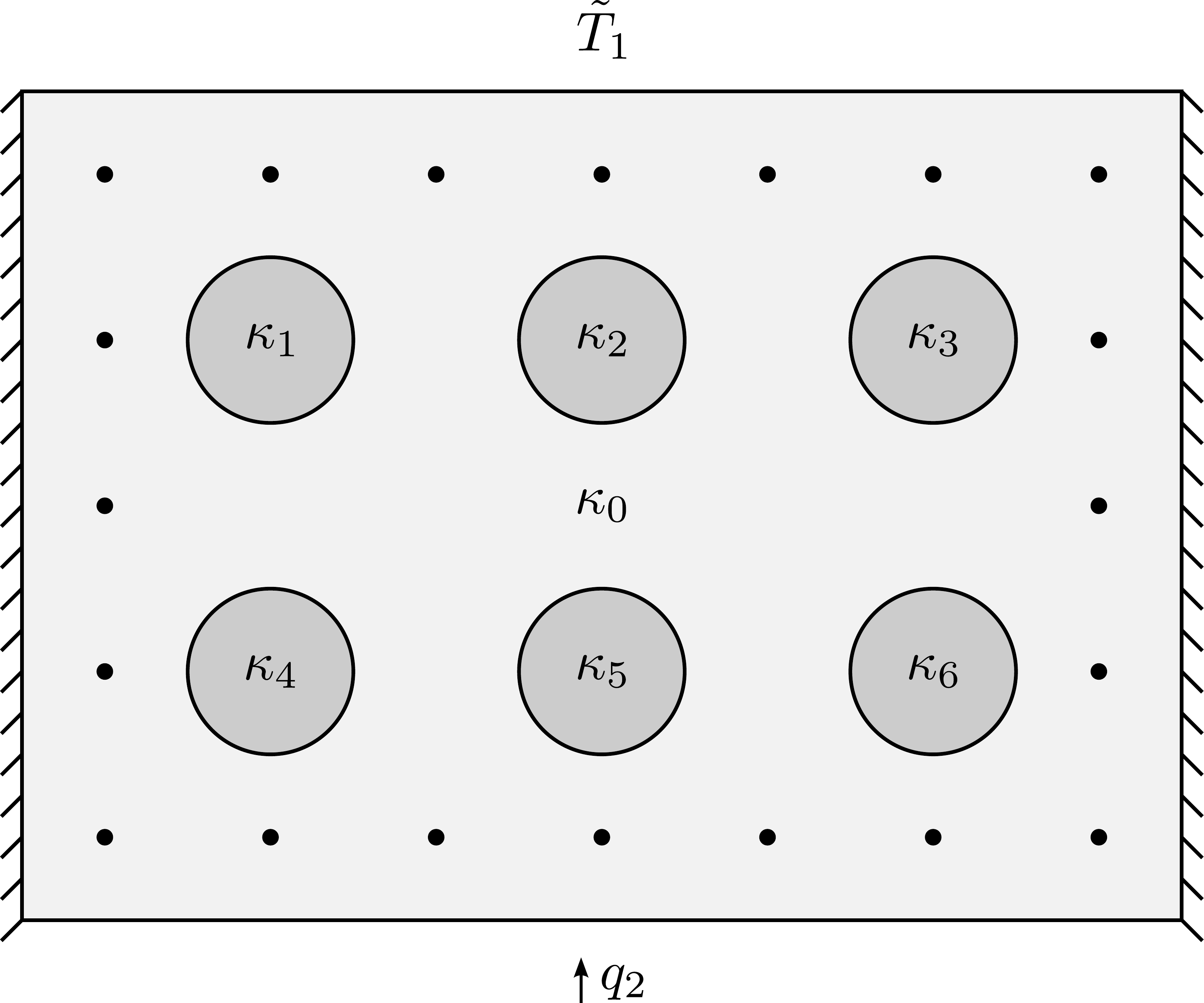}
			\vspace*{0.8cm}
		\end{minipage}
	}%
	\subfloat[Steady state solution]{
		\begin{minipage}{8cm}
			\label{fig:cs3:solution}
			\includegraphics[width=\linewidth]{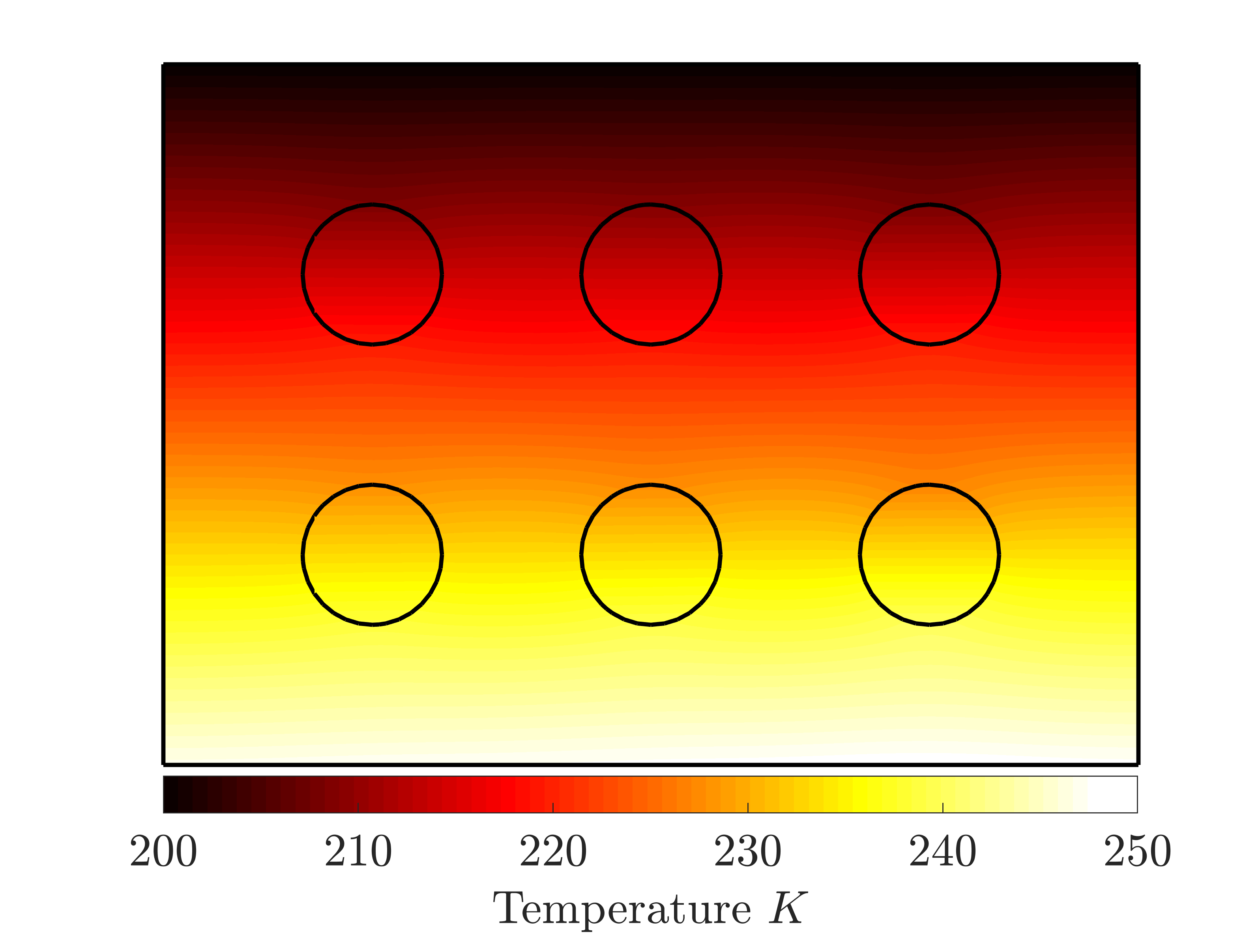}
		\end{minipage}
	}%
	\caption{\emph{Moderate-dimensional heat transfer problem}: 
		Model setup and exemplary solution.}
\end{figure}

Consider the diffusion-driven stationary heat transfer problem sketched in 
Figure~\subref*{fig:cs3:setup}. It models a 2D plate with a background matrix 
of 
constant thermal conductivity $\kappa_0$ and $6$ inclusions with  
conductivities 	
$\ve{\kappa}\eqdef(\kappa_1,\dots,\kappa_6)\trans$. The diffusion driven 
steady 
state 
heat distribution is described by a heat equation in Euclidean coordinates 
$\ve{r}\eqdef(r_1,r_2)\trans$ of the form
\begin{equation}
	\label{eq:ex3:heatTrans}
	\nabla\cdot (\kappa(\ve{r})\nabla\tilde{T}(\ve{r})) = 0,
\end{equation}
where the thermal conductivity field is denoted by $\kappa$ and the temperature 
field 
by $\tilde{T}$. The boundary conditions of the plate are given by a no-heat-flux 
Neumann boundary conditions on the \emph{left} and \emph{right} sides 
($\partial \tilde{T}/r_1=0$), a Neumann boundary condition on the \emph{bottom} 
(${\kappa_0\partial \tilde{T}/r_2=q_2}$) and a temperature $\tilde{T}_1$ 
Dirichlet boundary condition on the \emph{top}.

We employ a finite element (FE) solver to 
solve the weak form of Eq.~\eqref{eq:ex3:heatTrans} by discretizing the domain 
into approximately $10^5$ triangular elements. 
A sample solution returned by the FE-solver is shown in 
Figure~\ref{fig:cs3:solution}.

In this example we intend to infer the thermal conductivities $\ve{\kappa}$ 
of the inclusions. 
We assume the same problem constants as in 
\citet{NagelJCP2016} ({\em i.e.}, $q_2=2{,}000~\wmq$, $\tilde{T}_1=200~\K$, 
$\kappa_0=45\wmk$). The forward model $\cm$ takes as an input the 
conductivities of the inclusions $\ve{\kappa}$, solves the finite element 
problem and 
returns the steady state temperature 
$\tilde{\ve{T}}\eqdef(\tilde{T}_1,\dots,\tilde{T}_{20})\trans$ at the 
measurement points, {\em i.e.}, $\cm:\ve{\kappa}\mapsto\tilde{\ve{T}}$.

To solve the inverse problem, we assume a multivariate lognormal prior 
distribution with independent marginals on the inclusion conductivities, \ie 
$\pi(\ve{\kappa})=\prod_{i=1}^6\mathcal{LN}(\kappa_i\vert\mu=30~\wmk, 
\sigma=6~\wmk)$. 
We further assume an additive Gaussian discrepancy model, which yields the 
likelihood function
\begin{equation}
	\label{eq:ex3:likelihood}
	\cl(\ve{\kappa};\Bdata) = \frac{1}{2\pi\sigma^2}\exp 
	\left(-\frac{1}{2\sigma^2}\left(\ve{T}-\cm(\ve{\kappa})\right)\trans 
	\left(\ve{T}-\cm(\ve{\kappa})\right)\right),
\end{equation}
with a discrepancy standard deviation of $\sigma$.

As measurements, we generate one temperature field with 
$\hat{\ve{\kappa}}\eqdef(32,36,20,24,40,28)\trans~\wmk$ and collect its values 
at $20$ points indicated by black dots in 
Figure~\subref*{fig:cs3:setup}. We then perturb these temperature values with 
additive Gaussian noise and use them as the inversion data $\Bdata \eqdef 
\ve{T} = 
(T_1,\dots,T_{20})\trans$.

We look at two instances of this problem that differ only by the discrepancy parameter $\sigma$ from Eq.~\eqref{eq:ex3:likelihood}. The prior model response has a standard deviation of approximately $0.3~\K$, depending on the measurement point $T_i$. We therefore solve the 
problem first with a \emph{large} value $\sigma=0.25~\K$ and second with a 
\emph{small} value $\sigma=0.1~\K$. As the discrepancy standard deviation determines how peaked the 
likelihood function is, the first problem has a likelihood function with a 
much 
wider support and is in turn significantly easier to solve than the second 
one. It is noted here that in practice, the peakedness of the likelihood 
function is either increased by a smaller discrepancy standard deviation, or 
the inclusion of additional experimental data. 

To monitor the dependence of the algorithms on the number of likelihood 
evaluations, we solve both problems with a set of maximum likelihood 
evaluations $\Ntot=\{1{,}000; 2{,}000, 5{,}000; 10{,}000; 30{,}000\}$. The number of 
refinement samples is set to $\NRefine=1{,}000$. 

As a benchmark, we use reference posterior  
samples generated by the affine-invariant ensemble sampler
MCMC algorithm \citep{MCMC:Goodman2010} with $30{,}000$ steps and $50$ 
parallel chains, requiring a total of $\Ntot=1.5\cdot 10^6$ 
likelihood evaluations. 
Based on numerous heuristic convergence tests and due to the large number of MCMC steps, the resulting samples can be considered to accurately represent the true posterior distributions.

\begin{figure}
	\centering
	\subfloat[\emph{Large discrepancy}, $\sigma = 0.25~\K$]{
		\begin{minipage}{14cm}
			\includegraphics[width=\linewidth]{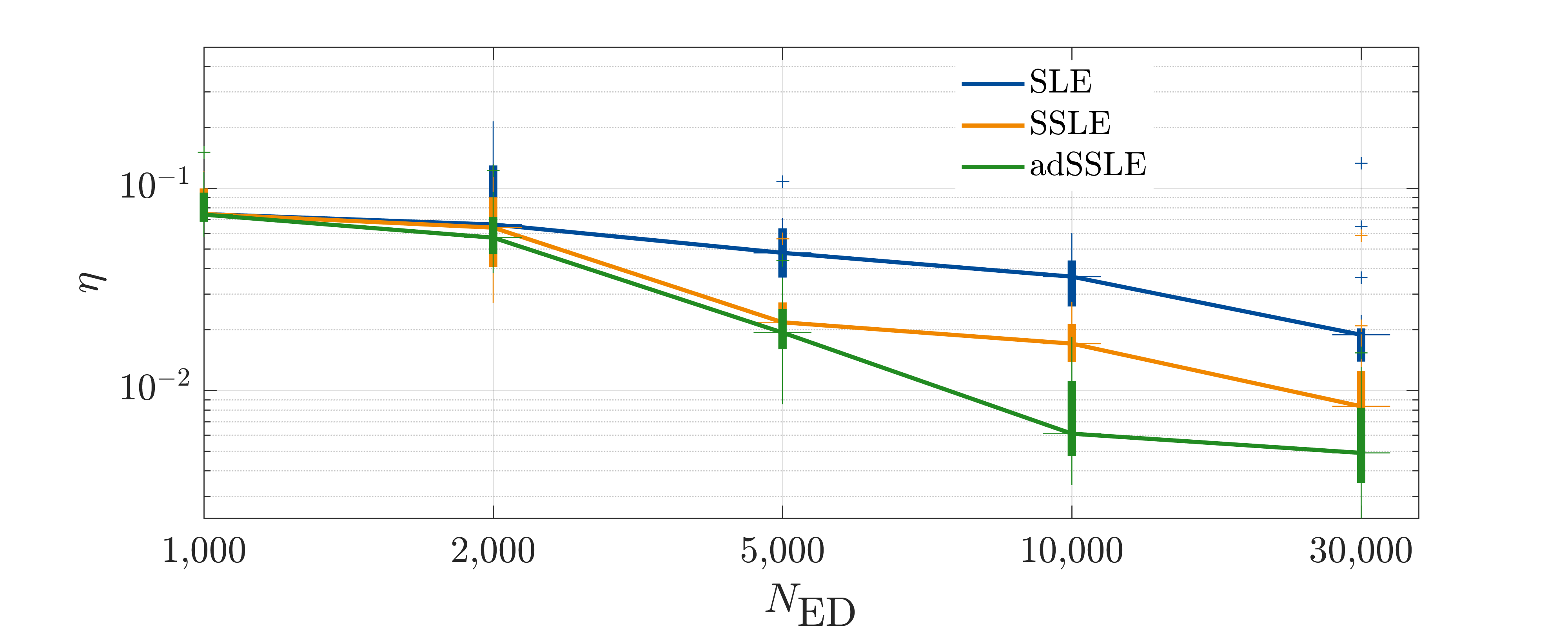}
		\end{minipage}
	}%
	\\
	\subfloat[\emph{Small discrepancy}, $\sigma = 0.1~\K$]{
		\begin{minipage}{14cm}
			\includegraphics[width=\linewidth]{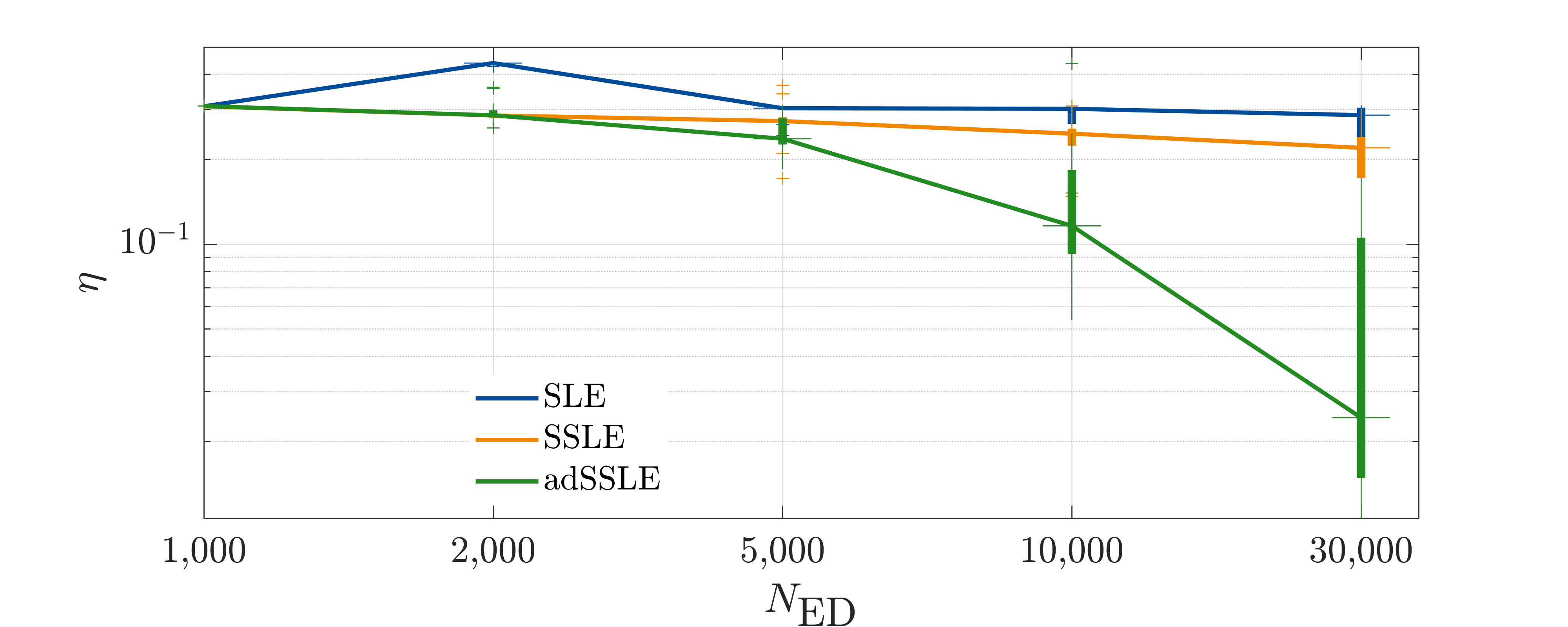}
		\end{minipage}
	}%
	\caption{\emph{Moderate-dimensional heat transfer problem}: Convergence of the $\eta$ error measure 
		(Eq.~\eqref{eq:eq3:errorMeasure}) as a function of the experimental design size $\Ntot$ in $20$ replications for SLE, SSLE with a 
		static experimental design and the proposed adSSLE approach. We 
		display the two discrepancy standard deviation cases $\sigma = 
		\{0.25,0.1\}~\K$.}%
	\label{fig:cs2:Convergence}%
\end{figure}

The results of the analyses are summarized in Figure~\ref{fig:cs2:Convergence}, where the error measure $\eta$ is plotted against the number of likelihood evaluations for the \emph{large} and \emph{small} standard deviation case. 
For the \emph{large} discrepancy standard deviation case, both SSLE approaches 
clearly outperform standard SLE {\em w.r.t.}\ the error measure $\eta$. This is 
most significant at mid-range experimental designs ($\Ntot={5{,}000; 10{,}000}$), 
where SLE does not reach the 
required high degrees and fails to accurately approximate the likelihood 
function. At larger experimental designs SLE catches up to non-adaptive SSLE but 
is still outperformed by the proposed adSSLE approach.
The real strength of the adaptive algorithm shows for the case of a \emph{small} discrepancy standard deviation, where the limitations of fixed experimental designs become obvious. When the likelihood function is nonzero
in a small subdomain of the prior, the global SLE and non-adaptive SSLE approach will fail in 
practice because of the insufficient number of samples placed in the 
informative regions. The adSSLE approach, however, works very well 
in these types of problems. It manages to identify the regions of interest and 
produces a likelihood approximation that accurately reproduces the posterior 
marginals.

Tables~\ref{tab:cs2:largeVar} and 
\ref{tab:cs2:smallVar} show the convergence of the adSSLE method moment 
estimate (mean and variance) to the reference solution for a single run. 
In brackets next to the moment estimates $\xi$, the relative error $\epsilon\eqdef\vert\xi_{\mathrm{MCMC}}-\xi_{\mathrm{SSLE}}\vert/\xi_{\mathrm{MCMC}}$ is also shown. 
Due to the non-strict positivity of the SSLE estimate, one variance estimate computed 
with Eq.~\eqref{eq:SSE:SSEforBI:QoI} is negative and is therefore omitted 
from Table~\ref{tab:cs2:smallVar}.

\renewcommand\arraystretch{0.8}
\setlength{\tabcolsep}{4pt} 
\begin{table}
	\caption{\emph{Moderate-dimensional heat transfer problem}: adSSLE results with \emph{large} discrepancy standard 
		deviation 
		$\sigma = 0.25~\K$. Relative errors {\em w.r.t.}\ MCMC reference solution 
		are shown in brackets.}
	\label{tab:cs2:largeVar}
	\centering
	\footnotesize
	\begin{tabular}{rl|cccccc}
		\hline&$\Esp{\kappa_i\vert\cy}$&$\kappa_{1}$&$\kappa_{2}$&$\kappa_{3}$&$\kappa_{4}$&$\kappa_{5}$&$\kappa_{6}$\\
		\hline\textbf{adSSLE}&$\Ntot =1{,}000$&$30.00 (0.7\%)$&$30.00 (7.1\%)$&$19.51 (5.8\%)$&$30.01 (7.1\%)$&$35.39 (2.9\%)$&$30.00 (14.6\%)$\\
		&$\Ntot =2{,}000$&$31.74 (6.5\%)$&$29.80 (7.7\%)$&$20.89 (0.9\%)$&$30.00 (7.1\%)$&$35.56 (2.4\%)$&$28.30 (8.1\%)$\\
		&$\Ntot =5{,}000$&$29.70 (0.4\%)$&$31.74 (1.7\%)$&$20.21 (2.4\%)$&$30.21 (6.5\%)$&$36.45 (0.0\%)$&$27.58 (5.4\%)$\\
		&$\Ntot =10{,}000$&$29.96 (0.5\%)$&$32.41 (0.3\%)$&$19.90 (3.9\%)$&$31.55 (2.3\%)$&$36.57 (0.3\%)$&$26.20 (0.1\%)$\\
		&$\Ntot =30{,}000$&$29.92 (0.4\%)$&$32.40 (0.3\%)$&$19.94 (3.7\%)$&$31.96 (1.1\%)$&$36.53 (0.2\%)$&$26.15 (0.1\%)$\\
		\hline\textbf{MCMC}&&$\mathbf{29.81}$&$\mathbf{32.30}$&$\mathbf{20.71}$&$\mathbf{32.31}$&$\mathbf{36.45}$&$\mathbf{26.18}$\\
		\hline\hline&$\sqrt{\Var{\kappa_i\vert\cy}}$&&&&&&\\
		\hline\textbf{adSSLE}&$\Ntot =1{,}000$&$6.00 (86.5\%)$&$6.01 (41.7\%)$&$6.00 (142.0\%)$&$5.99 (18.1\%)$&$4.44 (15.9\%)$&$5.99 (97.1\%)$\\
		&$\Ntot =2{,}000$&$6.09 (89.3\%)$&$4.78 (12.9\%)$&$3.44 (38.6\%)$&$6.00 (18.1\%)$&$4.34 (13.5\%)$&$5.38 (76.9\%)$\\
		&$\Ntot =5{,}000$&$2.19 (32.0\%)$&$5.61 (32.4\%)$&$3.09 (24.4\%)$&$6.06 (19.4\%)$&$3.62 (5.5\%)$&$4.78 (57.2\%)$\\
		&$\Ntot =10{,}000$&$2.62 (18.7\%)$&$4.40 (3.7\%)$&$2.82 (13.6\%)$&$5.15 (1.5\%)$&$3.96 (3.6\%)$&$3.14 (3.4\%)$\\
		&$\Ntot =30{,}000$&$2.40 (25.5\%)$&$4.38 (3.4\%)$&$2.70 (8.8\%)$&$5.23 (3.1\%)$&$3.87 (1.2\%)$&$3.19 (5.0\%)$\\
		\hline\textbf{MCMC}&&$\mathbf{3.22}$&$\mathbf{4.24}$&$\mathbf{2.48}$&$\mathbf{5.08}$&$\mathbf{3.83}$&$\mathbf{3.04}$\\
		\hline
	\end{tabular}
\end{table}

\begin{table}
	\caption{\emph{Moderate-dimensional heat transfer problem}: adSSLE results with \emph{small} discrepancy standard 
		deviation 
		$\sigma = 0.1~\K$. Relative errors {\em w.r.t.}\ MCMC reference solution 
		are shown in brackets. Field with an asterisk ($*$) indicates negative variance estimate.}
	\label{tab:cs2:smallVar}
	\centering
	\footnotesize
	\begin{tabular}{rl|cccccc}
		\hline&$\Esp{\kappa_i\vert\cy}$&$\kappa_{1}$&$\kappa_{2}$&$\kappa_{3}$&$\kappa_{4}$&$\kappa_{5}$&$\kappa_{6}$\\
		\hline\textbf{adSSLE}&$\Ntot =1{,}000$&$30.00 (3.8\%)$&$30.00 (10.7\%)$&$30.00 (62.2\%)$&$30.00 (5.7\%)$&$30.00 (22.8\%)$&$30.00 (17.6\%)$\\
		&$\Ntot =2{,}000$&$30.00 (3.8\%)$&$34.58 (2.9\%)$&$30.00 (62.2\%)$&$30.00 (5.7\%)$&$30.00 (22.8\%)$&$30.00 (17.6\%)$\\
		&$\Ntot =5{,}000$&$30.00 (3.8\%)$&$34.70 (3.2\%)$&$25.30 (36.8\%)$&$30.00 (5.7\%)$&$37.28 (4.0\%)$&$30.00 (17.6\%)$\\
		&$\Ntot =10{,}000$&$30.00 (3.8\%)$&$34.71 (3.3\%)$&$18.92 (2.3\%)$&$30.00 (5.7\%)$&$37.87 (2.5\%)$&$25.44 (0.3\%)$\\
		&$\Ntot =30{,}000$&$31.16 (0.0\%)$&$34.28 (2.0\%)$&$18.57 (0.4\%)$&$31.37 (1.4\%)$&$38.68 (0.4\%)$&$25.56 (0.2\%)$\\
		\hline\textbf{MCMC}&&$\mathbf{31.17}$&$\mathbf{33.61}$&$\mathbf{18.49}$&$\mathbf{31.83}$&$\mathbf{38.84}$&$\mathbf{25.51}$\\
		\hline\hline&$\sqrt{\Var{\kappa_i\vert\cy}}$&&&&&&\\
		\hline\textbf{adSSLE}&$\Ntot =1{,}000$&$6.00 (268.3\%)$&$6.00 (149.7\%)$&$6.00 (362.1\%)$&$6.00 (69.4\%)$&$5.99 (202.8\%)$&$5.99 (264.7\%)$\\
		&$\Ntot =2{,}000$&$6.00 (268.5\%)$&$4.53 (88.9\%)$&$6.00 (362.2\%)$&$6.00 (69.5\%)$&$6.00 (203.1\%)$&$6.00 (265.2\%)$\\
		&$\Ntot =5{,}000$&$6.00 (268.5\%)$&$4.43 (84.3\%)$&$2.87 (120.8\%)$&$6.00 (69.5\%)$&$4.29 (116.8\%)$&$6.00 (265.2\%)$\\
		&$\Ntot =10{,}000$&$4.49 (176.0\%)$&$*(*)$&$1.84 (42.1\%)$&$2.58 (27.2\%)$&$3.89 (96.6\%)$&$2.66 (61.8\%)$\\
		&$\Ntot =30{,}000$&$1.20 (26.6\%)$&$3.60 (49.8\%)$&$1.62 (25.2\%)$&$2.66 (24.7\%)$&$3.06 (54.6\%)$&$1.74 (5.9\%)$\\
		\hline\textbf{MCMC}&&$\mathbf{1.63}$&$\mathbf{2.40}$&$\mathbf{1.30}$&$\mathbf{3.54}$&$\mathbf{1.98}$&$\mathbf{1.64}$\\
		\hline
	\end{tabular}
\end{table}

\begin{figure}
	\centering
	
	\subfloat[\emph{Large discrepancy}, $\sigma = 0.25~\K$, adSSLE]{
		\begin{minipage}{8cm}
			\includegraphics[width=\linewidth]{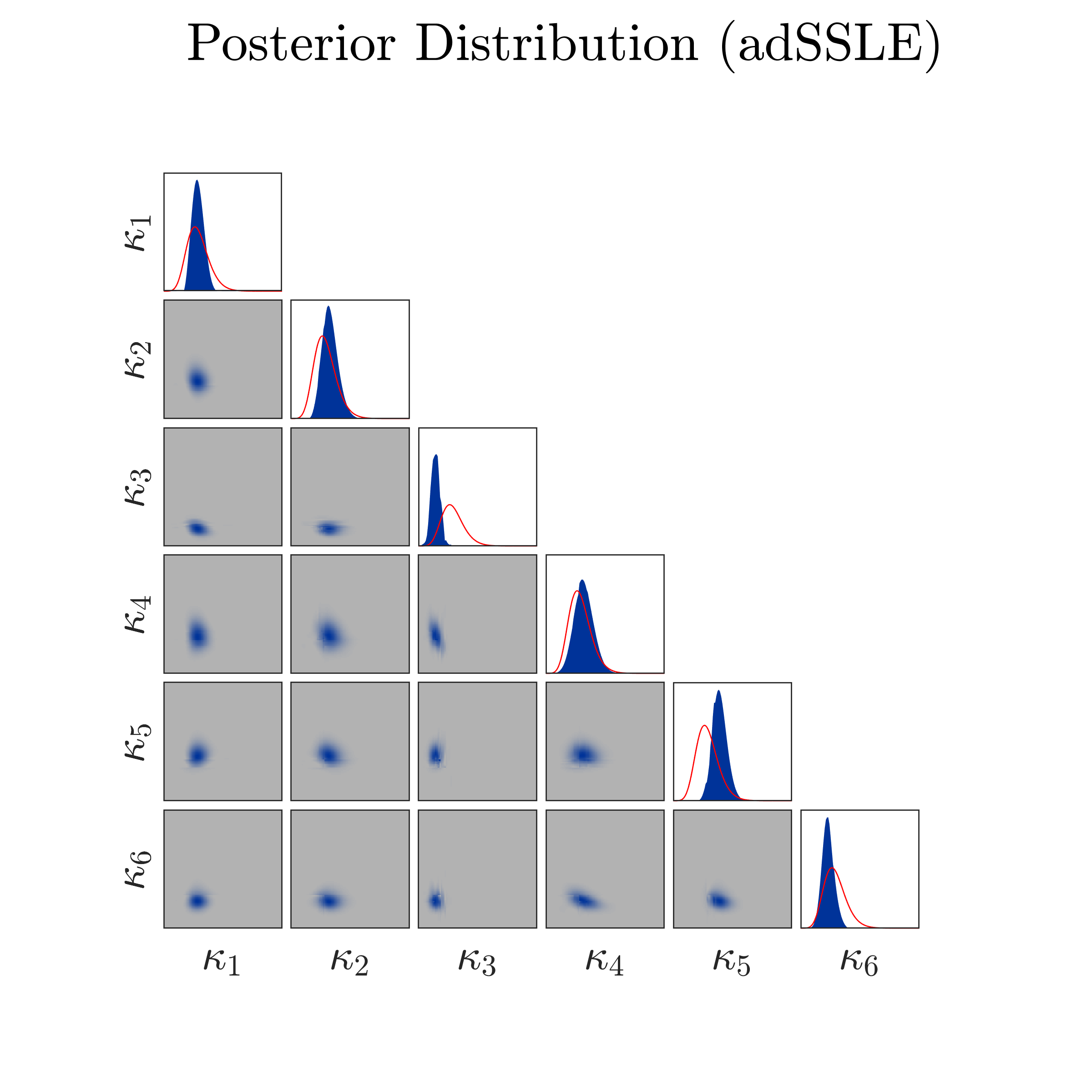}
		\end{minipage}
	}%
	\subfloat[\emph{Large discrepancy}, $\sigma = 0.25~\K$, MCMC]{
		\begin{minipage}{8cm}
			\includegraphics[width=\linewidth]{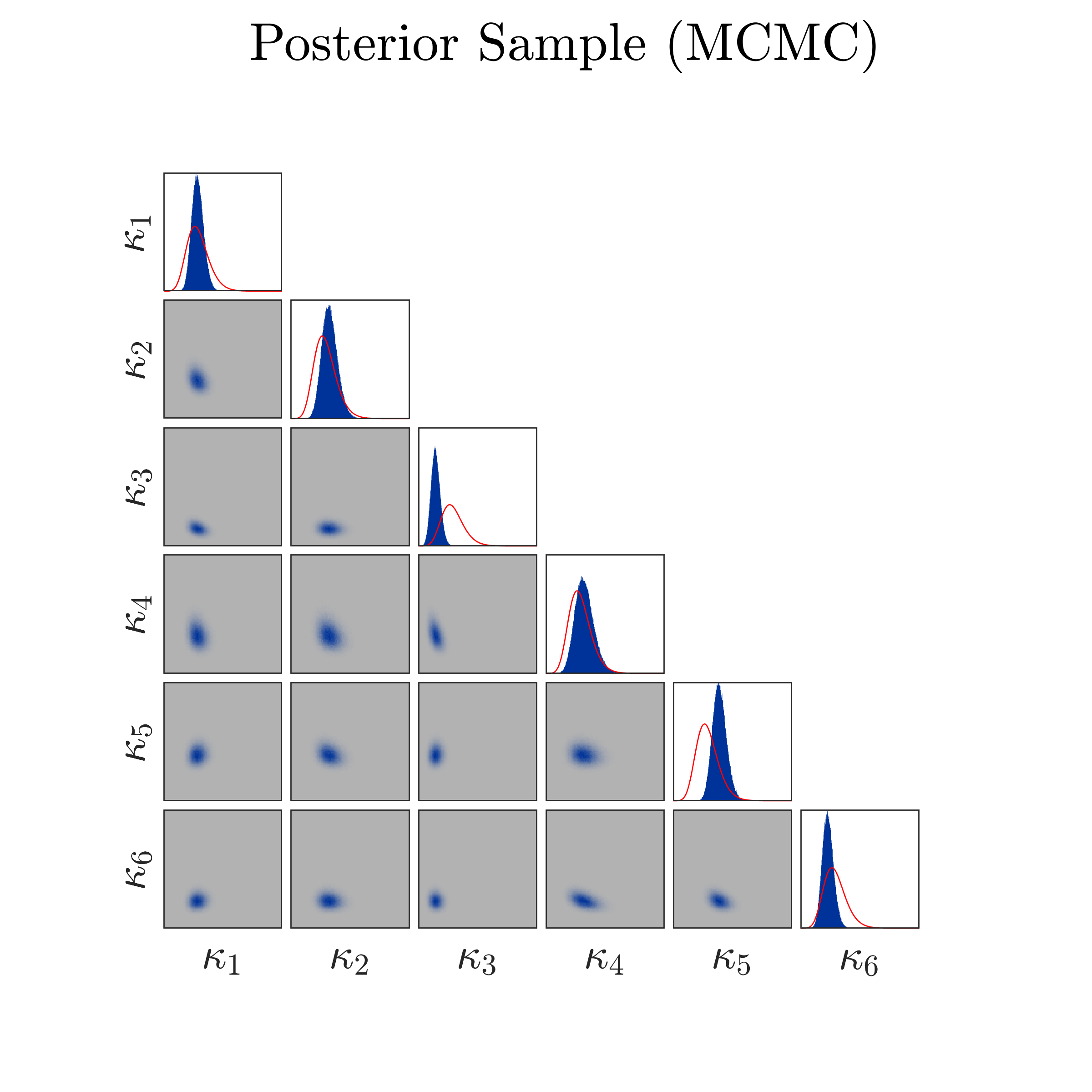}
		\end{minipage}
	}%
	\\
	\subfloat[\emph{Small discrepancy}, $\sigma = 0.1~\K$, adSSLE]{
		\begin{minipage}{8cm}
			\includegraphics[width=\linewidth]{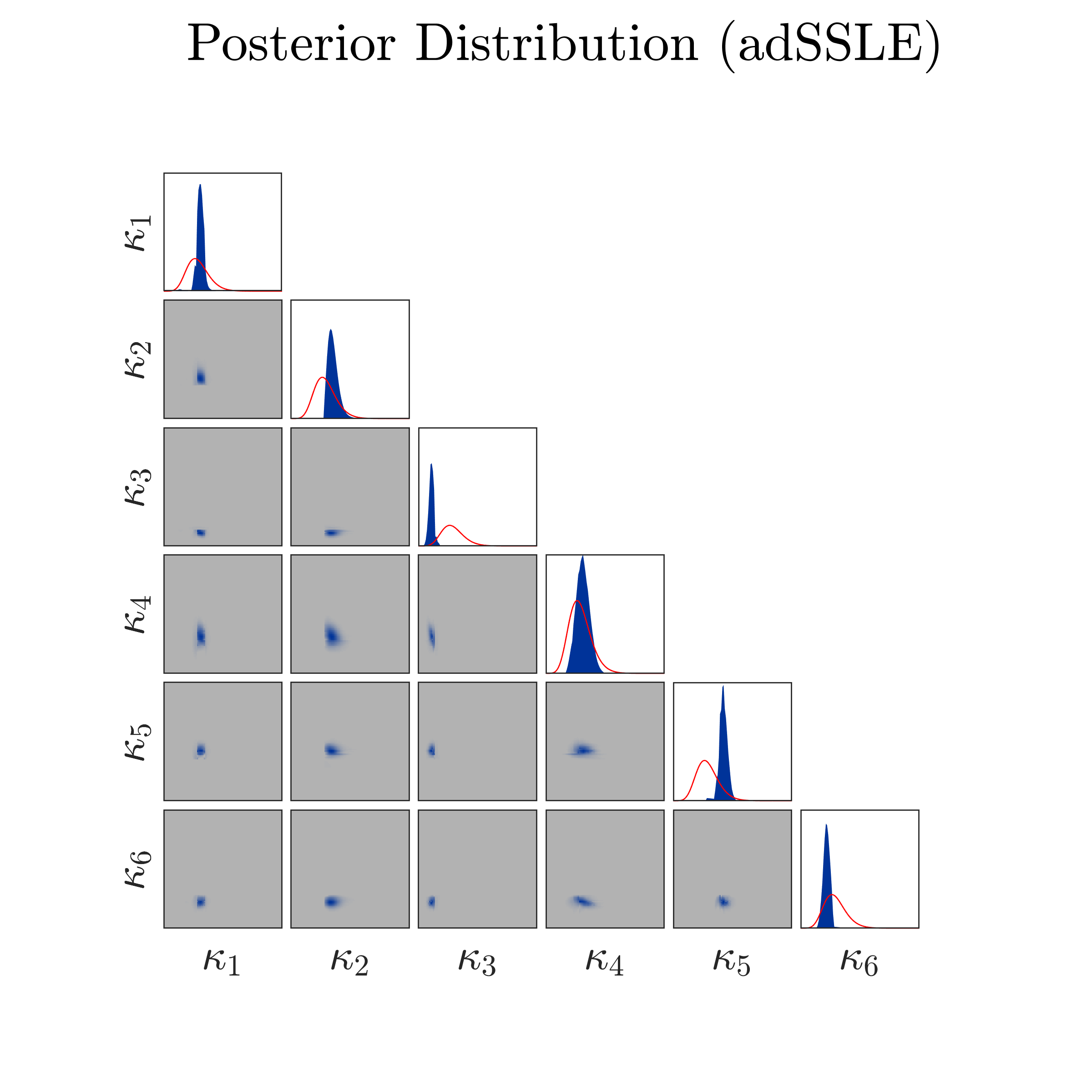}
		\end{minipage}
		\label{fig:cs2:Posterior:SSESmall}
	}%
	\subfloat[\emph{Small discrepancy}, $\sigma = 0.1~\K$, MCMC]{
		\begin{minipage}{8cm}
			\includegraphics[width=\linewidth]{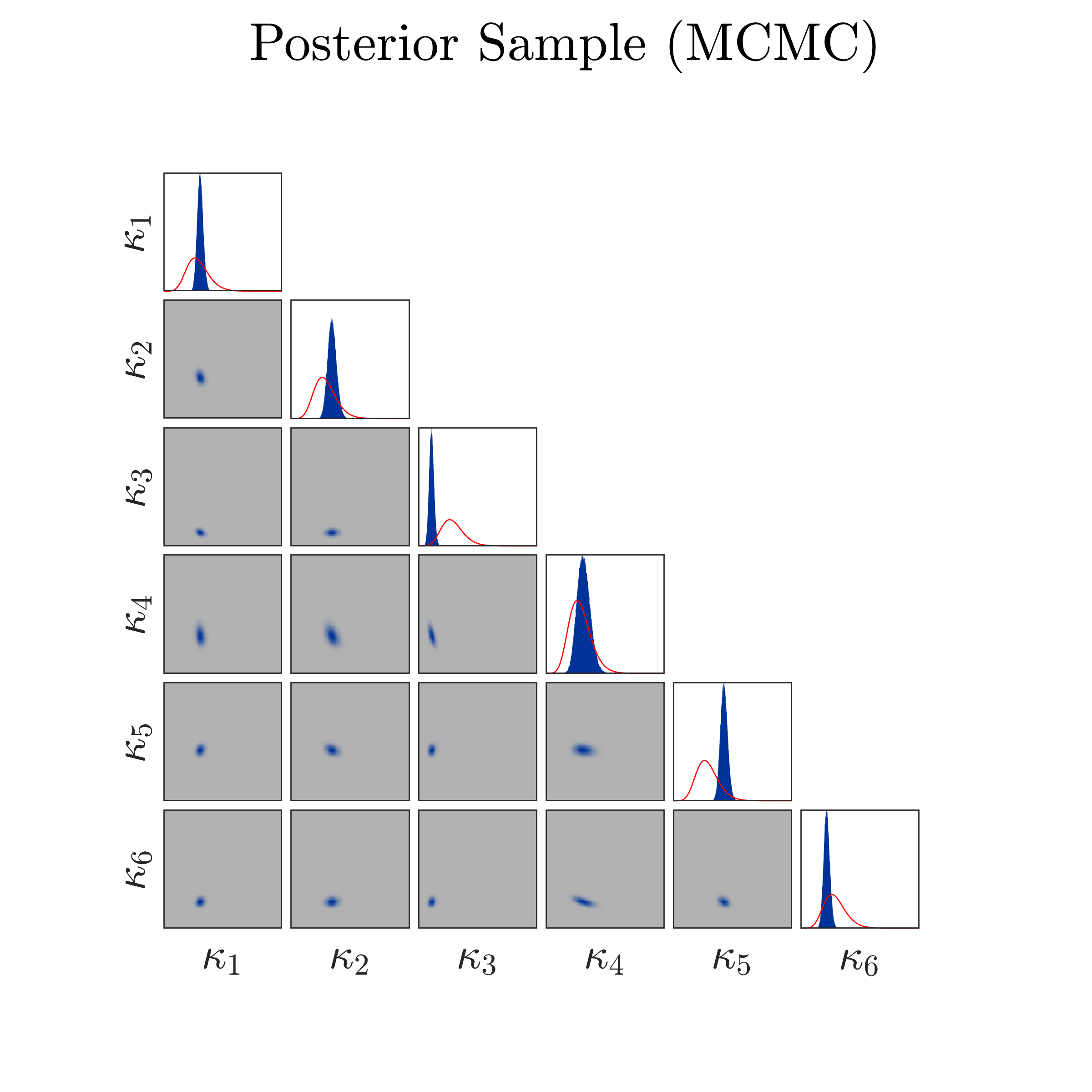}
		\end{minipage}
	}%
	\caption{\emph{Moderate-dimensional heat transfer problem}: Comparison plots of the posterior distribution marginals computed 
		from adSSLE ($\Ntot=30{,}000$) and MCMC ($\Ntot = 1.5\cdot 10^6$). The prior marginals are shown in red.}%
	\label{fig:cs2:Posterior}%
\end{figure}

The full posterior marginals obtained from one run of adSSLE with 
$\Ntot=30{,}000$ are also compared to those of the reference MCMC and 
displayed in Figure~\ref{fig:cs2:Posterior}. The individual plots show the univariate posterior marginals (\emph{i.e.} $\pi(x_i\vert\cY)$) on the main diagonal and the bivariate posterior marginals (\emph{i.e.}\ $\pi(\vx_{ij}\vert\cY)$) in the $i$-th row and $j$-th column. It can be clearly seen that the 
posterior characteristics are very well captured. However, the adSSLE approach sometimes fails to accurately represent the tails of the distribution. This is especially obvious in the small discrepancy case in Figure~\subref*{fig:cs2:Posterior:SSESmall} where the tail is sometimes cut off. We emphasize here that the SSLE marginals are obtained analytically as 1D and 2D surfaces for the univariate and bivariate marginals respectively. For the reference MCMC approach, on the other hand, they need to be approximated with histograms based on the available posterior sample.

\subsubsection{ Convergence of the posterior moments}
\label{sec:applications:ex2:convergence}
In practical inference applications, posterior moments are often one of the main quantities of interest. An estimator of these moments is readily available at every refinement step of adSSLE through Eq.~\eqref{eq:SSE:SSEforBI:QoI}. 

\begin{figure}
	\centering
	
	\subfloat[Mean]{
		\begin{minipage}{8cm}
			\includegraphics[width=\linewidth]{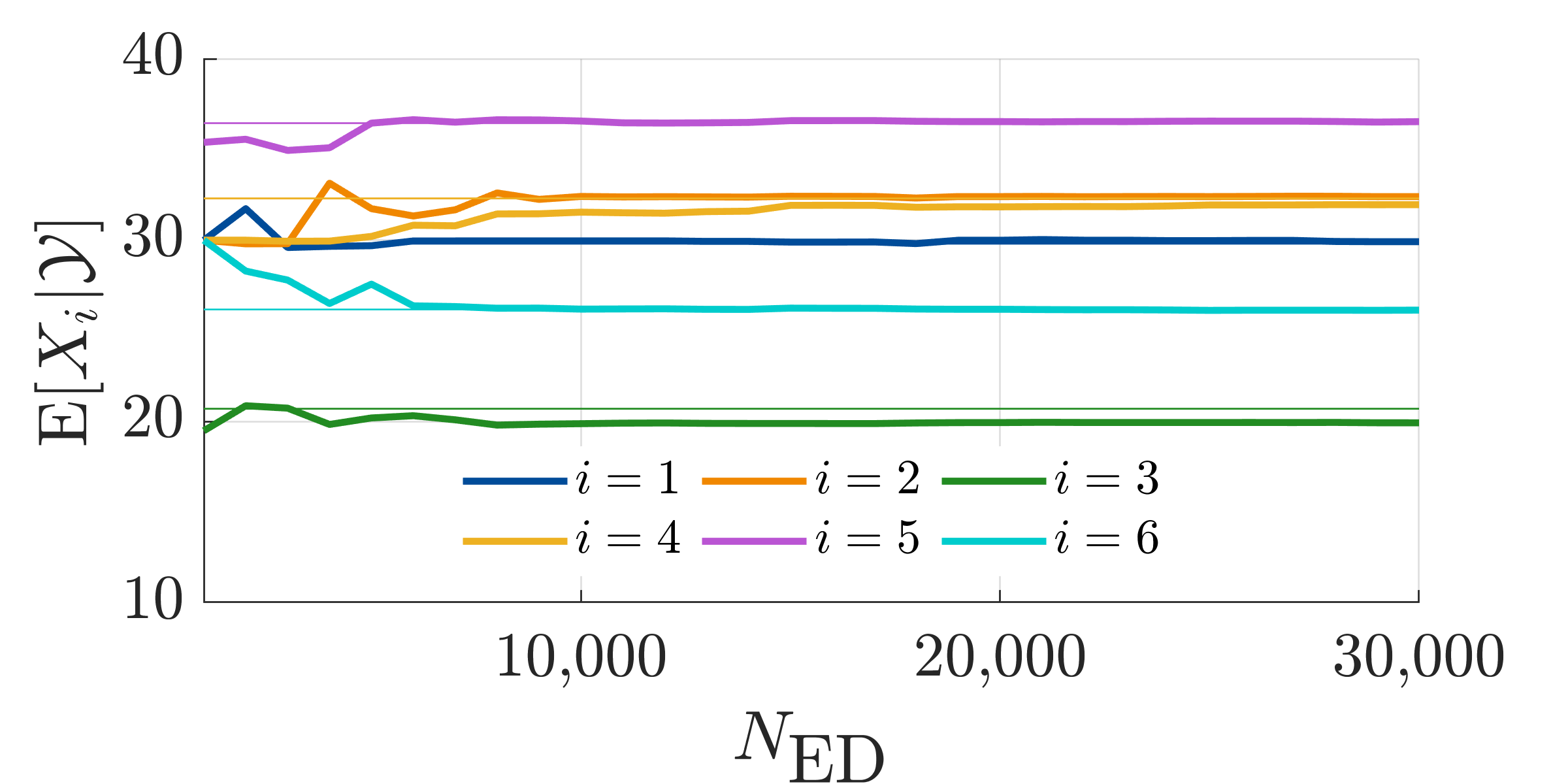}
		\end{minipage}
	}%
	\subfloat[Standard deviation]{
		\begin{minipage}{8cm}
			\includegraphics[width=\linewidth]{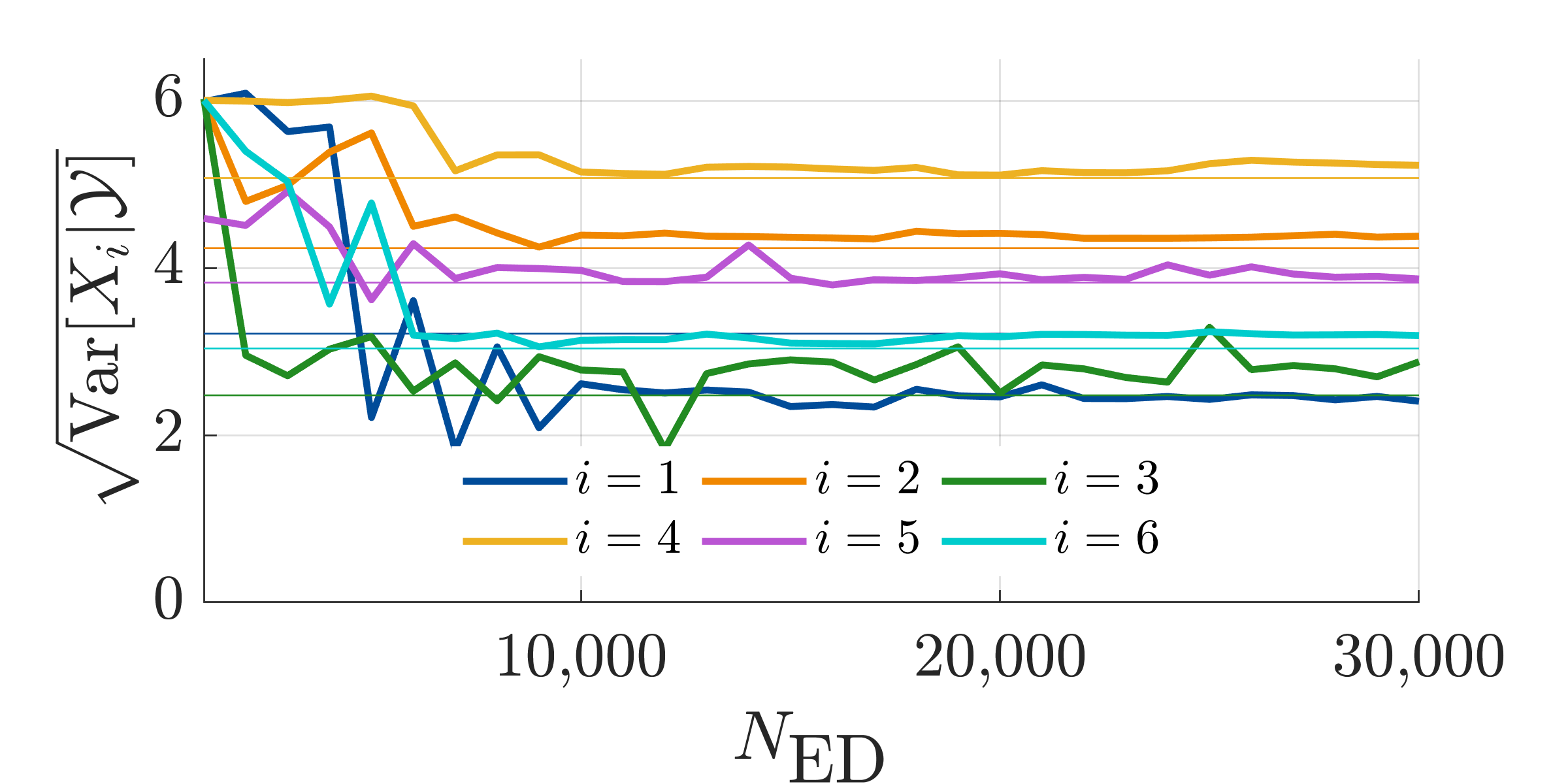}
		\end{minipage}
	}%
	\caption{\emph{Moderate-dimensional heat transfer problem}: Evolution of the posterior moment estimation for a typical run of the adSSLE algorithm on the \emph{small discrepancy problem}. The thin lines show the MCMC reference solution.}%
	\label{fig:cs2:ConvergenceMoments}%
\end{figure}

Tracking the evolution of the posterior moments throughout the adSSLE iterations can be used as a heuristic estimator of the convergence of the adSSLE algorithm. However, only the stability of the solution can be assessed, without guarantees on the bias.
As an example, we now consider the \emph{large discrepancy problem} and plot the evolution of the posterior mean and standard deviation for every $X_i$ as a function of the number of likelihood evaluations in Figure~\ref{fig:cs2:ConvergenceMoments}. It can be seen that after $\sim10{,}000$ likelihood evaluations, most moment estimators achieve convergence to a value close to the reference solution. This plot also reveals a small bias of the $\Esp{X_3\vert\cy}$ and $\sqrt{\Var{X_3\vert\cy}}$ estimators, that was previously highlighted in Table~\ref{tab:cs2:largeVar}.

\subsubsection{Influence of $\NRefine$}
\label{sec:applications:ex2:nref}
\begin{figure}
	\centering
	\subfloat[\emph{Large discrepancy}, $\sigma = 0.25~\K$]{
		\begin{minipage}{14cm}
			\includegraphics[width=\linewidth]{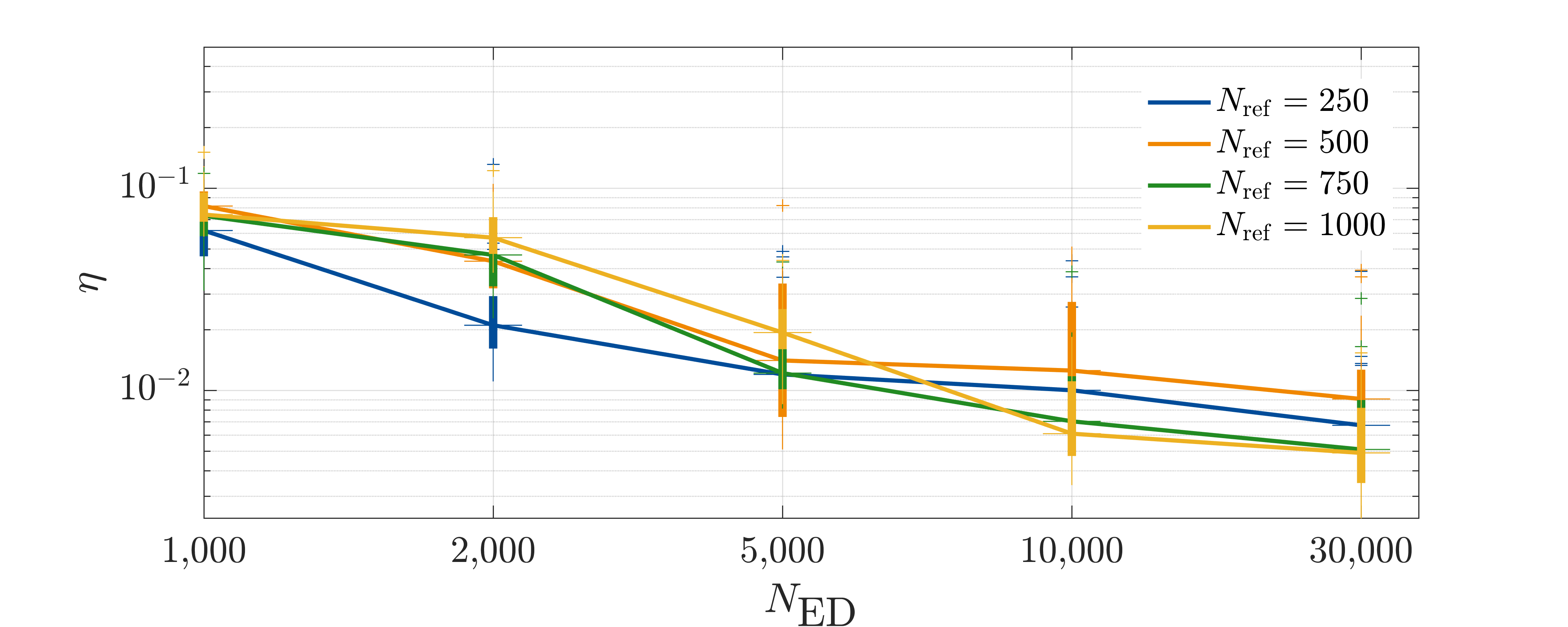}
		\end{minipage}
	}%
	\\
	\subfloat[\emph{Small discrepancy}, $\sigma = 0.1~\K$]{
		\begin{minipage}{14cm}
			\includegraphics[width=\linewidth]{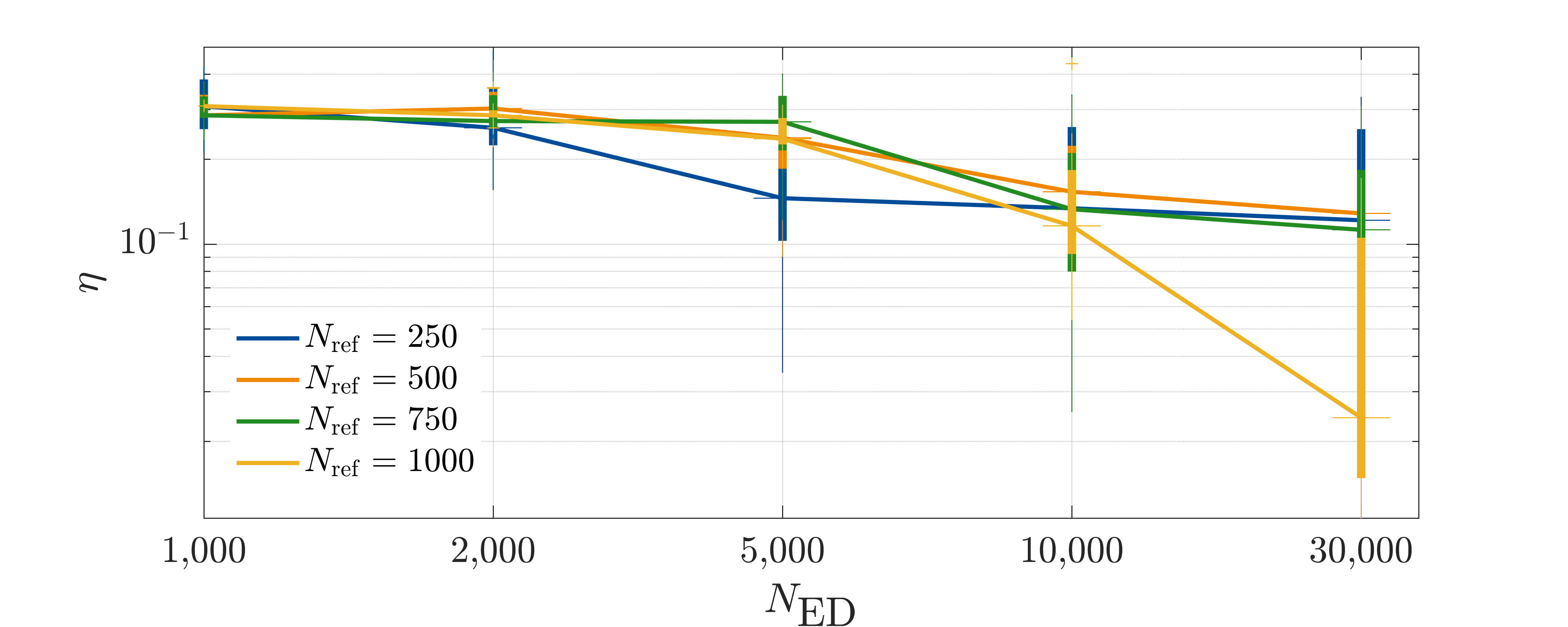}
		\end{minipage}
	}%
	\caption{\emph{Moderate-dimensional heat transfer problem}: Convergence of the $\eta$ error measure 
		(Eq.~\eqref{eq:eq3:errorMeasure}) as a function of the experimental design size $\Ntot$ in $20$ replications for the proposed adSSLE approach with different $\NRefine$ parameters. We display the two discrepancy 
		standard deviation cases $\sigma = \{0.25,0.1\}~\K$.}%
	\label{fig:cs2:ConvergenceNRefine}%
\end{figure}

The main hyperparameter of the proposed adSSLE algorithm is the number $\NRefine$, which corresponds to the number of sample points that are required at each PCE construction step (see Section~\ref{sec:SSE:modifications}). 
In Figure~\ref{fig:cs2:ConvergenceNRefine} we display the effect of different $\NRefine$ values on the convergence in the small and large discrepancy problems. 

$\NRefine$ influences the accuracy of the two error estimators used inside the adSSLE algorithm. They are: (i) the residual expansion accuracy $E_{\mathrm{LOO}}$ in Eq.~\eqref{eq:algo:error} and (ii) the splitting error $E_{\mathrm{split}}$ in Eq.~\eqref{eq:SSE:algorithmMod:splitError}. 

Small values of $\NRefine$ allow to quickly obtain a crude likelihood approximation with limited experimental design sizes $\Ntot$, but this comes at the cost of lower convergence rates at larger $\Ntot$. This behaviour can be partially attributed to the deterioration of residual expansion error $E_{\mathrm{LOO}}$ in Eq.~\eqref{eq:algo:error}. At small experimental design sizes, the overall number of terminal domains is relatively small and this effect is not as pronounced. At larger experimental designs and higher numbers of subdomains, however, the error estimators high variances can lead to difficulties in identifying the true high error subdomains.

Large values of $\NRefine$ lead to slower initial convergence rates because of the smaller number of overall subdomains. The algorithm stability, however, is increased because both error estimators have lower variance and thereby allow the algorithm to more reliably identify the true high error subdomains and choose the split directions that maximize Eq.~\eqref{eq:SSE:algorithmMod:splitDirection}.

\subsection{High-dimensional diffusion problem}
\label{sec:cs:ex3}

The last cast study shows that adSSLE for Bayesian model inversion remains feasible in high dimensional problems with low effective dimensionality. The considered forward model is often used as a standard benchmark in UQ computations \citep{Shin2016, Fajraoui2017}. It 
represents the 1-D diffusion along a domain with coordinate $\xi\in[0, 1]$ given by the following 
boundary value problem:
\begin{equation}
	-\frac{\partial}{\partial \xi}\left[\kappa(\xi)\frac{\partial 
		u}{\partial 
		\xi}(\xi)\right] = 1, \quad \text{with} \quad 
	\begin{cases}
		u(0)=0,\\
		\frac{\partial u}{\partial \xi}(1) = 1.
	\end{cases}
\end{equation}

\begin{figure}
	\centering
	\subfloat[Diffusion coefficient]{
		\begin{minipage}{8cm}
			\label{fig:cs4:diffusion}
			\includegraphics[width=\linewidth]{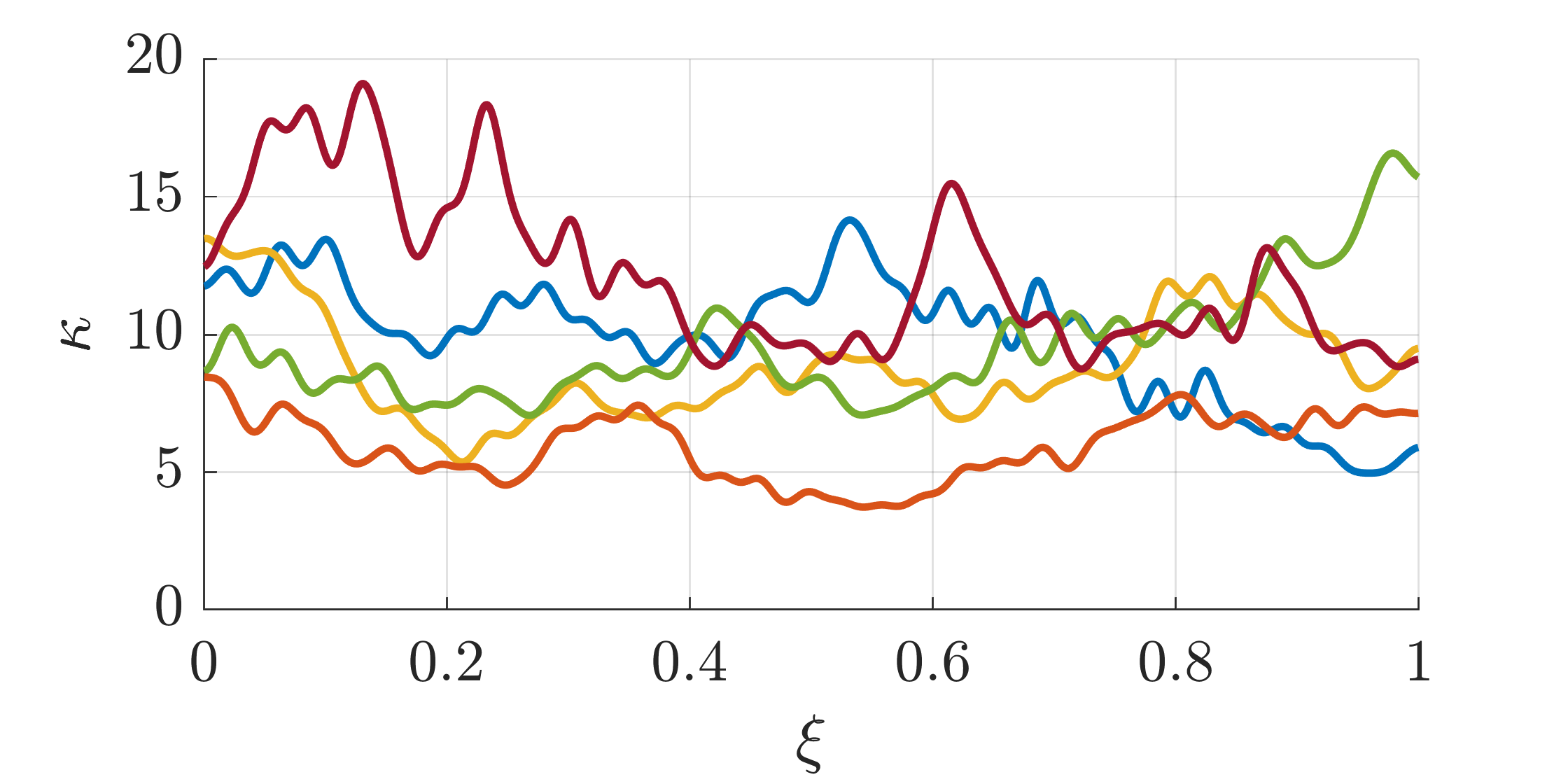}
		\end{minipage}
	}%
	\subfloat[Concentration]{
		\begin{minipage}{8cm}
			\label{fig:cs4:concentration}
			\includegraphics[width=\linewidth]{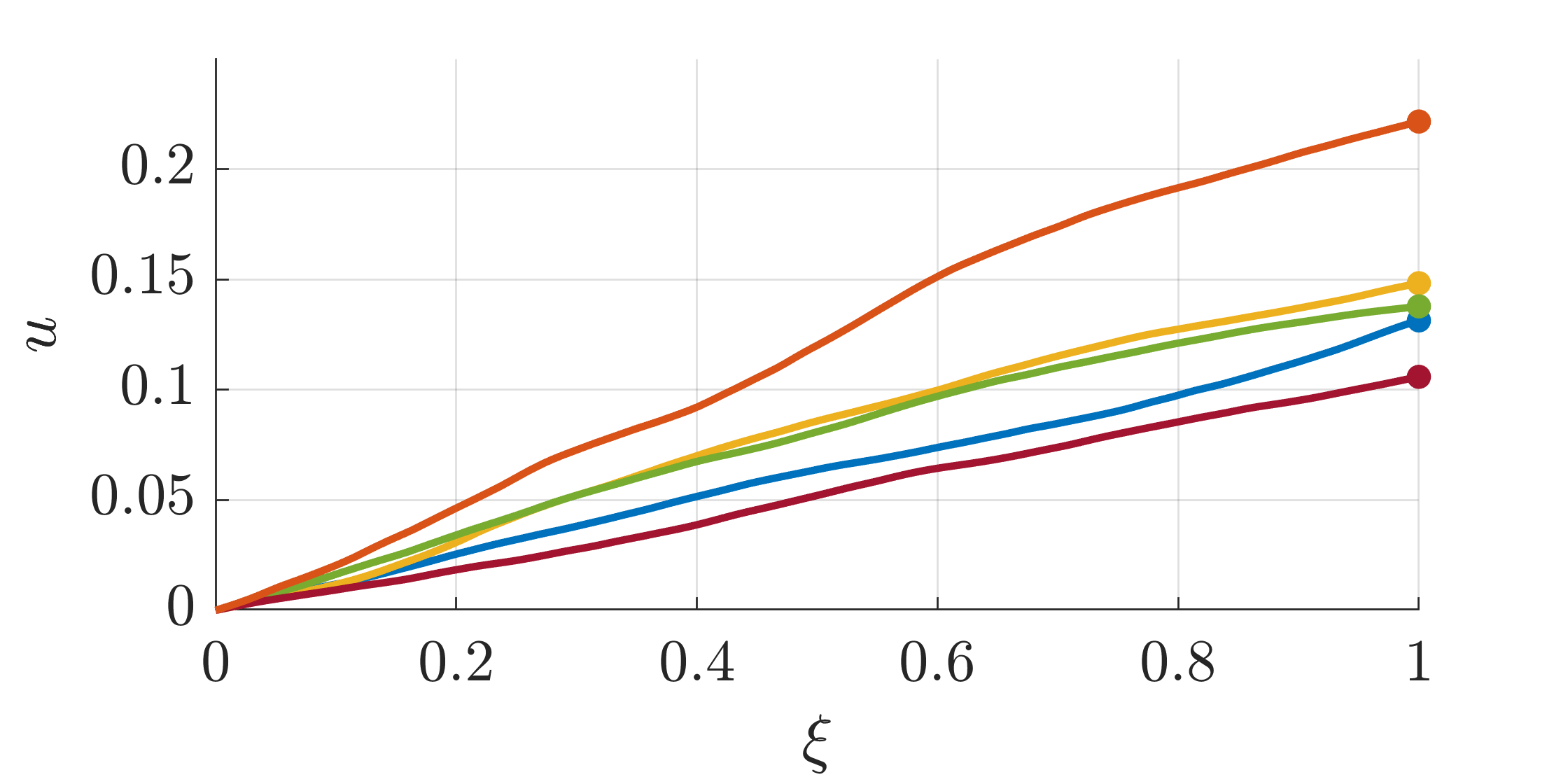}
		\end{minipage}
	}%
	\caption{\emph{High-dimensional diffusion problem}: $5$ 
		independent realizations of $\BParams$ and the resulting $\kappa$ and $u$ 
		with the model output $u(1)$ highlighted with a circle. 
		\label{fig:cs4:setup}}
\end{figure}

The concentration field $u$ can be used to describe any steady-state diffusion 
driven process ({\em e.g.}, heat diffusion, concentration diffusion, etc.). Assume that the 
diffusion coefficient $\kappa$ is a log-normal random field 
given by $\kappa(\xi, \omega) = \exp{(10 + 3g(\xi))}$ where $g$ is a 
standard normal stationary Gaussian random field with exponential 
autocorrelation function $\rho(\xi,\xi') = \exp{(-3\abs{\xi'-\xi})}$. Let $g$ be 
approximated through a \emph{truncated Karhunen-Lo\`{e}ve 
	expansion}
\begin{equation}
	g(\xi) \approx \sum_{k=1}^M X_ke_k(\xi),
\end{equation}
with the pairwise uncorrelated random variables $X_k$ denoting the \emph{field 
	coefficients} and the 
real valued function $e_k$ obtained from the solution of the Fredholm equation 
for $\rho$ \citep{Ghanembook1991}. The truncation variable is set to $M=62$ to 
explain $99\%$ of the variance. Some realizations of the random field and 
resulting concentrations are shown in Figure~\ref{fig:cs4:setup}.

In this example, the random vector of coefficients 
$\BParams=(X_1,\dots,X_{62})$ shall be 
inferred using a single measurement of the diffusion field at $u(\xi=1)$ 
given by $\Bdata=0.16$. The 
considered model therefore takes as an input a realization of that random 
vector, and returns the diffusion field at $\xi=1$, {\em i.e.}, 
$\cm:\Bparams\mapsto u(1)$. It is expected that due to the single measurement location at $\xi=1$, only very little information about the parameters will be recovered in the inverse problem.

We impose a standard normal prior on the field coefficients such that $\Bprior 
= \prod_{i=1}^{62}\cn(x_i\vert0,1)$ and assume the standard additive 
discrepancy model with known discrepancy variance $\sigma^2=10^{-6}$. This 
yields the likelihood function
\begin{equation}
	\label{eq:ex4:likelihood}
	\cl(\Bparams;\Bdata) = \frac{1}{2\pi\sigma^2}\exp 
	\left(-\frac{1}{2\sigma^2}\left(0.16-\cm(\Bparams)\right)^2\right).
\end{equation}

We proceed to compare the performance of standard SLE, non-adaptive SSLE and the 
proposed adSSLE approach on this example. We solve the problem with a set 
of maximum likelihood evaluations $\Ntot=\{2{,}000; 4{,}000; 6{,}000; 8{,}000; 
10{,}000\}$. 

In the present high-dimensional case, it is necessary to set $\NRefine$ to a \emph{relatively} large number ($\NRefine=2{,}000$). At smaller $\NRefine$ numbers the variance of the estimator of $E_{\mathrm{split}}$ in Eq.~\eqref{eq:SSE:algorithmMod:splitError} makes it difficult for the algorithm to correctly identify the splitting direction that maximizes Eq.~\eqref{eq:SSE:algorithmMod:splitDirection}.

To compare the results of the algorithms, they are compared to a reference 
MCMC solution obtained with the affine-invariant ensemble sampler
\citep{MCMC:Goodman2010} 
algorithm with $100{,}000$ steps and $100$ parallel chains at a total cost of 
$10^7$ likelihood evaluations.

\begin{figure}
	\centering
	\begin{minipage}{14cm}
		\includegraphics[width=\linewidth]{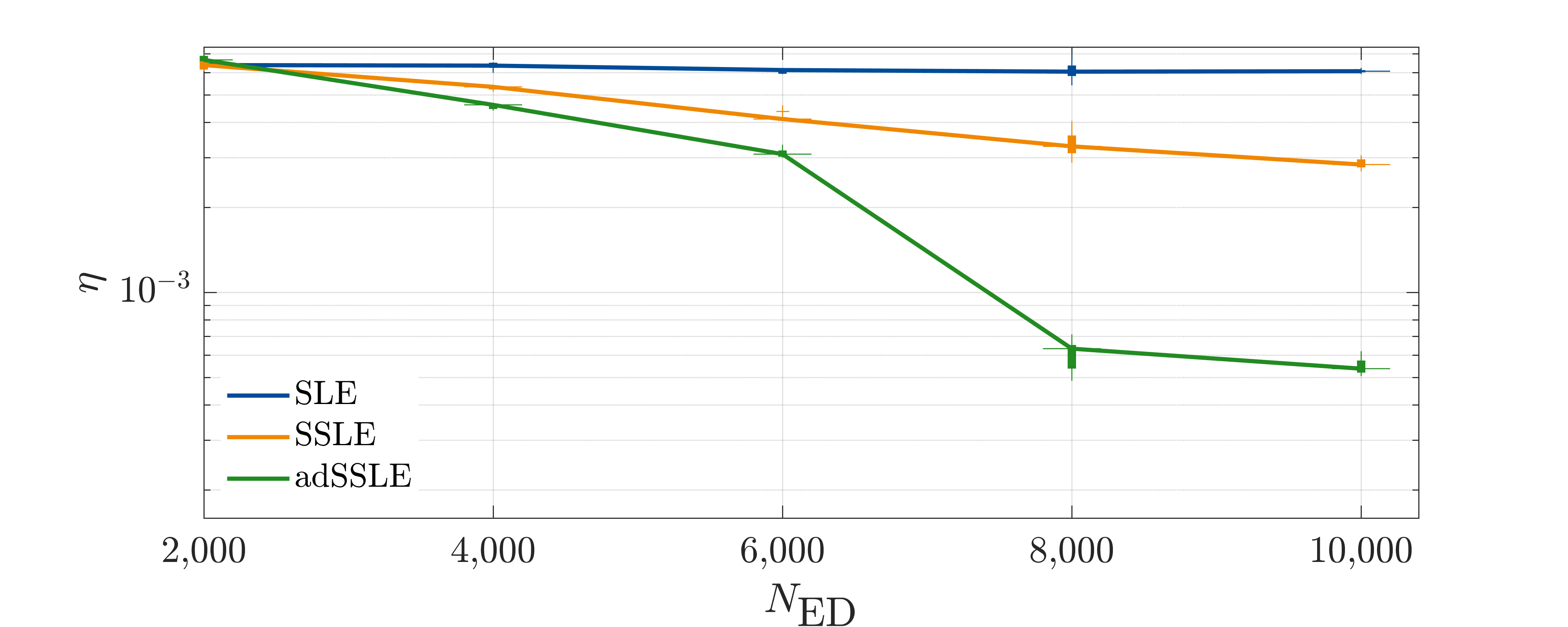}
	\end{minipage}
	\caption{\emph{High-dimensional diffusion problem}: Convergence of the $\eta$ error measure 
		(Eq.~\eqref{eq:eq3:errorMeasure}) over five replications for SLE, SSLE with 
		a static experimental design and the proposed adSSLE approach.}%
	\label{fig:cs3:Convergence}%
\end{figure}

To allow a quantitative comparison, we again use the error measure from 
Eq.~\eqref{eq:eq3:errorMeasure} with $M=62$. It is plotted for a set of maximum 
likelihood evaluations in Figure~\ref{fig:cs3:Convergence}. It is clear that 
both SSLE algorithms outperform SLE, while the adSSLE approach manages to 
improve the performance of SLE by an order of magnitude. 

The overall small magnitude of the error $\eta$ in Figure~\ref{fig:cs3:Convergence} can be attributed to the low \emph{active dimensionality} of this problem. 
Despite its high nominal dimensionality ($M=62$), this problem in fact has only very few active dimensions, as the first few variables are significantly more important than the rest. In physical terms, very local fluctuations of the conductivity do not influence the output $u(1)$ which results from an integration of these fluctuations.
Therefore, the biggest change between prior and posterior distribution happens in the first few parameters (see also Table~\ref{tab:cs4:Marginal}), while the other parameters remain unchanged by the updating procedure. 
This results in a small value of the Jensen-Shannon divergence for the inactive dimensions that lower the average value $\eta$ as defined in Eq.~\eqref{eq:eq3:errorMeasure}. 

\begin{figure}
	\centering
	\subfloat[adSSLE]{
		\begin{minipage}{8cm}
			\includegraphics[width=\linewidth]{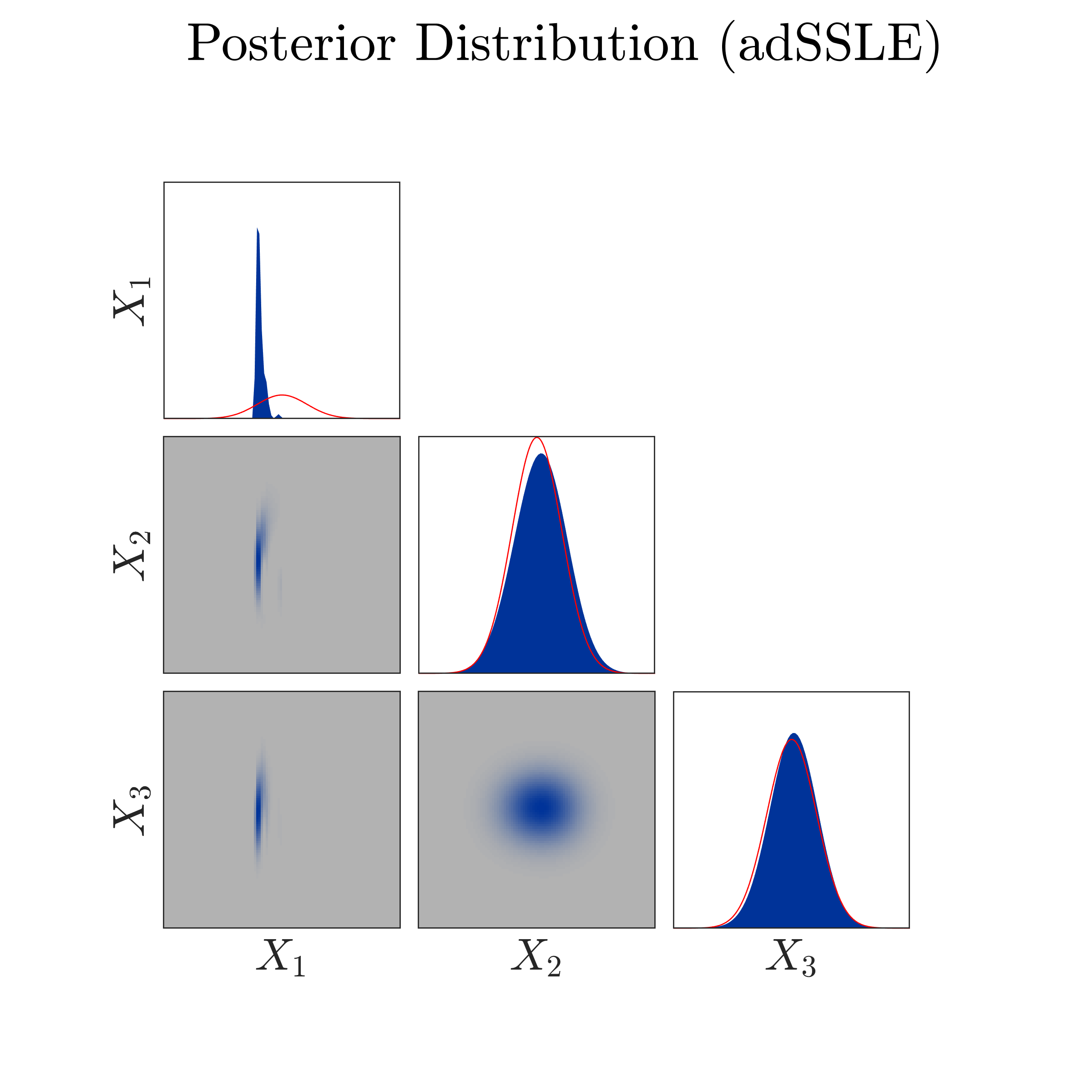}
		\end{minipage}
	}%
	\subfloat[MCMC]{
		\begin{minipage}{8cm}
			\includegraphics[width=\linewidth]{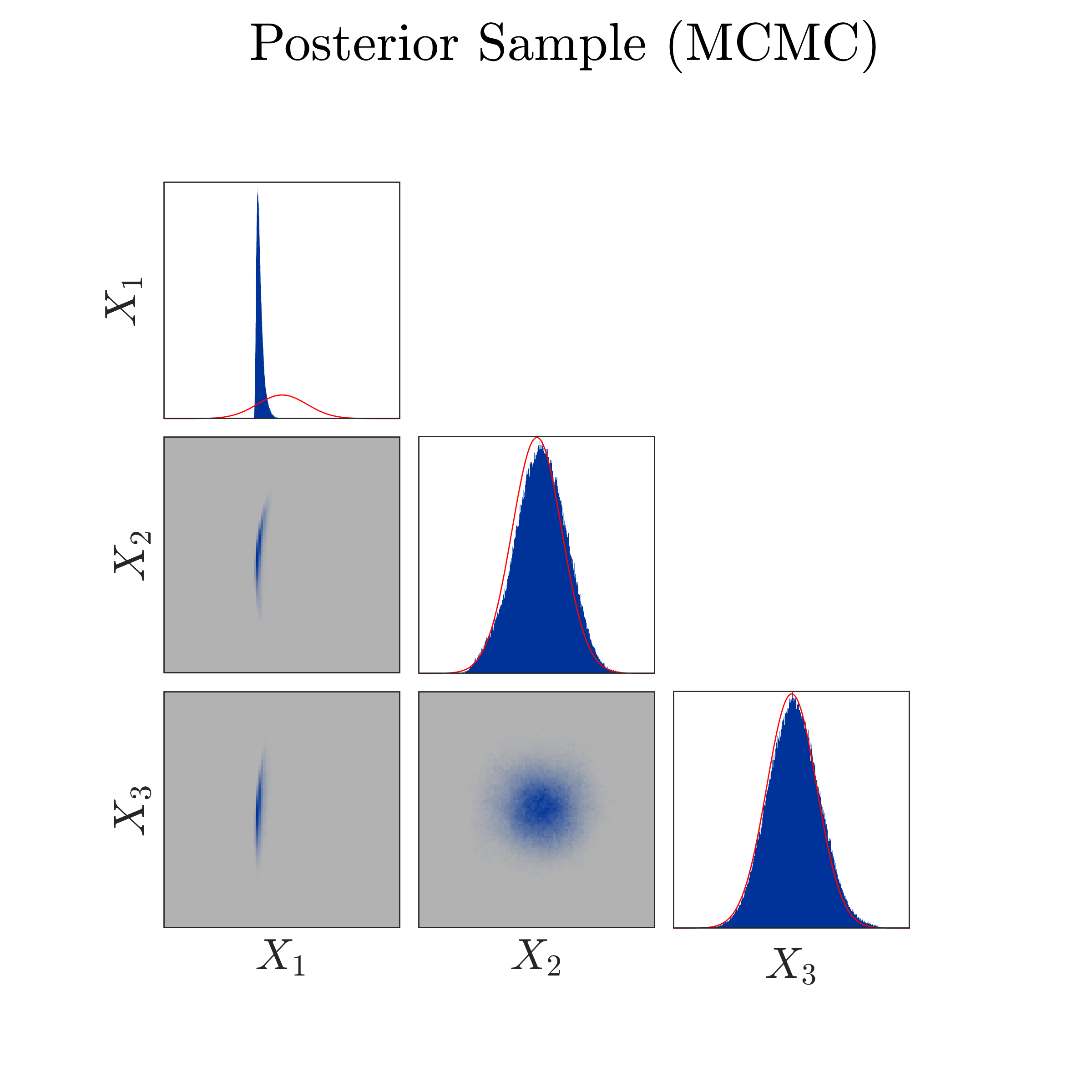}
		\end{minipage}
	}%
	\caption{\emph{High-dimensional diffusion problem}: Comparison plots of the posterior distribution marginals for the 
		first $3$ parameters $\{X_1, X_2, X_3\}$ computed 
		from adSSLE ($\Ntot=10{,}000$) and MCMC ($\Ntot = 10^7$). The prior marginals are shown in red.}%
	\label{fig:cs4:Posterior}%
\end{figure}

To highlight the results of one adSSLE algorithm instance with 
$\Ntot=10{,}000$, we 
display plots of the marginal posteriors in Figure~\ref{fig:cs4:Posterior}. Due 
to the low active dimensionality of the 
problem, we focus on the first $3$ parameters $\{X_1, X_2, X_3\}$. The 
remaining posterior 
parameters are not significantly influenced by the considered data. The 
comparative plots show a good 
agreement between the adSSLE and the reference solution, especially {\em w.r.t.} the interaction between $X_1$ and $\{X_2,X_3\}$. As can be expected from the single measurement location, only the first parameter $X_1$ is significantly influenced by the Bayesian updating procedure.

For the same instance, we also compute the first two posterior 
moments for all posterior marginals and compare them to the MCMC reference 
solution. The resulting values are presented in Table~\ref{tab:cs4:Marginal}.  
Keeping in mind that the prior distribution is a multivariate standard normal 
distribution ($\Esp{X_i}=0$ and $\sqrt{\Var{X_i\vert\cy}}=1$ for 
$i=1,\dots,62$) it is obvious from this table that the data most significantly 
affects the first three parameters. 

The adSSLE approach manages to accurately recover the first two posterior 
moments at the relatively low cost of $\Ntot=10{,}000$ likelihood evaluations. 
The average absolute error for $\Esp{X_i}$ and $\sqrt{\Var{X_i\vert\cy}}$ is 
approximately $0.02$.

\begin{table}
	\centering
	\caption{\emph{High-dimensional diffusion problem}: Posterior mean and standard deviation for selected marginals $X_i\vert\Bdata, i=1,\dots,6,10,20,\dots,50,62$. The values in brackets are computed from the MCMC reference 
		solution. The prior is a multivariate standard normal distribution 
		($\Esp{X_i}=0$ and $\sqrt{\Var{X_i\vert\cy}}=1$).}%
	\label{tab:cs4:Marginal}%
	\footnotesize
	\begin{tabular}{rccccccc}
		\hline &$X_{1}$&$X_{2}$&$X_{3}$&$X_{4}$&$X_{5}$&$X_{6}$\\
		\hline $\Esp{X_i\vert\cy}$&$-0.88 (-0.89)$&$0.14 (0.14)$&$0.07 (0.09)$&$0.04 (-0.03)$&$-0.04 (-0.00)$&$-0.04 (0.03)$\\
		$\sqrt{\Var{X_i\vert\cy}}$&$0.16 (0.14)$&$1.07 (1.04)$&$0.97 (1.03)$&$1.02 (1.00)$&$1.00 (1.00)$&$0.98 (1.03)$\\
		\hline &$X_{10}$&$X_{20}$&$X_{30}$&$X_{40}$&$X_{50}$&$X_{62}$\\
		\hline $\Esp{X_i\vert\cy}$&$0.00 (0.03)$&$0.00 (0.00)$&$-0.02 (-0.01)$&$-0.03 (-0.01)$&$-0.00 (0.01)$&$0.00 (-0.03)$\\
		$\sqrt{\Var{X_i\vert\cy}}$&$1.03 (1.01)$&$1.03 (0.98)$&$1.00 (1.00)$&$1.04 (1.02)$&$1.00 (0.99)$&$1.00 (0.98)$\\
		\hline 
	\end{tabular}
\end{table}
\section{Conclusions}

Motivated by the recently proposed spectral likelihood expansions (SLE) framework for Bayesian model inversion
presented in \citet{NagelJCP2016}, we showed that the same analytical 
post-processing capabilities can be derived when the novel SSE approach from \citet{Marelli2020} is applied to likelihood functions, giving rise to the proposed SSLE approach. 
Because SSE is designed for models with \emph{complex local characteristics}, 
it was expected to outperform SLE on practically relevant, highly localized likelihood functions. 
To further improve SSLE performance, we 
introduced a novel adaptive sampling procedure and modified partitioning strategy.

There are a few unsolved shortcomings of SSLE that will be 
addressed in future works. Namely, the discontinuities at the subdomain boundaries may be a source of error that should be addressed. Additionally, for the adSSLE algorithm, it is not possible at the moment to specify the optimal $\NRefine$ parameter a priori. In light of the considerable influence of that parameter as shown in Section~\ref{sec:applications:ex2:nref}, it might be necessary to adaptively adjust it, or decouple this parameter from the termination criterion. 

Approximating likelihood functions through local PCEs prohibits the enforcement
of strict positivity throughout the function domain. For visualization purposes
this is not an issue, as negative predictions can simply be set to $0$ in a 
post-processing step. When computing posterior expectations with 
Eq.~\eqref{eq:SSE:SSEforBI:QoI}, however, this can lead to erroneous results 
such as negative posterior variances. One way to enforce strict positivity is 
through an initial transformation of the likelihood function ({\em e.g.}, 
log-likelihood $\log{\cl}\approx\sum_{k\in\ck}f_k^{\mathrm{PCE}}(\BParams)$). 
This is avoided in the present work because it comes at a 
loss of the desirable analytical post-processing properties.

Another class of problems that can cause instability when focusing directly on the likelihood function (both for SLE and SSLE), is that of problems with a very sharp likelihood function.  This can happen, e.g. in the case of many available measurements and small discrepancy variances, that cause the likelihood function to reduce to the Dirac delta function, which has a very dense spectral representation.

In problems with few active dimensions (e.g. High-dimensional diffusion problem, Section~\ref{sec:cs:ex3}), SSLE performs well because the likelihood is constant in many dimensions. The local sparse PCE construction can exploit this and could therefore handle up to hundreds of input dimensions. However, if the number of active dimensions is high as well, the present SSLE algorithm will require prohibitively large experimental designs, thereby rendering it unfeasible.

The biggest advantage of SSLE, however, is that it poses the challenging Bayesian computation in a function approximation setting. This yields an analytical expression of the posterior distribution and preserves the analytical post-processing capabilities of SLE while delivering highly improved likelihood function approximations. As opposed to many existing algorithms, SSLE can even efficiently solve Bayesian problems with multiple posterior modes. As shown in the case studies, the proposed adaptive algorithm further capitalizes on the compact support nature of likelihood functions and leads to significant performance gains, especially at larger experimental designs. 


\section*{Acknowledgements}
The PhD thesis of the first author is supported by ETH grant \#44 17-1.

\bibliographystyle{chicago}
\bibliography{Biblio}
\end{document}